\documentclass[showpacs,eqsecnum,preprintnumbers]{revtex4}
\usepackage{amsmath}

\begin{document}

\title{Baryon wave function in large-$N_{c}$ QCD: Universality, nonlinear
evolution equation and asymptotic limit}
\author{P.V. Pobylitsa}
\affiliation{Institute for Theoretical Physics II, Ruhr University Bochum, D-44780
Bochum, Germany\\ {\em and} \\
Petersburg Nuclear Physics Institute, Gatchina, St. Petersburg, 188300,
Russia}

\begin{abstract}
The $1/N_{c}$ expansion is formulated for the baryon wave function in terms
of a specially constructed generating functional. The leading order of this
$1/N_{c}$ expansion is universal for all low-lying baryons [including the
$O(N_{c}^{-1})$ and $O(N_{c}^{0})$ excited resonances] and for baryon-meson
scattering states. A nonlinear evolution equation of Hamilton-Jacobi type is
derived for the generating functional describing the baryon distribution
amplitude in the large-$N_{c}$ limit. In the asymptotic regime this
nonlinear equation is solved analytically. The anomalous dimensions of the
leading-twist baryon operators diagonalizing the evolution are computed
analytically up to the next-to-leading order of the $1/N_{c}$ expansion.
\end{abstract}

\pacs{11.15.Pg, 11.10.Hi, 12.38.Lg, 14.20.-c}
\maketitle

\section{Introduction}

The limit of the large number of colors $N_{c}\rightarrow\infty$ and the
$1/N_{c}$ expansion in QCD are powerful methods of
theoretical analysis which provide us not only with a qualitative
understanding of nonperturbative phenomena of QCD but also sometimes lead to
interesting quantitative results. The history of the applications of the
$1/N_{c}$ expansion in QCD is almost as old
\cite{Hooft-74,Coleman-85,Witten-79-80,Witten-79}
as QCD itself and the roots of this method in statistical and many-body
physics are even deeper.

In spite of this long history some quantities important for the
phenomenological applications of QCD have for a long time remained out of
the scope of the large-$N_{c}$ community. In this paper we construct the $1/N_{c}$ expansion
for the baryon wave function. The light-cone version of the baryon wave
function (distribution amplitude) plays a crucial role in the QCD analysis
of hard exclusive phenomena \cite{LB-79,LB-80,CZ-84,BL-89}.

Apart from the importance for practical applications, the baryon
wave function is a quantity with a very interesting structure of the
$1/N_c$ expansion. The traditional methods
of the $1/N_{c}$ expansion usually deal with quantities that have a
\emph{power} large-$N_{c}$ behavior. For a vast amount of baryon properties
(masses, mass splittings, magnetic moments, form factors, structure
functions, etc.), the large-$N_{c}$ behavior can be described by a power
counting of $N_{c}$. On the contrary, the large-$N_{c}$ analysis of the
baryon wave function requires the development of methods that can be applied
to quantities which depend on $N_{c}$ \emph{exponentially}.

The main part of this paper is devoted to the light-cone case
which is especially interesting for practical applications. However, for
simplicity we prefer to start the analysis of the large-$N_{c}$ behavior
from the ``covariant quark wave function'' $\hat{\Psi}^{B}$ of a baryon $|B\rangle$,
\begin{equation}
\hat{\Psi}
_{(f_{1}s_{1})(f_{2}s_{2})\ldots(f_{N_{c}}s_{N_{c}})}
^{B}(z_{1},z_{2},\ldots
,z_{N_{c}})  
\equiv 
\frac{1}{N_{c}!}\varepsilon _{c_{1}\ldots c_{N_{c}}}\langle 0|{\cal T}\left[
q_{c_{1}f_{1}s_{1}}(z_{1})
q_{c_{2}f_{2}s_{2}}(z_{2})
\ldots
q_{c_{N_{c}}f_{N_{c}}s_{N_{c}}}(z_{N_{c}})\right] |B\rangle \,.
\label{Phi-B-def}
\end{equation}
Here $q_{cfs}(z)$ is the quark field taken at the space-time point $z$,
and $c,f,s$ are color, flavor and spin indices, respectively. The Levi-Civita
tensor $\varepsilon _{c_{1}\ldots c_{N_{c}}}$ is totally antisymmetric in
color indices. We use the notation $\hat{\Psi}$ with a hat in order to
distinguish the covariant wave function (\ref{Phi-B-def}) from the
light-cone case, which is the main subject of the paper.
Because of the time ordering $\cal T$ and Fermi statistics of quarks, the function
$\hat{\Psi}
_{(f_{1}s_{1})\ldots(f_{N_{c}}s_{N_{c}})}
^{B}(z_{1},\ldots ,z_{N_{c}})$
is symmetric with respect to permutations of $(z_{k},f_{k},s_{k})$.
Special care should be taken about the local gauge invariance. We assume an
insertion of Wilson lines on the RHS of Eq. (\ref{Phi-B-def}) in such a way
that function $\hat{\Psi}^{B}$ becomes gauge invariant but still
remains symmetric with respect to permutations of $(z_{k},f_{k},s_{k})$
(for example, we can choose straight
Wilson lines connecting each quark field with the ``center of mass'' of
points $z_{1},z_{2},\ldots ,z_{N_{c}}$ and antisymmetrize the $N_{c}$ color
indices at this center point).

Our aim is to understand the large-$N_{c}$ behavior of the baryon wave function
(\ref{Phi-B-def}). But if we try to apply the $1/N_{c}$ expansion to the
function (\ref{Phi-B-def}) directly, then we immediately meet a
problem. This function depends on $N_{c}$ arguments. From the mathematical
point of view it hardly makes sense to speak about the large-$N_{c}$
asymptotic behavior of a function depending on $N_{c}$ variables. For a
consistent treatment of the large-$N_{c}$ limit we need some intermediate
object which

1) contains the same information as the wave function
$\hat{\Psi}
_{(f_{1}s_{1})(f_{2}s_{2})\ldots(f_{N_{c}}s_{N_{c}})}
^{B}(z_{1},z_{2},\ldots,z_{N_{c}})$,

2) allows to take the large-$N_{c}$ limit and to construct the $1/N_{c}$
expansion.

To this aim we introduce the generating functional 
\begin{equation}
\hat{\Phi}_{B}(g)=\sum\limits_{f_{k}s_{k}}\int dz_{1}\int dz_{2}\ldots \int
dz_{N_{c}}g_{f_{1}s_{1}}(z_{1})g_{f_{2}s_{2}}(z_{2})\ldots
g_{f_{N_{c}}s_{N_{c}}}(z_{N_{c}})
\hat{\Psi}_{(f_{1}s_{1})(f_{2}s_{2})\ldots(f_{N_{c}}s_{N_{c}})}
^{B}(z_{1},z_{2},\ldots ,z_{N_{c}})  \label{Phi-f-1}
\end{equation}
depending on an arbitrary ``source'' function $g_{fs}(z)$. Since the function
$\hat{\Psi}^{B}$ is totally symmetric in permutations of $(z_{k},f_{k},s_{k})$,
at finite $N_{c}$ we can restore the wave function $\hat{\Psi}^{B}$ if we
know the functional $\hat{\Phi}_{B}(g)$: 
\begin{equation}
\hat{\Psi}
_{(f_{1}s_{1})(f_{2}s_{2})\ldots(f_{N_{c}}s_{N_{c}})}
^{B}(z_{1},z_{2},\ldots
,z_{N_{c}})=\frac{1}{N_{c}!}\left[ \prod\limits_{k=1}^{N_{c}}\frac{\delta }
{\delta g_{f_{k}s_{k}}(z_{k})}\right] \hat{\Phi}_{B}(g)\,.
\label{phi-phi-back}
\end{equation}
What can we say about the large-$N_{c}$ asymptotic behavior of the
functional $\hat{\Phi}_{B}(g)$? In principle, it is a dynamic problem and
without solving large-$N_{c}$ QCD we cannot give an absolutely reliable
answer. Nevertheless on general grounds it is natural to expect the behavior
of the type
\begin{equation}
\hat{\Phi}_{B}(g)=\left\{ N_{c}^{\nu _{B}}\hat{A}_{B}(g) \exp \left[ N_{c}
\hat{W}(g)\right] \right\} \left[ 1+O(N_{c}^{-1})\right] \,.  \label{W-def-0}
\end{equation}
Here the exponential behavior is controlled by the functional $\hat{W}(g)$.
The pre-exponential factor $\hat{A}_{B}(g)$ may be accompanied by a power
term $N_{c}^{\nu _{B}}$. In this paper we present a number of arguments in
favor of the large-$N_{c}$ behavior (\ref{W-def-0}). We shall also establish
some interesting properties of functionals $\hat{W}(g)$ and $\hat{A}_{B}(g)$.
In particular, we shall show that the functional $\hat{W}(g)$ is universal
for all low-lying baryons (see Sec. \ref{Universality-W-subsection} for
details).

\section{Main results}

\subsection{Notation}

In this paper we deal with the $1/N_{c}$ expansion for similar but different
versions of the baryon wave function. In all cases we introduce generating
functionals analogous to ${\hat{\Phi}}_B(g)$ (\ref{Phi-f-1}) and describe their
large-$N_{c}$ asymptotic behavior in terms of functionals similar to
$\hat{W}(g)$ and ${\hat{A}}_B(g)$ in Eq. (\ref{W-def-0}). In order to avoid confusion we
use a slightly different notation for various cases:
\begin{itemize}
\item covariant baryon wave function ${\hat{\Psi}}^B$ (\ref{Phi-B-def}) and the
associated functionals ${\hat{\Phi}}_B(g)$, $\hat{W}(g)$, ${\hat{A}}_B(g)$,
\item baryon wave function $\psi^B $ (\ref{psi-NRQM}) in a toy quark model and
functionals $\phi_B (g)$, $w(g)$, $a_B(g)$,
\item baryon light-cone wave function (distribution amplitude) $\Psi^B$ 
(\ref{BDA-def}) and functionals $\Phi_B (g)$, $W(g)$, $A_B(g)$,
\item asymptotic baryon distribution amplitude and other wave functions
${\tilde{\Psi}}^B$ diagonalizing the evolution kernel (\ref{H-Psi-Gamma-Psi}), and
functionals ${\tilde{\Phi}}_B(g)$, $W_{0}(g)$, ${\tilde{A}}_B(g)$.
\end{itemize}

On the other hand, the general structure of the $1/N_{c}$ expansion in all
these cases is almost identical. Therefore in the rest of \emph{this}
section describing the main results of the paper, we use the same notation
$\Phi_B(g)$, $W(g)$, $A_B(g)$ for all above cases.
The precise meaning of these functionals will be obvious from the context.

\subsection{General structure of the $1/N_{c}$ expansion and universality of
the functional $W(g)$}

One of the aims of this paper is to check the consistency of the large-$N_{c} $
structure (\ref{W-def-0}) of the functional $\Phi_{B}(g)$. The
traditional machinery of the large-$N_{c}$ analysis includes several
approaches:

\begin{itemize}
\item equivalence of the large-$N_{c}$ counting in QCD and in various models
imitating QCD
\cite{Hooft-74,Coleman-85,Witten-79-80,Witten-79,Bardakci-84,GS-84,Manohar-84}
(nonrelativistic quark model \cite{RG-75,KP-84,CT-86},
Skyrme model \cite{Skyrme-61,Witten-83,Adkins-83},
chiral quark-soliton model \cite{KRS-84,DP-86,DPP-88}, etc.),
\item direct counting of $N_{c}$ orders in perturbative Feynman diagrams
\cite{Hooft-74,Coleman-85,Witten-79-80,Witten-79},
\item methods based on the spin-flavor symmetry and consistency condition 
\cite{Bardakci-84,GS-84,DM-93,Jenkins-93,DJM-94}.
\end{itemize}

In Sec. \ref{NRQM-section} we compute the functional $\Phi_B(g)$ in a toy
quark model and show explicitly that the large-$N_{c}$ behavior of this
functional has the structure (\ref{W-def-0}). Using this toy model, we
describe the saddle point method which will be later applied to the analysis
of the large-$N_{c}$ behavior of the baryon distribution amplitude in QCD.

Now let us turn to the role of perturbative QCD in the analysis of the
large-$N_{c}$ behavior of the baryon wave function. \emph{Perturbative}
Feynman diagrams allow us to estimate the $N_{c}$ order of various 
\emph{nonperturbative} quantities even if the perturbation theory does not work
for these quantities. Unfortunately this method cannot be applied to
$\Phi_{B}(g)$ directly. Indeed the functional $\Phi _{B}(g)$ depends on $N_{c}$
exponentially so that arbitrarily high orders of $N_{c}$ can be met in
perturbative Feynman diagrams. Nevertheless perturbative QCD provides a
very important tool which allows us to check the large-$N_{c}$ structure of
$\Phi _{B}(g)$ described by Eq. (\ref{W-def-0}) --- we mean the evolution
equation for the baryon distribution amplitude. The main
part of this paper is devoted to the evolution of the baryon distribution
amplitudes at large $N_{c}$.

We stress that in this paper we are not interested in the physical
applications of the evolution of the baryon distribution amplitude. The
$O(N_{c})$ growth of the nucleon mass at large $N_{c}$ changes the
physical picture of hard processes in large-$N_{c}$ QCD. In fact, in
practical applications there is no need in the large-$N_{c}$ approximation
for the evolution.

Instead, we would like to use the evolution equation as a consistency test
for the large-$N_{c}$ structure (\ref{W-def-0}). In some sense the evolution equation plays
the same role in the analysis of the \emph{exponential} large-$N_{c}$
behavior as the perturbative Feynman diagrams in the analysis of quantities
with the \emph{power} large-$N_{c}$ asymptotic behavior. In physical
applications to hard processes the evolution equations are used for the
exponentiation of large logarithms of the hard momentum $\log Q$. In the
large-$N_{c}$ analysis of the baryon distribution amplitude, the evolution
equation is helpful due to its ability to exponentiate $N_{c}$.

An important property of the $1/N_{c}$ expansion  is the
universality of the functional $W(g)$: this functional is the same for all
low-lying baryons [including $O(N_{c}^{-1})$ and $O(N_{c}^{0})$ excited
resonances]. Moreover, the same functional $W(g)$ can be used for the quark
wave functions of baryon-meson scattering states. This universality property
is described in detail in Sec. \ref{Universality-W-subsection}. All methods
used for the analysis of the large-$N_{c}$ behavior in this paper (toy quark
model, evolution equations, asymptotic limit) confirm the universality of
the functional $W(g)$.

In contrast to the $W(g)$, the functional $A_{B}(g)$ appearing as a
pre-exponential factor in Eq.~(\ref{W-def-0}) depends on the baryon
(baryon-meson state) $B$. We find interesting factorization properties of
the functional $A_{B}(g)$ which are briefly described in Sections
\ref{A-factorization-section} and \ref{A-B-Nf-2-main-section}.  The complete analysis
of these factorization properties would require the techniques based on the
large-$N_{c}$ contracted $SU(2N_{F})$ spin-flavor symmetry
\cite{Bardakci-84,GS-84,DM-93,Jenkins-93,DJM-94}.
Here we do not touch upon this method, leaving it for a separate
paper~\cite{Pobylitsa-05}.

\subsection{Nonlinear evolution equation and its Hamilton-Jacobi structure}

In Sec. \ref{Baryon-DA-section} we introduce the light-cone version of the
functional $\Phi _{B}(g)$ describing the baryon distribution amplitude and
define the corresponding functionals $W(g)$ and $A_{B}(g)$ using the 
large-$N_{c}$ representation (\ref{W-def-0}). 
In Sec. \ref{Evolution-equation-section} we derive the evolution equations for $W(g)$
and $A_{B}(g)$. The resulting evolution equation (\ref{W-evolution-compact})
for the functional $W(g)$ is nonlinear. This nonlinear evolution equation
has the Hamilton-Jacobi form (Sec. \ref{Hailton-Jacobi-subsection}). The
asymptotic limit of the large normalization scale $\mu \rightarrow \infty $
is studied in Sections \ref{Asymptotic-limit-section}, 
\ref{W0-calculation-section}. In Sec. \ref{Anomalous-dimensions-section} we
solve the problem of the diagonalization of the evolution of the baryon
distribution amplitude at large $N_{c}$ and compute the corresponding
anomalous dimensions in two orders of the $1/N_{c}$ expansion.

\subsection{Conformal symmetry and integrability}

An essential part of this paper is devoted to the evolution equation and to
the problem of the diagonalization of anomalous dimensions of baryon
operators at large $N_{c}$. At $N_c=3$ this problem has attracted
much attention
\cite{Peskin-79,Kremer-80,Ohrndorf-82,Tesima-82,Nyeo-92,Stefanis-94}
and was studied with various methods ranging from straightforward
numerical attacks to the discovery of the integrability of the evolution
equation for baryons with helicity 3/2 \cite{BDM-98,BDKM-99}.

The conformal symmetry of the evolution equation is known to play an important
role in this problem \cite{LB-80,Ohrndorf-82,ER-80,Makeenko-81,BFKL-85,BB-89}.
It is well known that the large-$N_{c}$ limit is often
helpful for solving problems inspired by high-energy QCD
\cite{Lipatov-94,FK-94} where
the dynamics is determined by the \emph{gluon} sector. In these problems the
large-$N_{c}$ limit leads to the ``spin chains'' with the interaction only
between the next neighbors \cite{Lipatov-93}. The case of the \emph{quark}
distribution amplitudes of baryons is different. In this case we have
equally important pair interactions between all $N_{c}$ quarks. We derive a
variational equation for the asymptotic limit of the functional $W(g)$ and
solve this equation analytically in Sec. \ref{W0-calculation-section}.
Technically the solution is found in terms of a successfully guessed ansatz
but the role of the conformal symmetry standing behind this ansatz is rather
visible. Another problem solved analytically in this paper is the
diagonalization of the anomalous dimensions of baryon operators of leading
twist in two orders of the $1/N_{c}$ expansion [$O(N_{c})$ and $O(N_{c}^{0})$].
The calculation of these anomalous dimensions in the next-to-leading
$O(N_{c}^{0})$ order is based on the solution of rather nontrivial functional
equations. The analytical solutions of these equations are found in
Sec. \ref{Calculation-Omega-section}.

\subsection{Phenomenological applications}

On the phenomenological side one is certainly interested in the calculation
of the functionals $W(g)$ and $A_{B}(g)$ appearing in the $1/N_{c}$
expansion (\ref{W-def-0}). This challenging problem belongs to the same
class of difficulty as solving large-$N_{c}$ QCD. Although we do not
know the functionals $W(g)$ and $A_{B}(g)$, in this paper we establish
several properties which may be of interest for the phenomenological
applications. One of them is the universality of the functional $W(g)$ which
allows a unified description of different baryons and meson-baryon
scattering states.

The large-$N_{c}$ arguments are sometimes used \cite{PP-03,DP-04} for the
justification of various approximations in phenomenological models of
baryons. Our results impose certain theoretical restrictions on these
models. As was explained above, it is probably impossible to construct a
systematic $1/N_{c}$ expansion for the baryon wave function directly, without
using auxiliary objects like the generating functional (\ref{Phi-f-1}).
Therefore the models compatible with the $1/N_{c}$ expansion can access only
the weighted integrals of the baryon wave function rather than the wave
function itself. One should keep in mind that the one-to-one correspondence
(\ref{Phi-f-1}), (\ref{phi-phi-back}) between the wave function $\Psi ^{B}$
and the functional $\Phi _{B}(g)$ holds only at finite $N_{c}$. We cannot
compute the wave function $\Psi ^{B}$ at finite $N_c$
by applying the variational derivatives of Eq.
(\ref{phi-phi-back}) to the large-$N_{c}$ asymptotic expression (\ref{W-def-0})
for $\Phi _{B}(g)$.

Another theoretical restriction imposed on the phenomenological analysis by
the systematic $1/N_{c}$ expansion is the necessity to work with the
$N_{c}$-point baryon wave function [via the generating functional $\Phi_{B}(g)$].
The arguments of the $1/N_{c}$ expansion cannot be used in models explicitly
dealing with baryons ``made of three quarks''. Although one may think that
working with the 3-point baryon wave function is equivalent to a partial
resummation of higher $1/N_{c}$ corrections, the restriction to the 3-point
baryon wave functions is incompatible with the systematic
$1/N_{c}$-expansion.

Obviously the universality of the exponential large-$N_c$ behavior for
all low-lying baryons is not sufficient for serious practical
applications. One also needs a good control of the pre-exponential
non-universal functionals $A_B(g)$. The factorizable structure of these
functionals is studied in detail in Ref.~\cite{Pobylitsa-05} where an
interesting relation connecting the cases of nucleon and $\Delta$
resonance is derived.

\section{Universality of the generating functional $\hat{W}(g)$}

\label{Universality-W-subsection}

\subsection{Universality of $\hat{W}(g)$ for baryons}

As was already mentioned, the functional $\hat{W}(g)$ appearing in the
large-$N_{c}$ decomposition (\ref{W-def-0}) is universal for all low-lying
baryons. This property is rather general: it also holds in the case of the
light-cone distribution amplitude which is the main subject of the paper and
in the exactly solvable toy quark model that will be considered in 
Sec.~\ref{NRQM-section}. In order to be specific, we describe this universality
property for the case of the covariant baryon wave function (\ref{Phi-B-def})
keeping in mind that all statements made in this section are also valid in
the light-cone case.

First we must specify the baryons for which the universality of the
functional $\hat{W}(g)$ is expected. According to the standard picture of
baryons in large-$N_{c}$ QCD with $N_{f}=2$ flavors, the lowest baryons
have equal spin $J$ and isospin $T$ \cite{Witten-83,Balachandran-83}:
\begin{equation}
T=J=\left\{ 
\begin{array}{ll}
\frac{1}{2},\frac{3}{2},\ldots & \mathrm{for\,odd}\,N_{c}\,, \\ 
0,1,2,\ldots & \mathrm{for\,even}\,N_{c}\,.
\end{array}
\right.  \label{TS-sequence}
\end{equation}
The masses of these baryons have a $1/N_{c}$ suppressed splitting:
\begin{equation}
M_{T=J}=d_{1}N_{c}+d_{0}+N_{c}^{-1}\left[ d_{-1}^{(1)}
+d_{-1}^{(2)}J(J+1)\right] +O(N_{c}^{-2})\,.  \label{M-T-J}
\end{equation}
These baryons belong to the same representation of the spin-flavor symmetry
group
\cite{Bardakci-84,GS-84,DM-93,Jenkins-93,DJM-94} which becomes asymptotically exact at
large $N_{c}$. As a consequence, the functional $\hat{W}(g)$ introduced in
Eq. (\ref{W-def-0}) is the same for all low-lying baryons (\ref{TS-sequence}).
The dependence on the type of baryons $B$ appears only in the
pre-exponential factor $\hat{A}_{B}(g)$.

Moreover, at large $N_{c}$ one can also consider higher baryon resonances
with the excitation energy $O(N_{c}^{0})$. Their mass spectrum is described
by the formula
\begin{equation}
M=d_{1}N_{c}+\left( d_{0}+\sum_{i}\omega _{i}n_{i}\right)
+O(N_{c}^{-1})\,,  \label{M-harmonic}
\end{equation}
where $n_{i}=0,1,2,\ldots $ are integer numbers, and the $\omega _{i}$ are $N_{c}$
independent coefficients [we have omitted the higher $O(N_{c}^{-1})$
correction depending on additional quantum numbers]. At large $N_{c}\gg
n_{i} $ the functional $\hat{\Phi}(g)$ for these baryons also has the
structure (\ref{W-def-0}) with the same ``universal'' functional $\hat{W}(g)$
but with baryon-dependent pre-exponential functionals $\hat{A}_{B}(g)$.

\subsection{Baryon-meson scattering states}

Strictly speaking, the existence of the states (\ref{M-harmonic}) depends on
the mass thresholds which control the decays of excited baryons into lower
baryons and mesons. If these decays are possible, then the baryons
(\ref{M-harmonic}) have a parametrically large width $O(N_{c}^{0})$. In this case
instead of large-width baryons it is better to work directly with the
scattering states containing one baryon and one (or several) meson
\cite{HEHW-84,MP-85,MM-88,Mattis-89,MK-85,MB-89}. We still can use expression (\ref{M-harmonic})
for the description of the energy of these scattering states identifying
parameters $\omega_i$ with the energies of single mesons in the rest frame
of the $O(N_c)$ heavy baryon.

For these baryon-meson scattering states one can also define the wave function
(\ref{Phi-f-1}) with the large-$N_{c}$ behavior (\ref{W-def-0}). This
representation (\ref{W-def-0}) will contain the same ``universal''
functional $\hat{W}(g)$ as before (we assume that the number of mesons in
the scattering state is kept fixed in the limit of large $N_{c}$).

The spectrum of applications of the universality of the functional $\hat{W}(g)$
can be extended. For example, instead of the scattering states with one
baryon and on-shell mesons we can consider the matrix elements with
additional insertions of off-shell quark-antiquark color-singlet currents
\begin{equation}
J_{k}(y)=\sum\limits_{c=1}^{N_{c}}\bar{q}_{c}(y){\cal S}_{k}q_{c}(y)\,,
\label{J-current}
\end{equation}
where ${\cal S}_{k}$ is some spin-flavor matrix. If the number $n$ of these
insertions is kept fixed in the large-$N_{c}$ limit, then the large-$N_{c}$
behavior
\begin{gather}
\frac{1}{N_{c}!}\varepsilon _{c_{1}\ldots
c_{N_{c}}}\sum\limits_{f_{k}s_{k}}\int dz_{1}\int dz_{2}\ldots \int
dz_{N_{c}}g_{f_{1}s_{1}}(z_{1})g_{f_{2}s_{2}}(z_{2})\ldots
g_{f_{N_{c}}s_{N_{c}}}(z_{N_{c}})  \notag \\
\times \langle 0|{\cal T}\left\{ \left[ \prod_{k=1}^{n}J_{k}(y_{k})\right]
q_{c_{1}f_{1}s_{1}}(z_{1})
q_{c_{2}f_{2}s_{2}}(z_{2})
\ldots
q_{c_{N_{c}}f_{N_{c}}s_{N_{c}}}(z_{N_{c}})\right\} |B\rangle 
\notag \\
=
N_c^{\nu_{B,\{J_{k}\}}}
\hat{A}
_{B,\{J_{k}(y_{k})\}}(g)\exp \left[ N_{c}\hat{W}(g)\right]  \label{WF-J-k}
\end{gather}
is described by Eq. (\ref{W-def-0}) with the same functional $\hat{W}(g)$ as
in the case of the baryon wave function.

Let us consider QCD with the exact $SU(2)$ flavor invariance. At finite
$N_{c}$ the functional ${\hat{\Phi}}_B(g)$ is invariant if we simultaneously
perform a flavor rotation of the quark fields and of the baryon state. But
due to the universality of the functional $\hat{W}(g)$ the flavor rotation
of the baryon state is not needed. Thus we conclude that functional $\hat{W}(g)$
is invariant under flavor rotations of $g$. Similarly, the universality
of $\hat{W}(g)$ for baryons with different polarizations leads to the
invariance of $\hat{W}(g)$ with respect to the spacial rotations of $g$
[including both the rotation of the space argument $\mathbf{z}$
and the transformation of the spin index $s$ of $g_{fs}(z^{0},\mathbf{z})$].

\subsection{Universality and $SU(N_{f})$ symmetry}

Note that the universality of $W(g)$ is based on certain assumptions about
the large-$N_{c}$ behavior of the quantum numbers of baryons. For example,
in the case of the $SU(N_{f})$ flavor symmetry with $N_{f}=2$, the
functional $\hat{W}(g)$ is universal only for those baryons whose spin $J$
and isospin $T$ do not grow with $N_{c}$:
\begin{equation}
I,J=O(N_{c}^{0})\,.
\end{equation}
In the case of $N_{f}=3$ we must also keep the strangeness $S$ bounded in
the limit $N_{c}\rightarrow \infty $
\begin{equation}
S=O(N_{c}^{0})\,,
\end{equation}
if we want to have the universality of $\hat{W}(g)$.

\subsection{Factorization of the pre-exponential factors}
\label{A-factorization-section}

In contrast to the universality of the $\hat{W}(g)$, the functional
$\hat{A}_{B,\{J_{k}(y_{k})\}}(g)$ depends both on the type of the baryon and on the
currents $J_{k}(y_{k})$. An important property of the functional
$\hat{A}_{B,\{J_{k}(y_{k})\}}(g)$ is its factorization. In an oversimplified
form this factorization is
\begin{equation}
\hat{A}_{B,\{J_{k}(y_{k})\}}(g)=\hat{A}_{0}(g)\prod_{i}\left[ \hat{\xi}
_{i}(g)\right] ^{n_{i}}\prod_{k}\left[ \hat{\eta}_{J_{k}}(g)\right] \,,
\label{A-factorization}
\end{equation}
where the functional $\hat{A}_{0}(g)$ corresponds to the wave function of
the lowest baryon, the factors $\hat{\xi}_{i}(g)$ are associated with the
elementary $\omega _{i}$ excitations in Eq. (\ref{M-harmonic}), and the
factors $\hat{\eta}_{J_{k}}(g)$ correspond to the insertions of currents
$J_{k}$ in Eq. (\ref{WF-J-k}). The precise expression for
$\hat{A}_{B,\{J_{k}(y_{k})\}}(g)$ must also contain functional factors 
which come from the
zero modes corresponding to the spin-flavor rotations.
The role of zero modes is discussed in Sec. \ref{Zero-modes-section} and
in Ref. \cite{Pobylitsa-05}.

\section{Toy quark model}

\label{NRQM-section}

\subsection{Model}

We want to compute the analog $\phi_{B}(g)$ of the functional $\Phi_{B}(g)$
(\ref{Phi-f-1}) in a simple toy model and to show explicitly how the
large-$N_{c}$ asymptotic behavior (\ref{W-def-0}) appears. This can be done in two
ways. First we compute the functional $\phi_{B}(g)$ exactly and take the
large-$N_{c}$ limit of this exact result. In
Sec. \ref{Saddle-point-NRQM-section} we describe another approach based on the saddle
point method. The results obtained in the toy model will also be used in
Sec. \ref{Asymptotic-solution-method-2} where we study the asymptotic
solutions of the nonlinear evolution equation.

Our toy quark model deals with baryons made of $N_{c}$ quarks. We assume
that quarks with flavor $f$, spin $s$ and color $c$ are pinned at one point
so that the spacial motion is ignored. We consider the case of $N_{f}=2$
flavors.

In this simple model, the spin $J$ of a baryon coincides with its isospin $T$
\begin{equation}
T=J=\left\{ 
\begin{array}{ll}
\frac{1}{2},\frac{3}{2},\ldots ,\frac{N_{c}}{2} & \mathrm{for\,odd}\,N_{c}\,,
\\ 
0,1,2,\ldots ,\frac{N_{c}}{2} & \mathrm{for\,even}\,N_{c}\,,
\end{array}
\right.
\end{equation}
and the quark spin-flavor wave function of a baryon can be written in the
form
\begin{equation}
\psi _{(f_{1}s_{1})(f_{2}s_{2})\ldots
(f_{N_{c}}s_{N_{c}})}^{TT_{3}J_{3}}=c_{TN_{c}}\int
dRD_{J_{3}T_{3}}^{T}(R^{-1})\prod\limits_{k=1}^{N_{c}}R_{f_{k}s_{k}}\,.
\label{psi-NRQM}
\end{equation}
Here $D_{mm^{\prime }}^{j}(R)$ are Wigner functions for the group $SU(2)$.
The integral runs over the $SU(2)$ matrices $R$ with the Haar measure
normalized by the condition
\begin{equation}
\int dR=1\,.
\end{equation}
The normalization coefficient $c_{TN_{c}}$ is chosen in Eq.~(\ref{psi-NRQM})
so that
\begin{equation}
\sum_{f_{k}s_{k}}\left| \psi _{(f_{1}s_{1})(f_{2}s_{2})\ldots
(f_{N_{c}}s_{N_{c}})}^{TT_{3}J_{3}}\right| ^{2}=1\,.
\label{psi-normalization}
\end{equation}
This normalization constant is computed in Appendix \ref{NRQM-appendix} [see
Eq. (\ref{c-T-res-0})]
\begin{equation}
c_{TN_{c}}=\sqrt{\frac{2T+1}{N_{c}!}\left( \frac{N_{c}}{2}+T+1\right)
!\left( \frac{N_{c}}{2}-T\right) !}\,\,.  \label{c-T-res-again}
\end{equation}

\subsection{Generating function $\protect\phi_{TT_{3}J_{3}}(g)$}

By analogy with Eq. (\ref{Phi-f-1}) we introduce the generating function
\begin{align}
\phi_{TT_{3}J_{3}}(g) & =\sum_{f_{k}s_{k}}g_{f_{1}s_{1}}\ldots
g_{f_{N_{c}}s_{N_{c}}}\psi_{(f_{1}s_{1})
\ldots(f_{N_{c}}s_{N_{c}})}^{TT_{3}J_{3}}  \notag \\
& =c_{TN_{c}}\int dRD_{J_{3}T_{3}}^{T}(R^{-1})\left[ \mathrm{Tr}\left( Rg^{
\mathrm{tr}}\right) \right] ^{N_{c}}\,,  \label{Phi-start-0}
\end{align}
where $g_{fs}$ is an arbitrary matrix. The superscript tr in $g^{\mathrm{tr}} $
stands for the matrix transposition.

In Appendix \ref{NRQM-appendix} we compute $\phi(g)$ at finite $N_{c}$ [see
Eq. (\ref{Phi-NRQM-res-0})]
\begin{equation}
\phi_{TT_{3}J_{3}}(g)=\frac{2T+1}{c_{TN_{c}}}
\left( \det g\right) ^{N_{c}/2}D_{T_{3}J_{3}}^{T}\left[ \frac{g}{\left( \det g\right) ^{1/2}}
\right] \,.  \label{Phi-NRQM-res}
\end{equation}
Here $g\left( \det g\right) ^{-1/2}$ is an $SL(2,C)$ matrix. Therefore the
Wigner function $D_{T_{3}J_{3}}^{T}$ should be understood in the sense of
the complexification of $SU(2)$ to $SL(2,C)$.

At large $N_{c}$ we find from Eq. (\ref{c-T-res-again})
\begin{equation}
c_{TN_{c}}=2^{-N_{c}/2}\sqrt{2T+1}\left( \pi N_{c}^{3}/8\right) ^{1/4}\quad
(N_{c}\gg T)\,.  \label{CT-Nc-Large-Nc}
\end{equation}
The large-$N_{c}$ asymptotic behavior of $\phi _{TT_{3}J_{3}}(g)$ has the
standard form (\ref{W-def-0})
\begin{equation}
\phi _{TT_{3}J_{3}}(g)=\left\{ N_{c}^{\nu _{B}}a_{TT_{3}J_{3}}(g)\exp \left[
N_{c}w(g)\right] \right\} \left[ 1+O(N_{c}^{-1})\right]
\end{equation}
with
\begin{equation}
w(g)=\frac{1}{2}\ln \left( 2\det g\right) \,,  \label{W-NRQM-res}
\end{equation}
\begin{equation}
\nu _{B}=-3/4\,,
\label{nu-B-toy}
\end{equation}
\begin{equation}
a_{TT_{3}J_{3}}(g)=\left( \frac{8}{\pi }\right) ^{1/4}\sqrt{2T+1}
D_{T_{3}J_{3}}^{T}\left[ \frac{g}{\left( \det g\right) ^{1/2}}\right] \,.
\label{A-NRQM-res}
\end{equation}
We see from Eq. (\ref{W-NRQM-res}) that function $w(g)$ is independent of
$T=J,T_{3},J_{3}$. This is a manifestation of the ``universality'' property
of $w(g)$ which was discussed in Sec. \ref{Universality-W-subsection}.

\subsection{Saddle point method}

\label{Saddle-point-NRQM-section}

Although the above results (\ref{W-NRQM-res}) and (\ref{A-NRQM-res}) for
$w(g)$ and $a_{TT_{3}J_{3}}(g)$ can be read directly from the exact
finite-$N_{c}$ expression (\ref{Phi-NRQM-res}), it is instructive to derive these
results using the large-$N_{c}$ limit from the very beginning. At large
$N_{c}$ we can apply the saddle point method to the calculation of the
integral on the RHS of Eq. (\ref{Phi-start-0}). Indeed, the integrand
contains the factor $\left[ \mathrm{Tr}\left( Rg^{\mathrm{tr}}\right) \right]
^{N_{c}}$ which leads to the saddle point equation
\begin{equation}
\delta_{R}\mathrm{Tr}\left( Rg^{\mathrm{tr}}\right) =0\,.  \label{NRQM-SP-eq}
\end{equation}
The original integral runs in Eq.~(\ref{Phi-start-0}) over the $SU(2)$ matrices $R$. But the saddle point
method leads to a deformation of the ``integration contour'' so that we
must look for solutions $R$ and their variation $\delta_{R}$ in the
complexification of $SU(2)$, i.e. in $SL(2,C)$. Taking the variation
$\delta_{R}$ in the form of an infinitesimal $SL(2,C)$ rotation
\begin{equation}
R^{\prime}=(1+i\delta\omega^{a}\tau^{a})R
\end{equation}
with arbitrary complex infinitesimal parameters $\delta\omega^{a}$, we can
rewrite the saddle point equation (\ref{NRQM-SP-eq}) in the form
\begin{equation}
\sum\limits_{a}\delta\omega^{a}\mathrm{Tr}
\left( \tau^{a}Rg^{\mathrm{tr}}\right) =0\,.
\end{equation}
Thus
\begin{equation}
\mathrm{Tr}\left( \tau^{a}Rg^{\mathrm{tr}}\right) =0\,.
\end{equation}
We see that
\begin{equation}
\left( Rg^{\mathrm{tr}}\right) _{ij}=c\delta_{ij}\,,  \label{R-g-c}
\end{equation}
where $c$ is some complex number. Taking the determinant of this equation,
we find
\begin{equation}
\left( \det R\right) \left( \det g\right) =c^{2}\,.
\end{equation}
Since $R$ belongs to $SL(2,C)$, we have
\begin{equation}
\det R=1\,.
\end{equation}
Hence
\begin{equation}
\det g=c^{2}\,.
\end{equation}
Now we find two saddle points from Eq. (\ref{R-g-c}):
\begin{equation}
R_{\pm }^{-1}=\pm \frac{g^{\mathrm{tr}}}{\left( \det g\right) ^{1/2}}\,.
\label{R-via-g-saddle-point}
\end{equation}
The calculation of the Jacobian of fluctuations around these saddle points
leads to the following general result
\begin{equation}
\int dRf(R)\left[ \mathrm{Tr}\left( Rg^{\mathrm{tr}}\right) \right] ^{N_{c}}
\overset{N_{c}\rightarrow \infty }{\longrightarrow}\sum\limits_{\pm }\frac{2}{\sqrt{2\pi
N_{c}^{3}}}\left[ \mathrm{Tr}\left( R_{\pm }g^{\mathrm{tr}}\right) \right]
^{N_{c}}f(R_{\pm })\,.  \label{R-int}
\end{equation}
Inserting expression (\ref{R-via-g-saddle-point}) for $R_{\pm }$ and taking
$f(R)=$ $D_{J_{3}T_{3}}^{T}(R^{-1})$, we derive from (\ref{R-int})
\begin{equation}
\int dRD_{J_{3}T_{3}}^{T}(R^{-1})\left[ \mathrm{Tr}\left( Rg^{\mathrm{tr}
}\right) \right] ^{N_{c}}\overset{N_{c}\rightarrow \infty }{\longrightarrow}\,\frac{4}{
\sqrt{2\pi N_{c}^{3}}}\left[ 2\left( \det g\right) ^{1/2}\right]
^{N_{c}}D_{J_{3}T_{3}}^{T}\left[ \frac{g^{\mathrm{tr}}}{\left( \det g\right)
^{1/2}}\right] \,.
\end{equation}
Combining this with the asymptotic expression (\ref{CT-Nc-Large-Nc}) for
$c_{TN_{c}}$ and using the property of Wigner functions (\ref{D-R-transposed}),
we find the saddle point result for the integral (\ref{Phi-start-0}):
\begin{equation}
c_{TN_{c}}\int dRD_{J_{3}T_{3}}^{T}(R^{-1})\left[ \mathrm{Tr}\left( Rg^{
\mathrm{tr}}\right) \right] ^{N_{c}}\overset{N_{c}\rightarrow \infty }{\longrightarrow}\,
\sqrt{2T+1}\left( \frac{8}{\pi N_{c}^{3}}\right) ^{1/4}\left( 2\det g\right)
^{N_{c}/2}D_{T_{3}J_{3}}^{T}\left[ \frac{g}{\left( \det
g\right) ^{1/2}}\right] \,.
\end{equation}
We see that we have reproduced the above results (\ref{W-NRQM-res}),
(\ref{A-NRQM-res}) for $w(g)$ and $a_{TT_{3}J_{3}}(g)$.

\section{Generating functional for the baryon distribution amplitude}
\subsection{Baryon distribution amplitude}

\label{Baryon-DA-section}

The definition of the baryon distribution amplitude uses an auxiliary light-cone
vector  $n$. This vector $n$ can be used to impose light-cone gauge
\begin{equation}
(n\cdot A)=0\,.
\end{equation}
We are interested in the leading-twist baryon distribution amplitude which
is determined in terms of the ``good'' components of the quark field $q$,
\begin{equation}
(n\cdot \gamma )q_{cf}(\lambda n)\,.
\end{equation}
In order to separate the ``good'' components we 
introduce Dirac spinors $u_{s}$ associated with the vector $n$,
\begin{equation}
u_{s}\otimes \bar{u}_{s}=\frac{1}{2}(n\cdot \gamma )(1-2s\gamma _{5})\quad
\left( s=\pm \frac{1}{2}\right)\,,  \label{u-n-density}
\end{equation}
and define the fields
\begin{equation}
\chi _{cfs}(x)=(nP)^{1/2}\int_{-\infty }^{\infty }\frac{d\lambda }{2\pi }
\bar{u}_{-s}q_{cf}(\lambda n)\exp \left[ i\lambda x(nP)\right] \,.
\label{chi-psi}
\end{equation}
Now we can define the baryon distribution amplitude as the transition matrix
element between the vacuum and the baryon $B$ with momentum $P$:
\begin{align}
& \Psi _{(f_{1}s_{1})(f_{2}s_{2})\ldots
(f_{N_{c}}s_{N_{c}})}(x_{1},x_{2},\ldots ,x_{N_{c}})  \notag \\
& =\frac{1}{N_{c}!}\varepsilon _{c_{1}c_{2}\ldots c_{N_{c}}}\langle 0|\chi
_{c_{1}f_{1}s_{1}}(x_{1})\chi _{c_{2}f_{2}s_{2}}(x_{2})\ldots \chi
_{c_{N_{c}}f_{N_{c}}s_{N_{c}}}(x_{N_{c}})|B(P)\rangle \,.  
\label{BDA-def}
\end{align}
This definition of the distribution amplitude $\Psi $ is independent of the vector $n$.
Usually it is convenient to normalize $n$ by the condition $(nP)=1$
but we do not impose this constraint.
Indeed, at large $N_{c}$ we have $P=O(N_{c})$. We keep $n$ fixed at
large $N_{c}$, therefore $(nP)$ grows as $O(N_{c})$:
\begin{equation}
n=O(N_{c}^{0})\,,\quad (nP)=O(N_{c})\,.
\end{equation}

At large $N_{c}$ the distribution amplitude is concentrated at 
\begin{equation}
x_{k}\sim1/N_{c}\,.
\end{equation}
Therefore it is convenient to introduce new variables
\begin{equation}
y_{k}=N_{c}x_{k}
\end{equation}
which behave as $O(N_{c}^{0})$.

\subsection{Generating functional}

Now we define
\begin{align}
\Phi(g) &
=N_{c}^{-N_{c}/2}\int_{0}^{\infty}dy_{1}\int_{0}^{
\infty}dy_{2}\ldots\int_{0}^{
\infty}dy_{N_{c}}g_{f_{1}s_{1}}(y_{1})g_{f_{2}s_{2}}(y_{2})\ldots
g_{f_{N_{c}}s_{N_{c}}}(y_{N_{c}})  \notag \\
& \times\Psi_{(f_{1}s_{1})(f_{2}s_{2})\ldots(f_{N_{c}}s_{N_{c}})}\left( 
\frac{y_{1}}{N_{c}},\frac{y_{2}}{N_{c}},\ldots,\frac{y_{N_{c}}}{N_{c}}
\right) \,.  \label{Phi-def-1}
\end{align}
The factor $N_{c}^{-N_{c}/2}$ is inserted here in order to compensate the
$N_{c}$ growth of the contribution of the kinematical factor $(nP)^{N_{c}/2}$
coming to Eq. (\ref{BDA-def}) from the product of $N_{c}$ fields $\chi $ (\ref{chi-psi}).

At large $N_{c}$ we can write the general decomposition (\ref{W-def-0})
\begin{equation}
\Phi(g)=N_{c}^{\nu}A(g)\exp\left[ N_{c}W(g)\right] \left[ 1+O\left(
N_{c}^{-1}\right) \right] \,.  \label{W-def-1}
\end{equation}
From the definition (\ref{Phi-def-1}) of $\Phi(g)$ it is obvious that for
any constant $\lambda$ we have
\begin{equation}
\Phi(\lambda g)=\lambda^{N_{c}}\Phi(g)\,.
\end{equation}
Therefore
\begin{equation}
W(\lambda g)=W(g)+\ln\lambda\,,  \label{W-scaling}
\end{equation}
\begin{equation}
A(\lambda g)=A(g)\,.
\end{equation}
Differentiating identity (\ref{W-scaling}) with respect to $\lambda$, we find
\begin{equation}
\sum\limits_{fs}\int_{0}^{\infty}dyg_{fs}(y)\frac{\delta W(g)}{\delta
g_{fs}(y)}=1\,.  \label{W-scaling-2}
\end{equation}
The distribution amplitude contains the momentum conserving delta function: 
\begin{equation}
\Psi_{(f_{1}s_{1})(f_{2}s_{2})\ldots(f_{N_{c}}s_{N_{c}})}(x_{1},x_{2},
\ldots,x_{N_{c}})\sim\delta(x_{1}+x_{2}+\ldots+x_{N_{c}}-1)\,.
\label{Psi-delta-x}
\end{equation}
Because to the presence of this delta function, the functional $\Phi(g)$ has the
following property. For functions
\begin{equation}
g_{fs}^{(y_{0},\beta)}(y)=\beta_{fs}\delta(y-y_{0})  \label{g-delta}
\end{equation}
we have
\begin{equation}
\left| \Phi\left[ g^{(y_{0},\beta)}\right] \right| =\left\{ 
\begin{array}{cc}
0 & \mathrm{if}\,y_{0}\neq1 \,, \\ 
\infty & \mathrm{if}\,y_{0}=1 \,.
\end{array}
\right. 
\end{equation}
This corresponds to the following singularities of $W(g)$
\begin{equation}
\mathrm{Re}\,W\left[ g^{(y_{0},\beta)}\right] =\left\{ 
\begin{array}{cc}
-\infty & \mathrm{if}\,y_{0}\neq1\,, \\ 
+\infty & \mathrm{if}\,y_{0}=1\,.
\end{array}
\right.
\label{W-momentum}
\end{equation}

\section{Evolution equation}

\label{Evolution-equation-section}

\subsection{Evolution equation for the baryon distribution amplitude}

The evolution equation for the baryon distribution amplitude at arbitrary
$N_{c}$ can be easily read from the literature starting from the original
papers \cite{LB-79,LB-80}. However, one should be careful generalizing
the standard equations written for the 3-quark distribution amplitude
to the $N_{c}$-quark case. In large-$N_c$ QCD the
baryon distribution amplitude $\Psi(x_{1},\ldots, x_{N_{c}})$ depends on
on $N_{c}$ variables $x_k$ [one of them can be eliminated using the conservation
of the momentum (\ref{Psi-delta-x})]. In QCD with $N_{c}$ colors,
the dependence of the leading-twist baryon distribution amplitude $\Psi^\mu$ on the
normalization point $\mu$ is described by the evolution equation
\begin{equation}
\mu\frac{\partial}{\partial\mu}\Psi^{\mu}(x_{1},\ldots, x_{N_{c}})
=-\frac{N_{c}+1}{2N_{c}}\frac{\alpha_{s}(\mu)}{\pi}\sum\limits_{1\leq i<j\leq
N_{c}}K_{ij}\Psi^{\mu}(x_{1},\ldots, x_{N_{c}})\,.  \label{evolution-2}
\end{equation}
We work in the leading order in the strong coupling $\alpha_{s}(\mu)$. We
have omitted the flavor and spin indices concentrating on pair-interaction
structure of the evolution kernel. The indices $i,j$ of $K_{ij}$ imply that
this operator acts on the variables $x_{i},x_{j}$. For example, the compact
equation
\begin{equation}
\phi=K_{12}\psi
\end{equation}
should be expanded as follows
\begin{equation}
\phi^{(f_{1}s_{1})(f_{2}s_{2})}(x_{1},x_{2})  =\sum\limits_{f_{1}^{\prime
}s_{1}^{\prime}f_{2}^{\prime}s_{2}^{\prime}}\int
dx_{1}^{\prime}dx_{2}^{\prime}K_{(f_{1}^{\prime}s_{1}^{\prime})(f_{2}^{
\prime}s_{2}^{\prime})}^{(f_{1}s_{1})(f_{2}s_{2})}(x_{1},x_{2};x_{1}^{
\prime},x_{2}^{\prime })
\psi^{(f_{1}^{\prime}s_{1}^{\prime})(f_{2}^{\prime}s_{2}^{
\prime})}(x_{1}^{\prime},x_{2}^{\prime})\,.  \label{K-action-detailed}
\end{equation}
The kernel $K$ is diagonal in both flavor $f_{k}$ and helicity $s_{k}$:
\begin{equation}
K_{(f_{1}^{\prime}s_{1}^{\prime})(f_{2}^{\prime}s_{2}^{
\prime})}^{(f_{1}s_{1})(f_{2}s_{2})}(x_{1},x_{2};x_{1}^{\prime},x_{2}^{
\prime})=\delta
_{f_{1}^{\prime}}^{f_{1}}\delta_{f_{2}^{\prime}}^{f_{2}}\delta_{s_{1}^{
\prime }}^{s_{1}}\delta_{s_{2}^{\prime}}^{s_{2}}\tilde{K}
^{s_{1}s_{2}}(x_{1},x_{2};x_{1}^{\prime},x_{2}^{\prime})\,.
\label{K-tilde-def}
\end{equation}
More information on the evolution kernel and its properties can be found in
Appendix \ref{Evolution-kernel-Appendix}.

\subsection{Evolution equation for the generating functional}

Now we want to consider the large-$N_{c}$ limit. As was explained above,
a consistent analysis of the large-$N_{c}$ limit is possible in terms of the
generating functional $\Phi (g)$ (\ref{Phi-def-1}). Therefore the first step
is to rewrite the evolution equation (\ref{evolution-2}) in terms of the
generating functional $\Phi (g)$. Let us contract Eq.~(\ref{evolution-2})
with the product of functions $g_{f_{k}s_{k}}(y_{k})$
\begin{align}
& \mu \frac{\partial }{\partial \mu }\left[ \prod_{k=1}^{N_{c}}\int
_{0}^{\infty }dy_{k}g_{f_{k}s_{k}}(y_{k})\right] \Psi
_{(f_{1}s_{1})\ldots (f_{N_{c}}s_{N_{c}})}^{\mu }\left( \frac{
y_{1}}{N_{c}},\ldots ,\frac{y_{N_{c}}}{N_{c}}\right)  \notag \\
& =-\frac{N_{c}+1}{2N_{c}}\frac{\alpha _{s}(\mu )}{\pi }\sum\limits_{1\leq
i<j\leq N_{c}}\prod_{k=1}^{N_{c}}\left[ \int_{0}^{\infty
}dy_{k}g_{f_{k}s_{k}}(y_{k})\right]  \notag \\
& \times \left[ K_{ij}\Psi ^{\mu }\right] _{(f_{1}s_{1})\ldots
(f_{N_{c}}s_{N_{c}})}\left( \frac{y_{1}}{N_{c}},\ldots ,\frac{y_{N_{c}}}{
N_{c}}\right) \,.  \label{evolution-3}
\end{align}
Note that the operators $K_{ij}$ are invariant under dilations. Therefore
the transition from $x_{k}$ to $y_{k}=N_{c}x_{k}$ does not change the form
of these operators. All the $N_{c}(N_{c}-1)/2$ terms of the sum over $i<j$
on the RHS are identical and can be written in terms of the generating
functional $\Phi (g)$
\begin{align}
& \sum\limits_{1\leq i<j\leq N_{c}}\sum\limits_{f_{m}s_{m}}\left[
\prod_{k=1}^{N_{c}}\int_{0}^{\infty }dy_{k}g_{f_{k}s_{k}}(y_{k})
\right] \left[ K_{ij}\Psi ^{\mu }\right] _{(f_{1}s_{1})\ldots
(f_{N_{c}}s_{N_{c}})}\left( \frac{y_{1}}{N_{c}},\ldots ,\frac{y_{N_{c}}}{
N_{c}}\right)  \notag \\
& =\frac{N_{c}(N_{c}-1)}{2}\sum\limits_{f_{m}s_{m}}\left[
\prod_{k=1}^{N_{c}}\int_{0}^{\infty }dy_{k}g_{f_{k}s_{k}}(y_{c})
\right] \left[ K_{12}\Psi ^{\mu }\right] _{(f_{1}s_{1})\ldots
(f_{N_{c}}s_{N_{c}})}\left( \frac{y_{1}}{N_{c}},\ldots ,\frac{y_{N_{c}}}{
N_{c}}\right)  \notag \\
& =\frac{1}{2}\int_{0}^{\infty }dy_{1}\int_{0}^{\infty
}dy_{2}\sum
\limits_{f_{1}f_{2}s_{1}s_{2}}g_{f_{1}s_{1}}(y_{1})g_{f_{2}s_{2}}(y_{2})
\left[ K_{12}\frac{\delta }{\delta g_{f_{1}s_{1}}(y_{1})}\frac{\delta }{
\delta g_{f_{2}s_{2}}(y_{2})}\Phi _{\mu }(g)\right] \,.
\end{align}
Now the evolution equation (\ref{evolution-3}) takes the form
\begin{equation}
\mu \frac{\partial }{\partial \mu }\Phi _{\mu }(g)=-\frac{N_{c}+1}{2N_{c}}
\frac{\alpha _{s}(\mu )}{2\pi }\left\{ \left( g\otimes g\right) \cdot K\cdot 
\left[ \left( \frac{\delta }{\delta g}\otimes \frac{\delta }{\delta g}
\right) \Phi _{\mu }(g)\right] \right\}\,,  \label{evolution-4}
\end{equation}
where we use the short notation
\begin{align}
& \left( g\otimes g\right) \cdot K\cdot \left[ \left( \frac{\delta }{\delta g
}\otimes \frac{\delta }{\delta g}\right) \Phi _{\mu }(g)\right]  \notag \\
& =\int_{0}^{\infty }dy_{1}\int_{0}^{\infty
}dy_{2}\sum
\limits_{f_{1}f_{2}s_{1}s_{2}}g_{f_{1}s_{1}}(y_{1})g_{f_{2}s_{2}}(y_{2}) 
\left[ K_{12}\frac{\delta }{\delta g_{f_{1}s_{1}}(y_{1})}\frac{\delta }{
\delta g_{f_{2}s_{2}}(y_{2})}\Phi _{\mu }(g)\right]  \notag \\
& =\int_{0}^{\infty }dy_{1}\int_{0}^{\infty
}dy_{2}\int_{0}^{\infty }dy_{1}^{\prime }\int_{0}^{\infty
}dy_{2}^{\prime }\sum\limits_{f_{1}f_{2}s_{1}s_{2}}  \notag \\
& \times g_{f_{1}s_{1}}(y_{1})g_{f_{2}s_{2}}(y_{2})\tilde{K}
^{s_{1}s_{2}}(y_{1},y_{2};y_{1}^{\prime },y_{2}^{\prime })\frac{\delta }{
\delta g_{f_{1}s_{1}}(y_{1}^{\prime })}\frac{\delta }{\delta
g_{f_{2}s_{2}}(y_{2}^{\prime })}\Phi _{\mu }(g)\,.
\end{align}
Here we used Eq. (\ref{K-tilde-def}).

\subsection{Evolution equation in the leading order of the large-$N_{c}$
limit}

Now we can study the large-$N_{c}$ limit of the evolution equation 
(\ref{evolution-4}). We insert the large-$N_{c}$ ansatz (\ref{W-def-1}) into
Eq. (\ref{evolution-4}) and obtain in the leading order of the $1/N_{c}$
expansion:
\begin{equation}
\mu \frac{\partial }{\partial \mu }W_{\mu }(g)=-a(\mu )\left\{ \left(
g\otimes g\right) \cdot K\cdot \left[  \frac{\delta W_{\mu }(g)}{
\delta g}\otimes \frac{\delta W_{\mu }(g)}{\delta g} \right] \right\}\,,
\label{W-evolution-mu}
\end{equation}
where
\begin{equation}
a(\mu )=\lim_{N_{c}\rightarrow \infty }\frac{\alpha _{s}(\mu )N_{c}}{4\pi }
\,.  \label{a-mu-def}
\end{equation}
This limit exists since
\begin{equation}
\alpha _{s}(\mu )=O(N_{c}^{-1})\,.
\end{equation}
Eq.~(\ref{W-evolution-mu}) is obviously compatible with the constraint
(\ref{W-scaling}) on $W_{\mu }(g)$. The evolution equation
(\ref{W-evolution-mu}) is nonlinear in $W_{\mu }(g)$. Being a first order
differential equation in $\mu $, it allows us (in principle) to find $W_{\mu }$ at
any normalization point $\mu $ starting from the initial value for $W_{\mu _{0}}$
at some point $\mu _{0}$.

Instead of $\mu$ it is convenient to introduce the new variable $t$ such that
\begin{equation}
dt=2a(\mu )\frac{d\mu }{\mu }\,.  \label{dt-d-mu}
\end{equation}
Then
\begin{equation}
\frac{\partial }{\partial t}W(g,t)=-\frac{1}{2}\left\{ \left( g\otimes
g\right) \cdot K\cdot \left[ \frac{\delta W(g,t)}{\delta g}\otimes 
\frac{\delta W(g,t)}{\delta g} \right] \right\} \,.
\label{W-evolution-compact}
\end{equation}

Note that the evolution equation (\ref{W-evolution-compact}) does not depend
on the quantum numbers of the baryon. This agrees with the statement that
functional $W_{\mu }(g)$ is universal for all low-lying baryons. The
difference between baryons appears in the pre-exponential factor 
$A(g)$ (\ref{W-def-1}).

\subsection{Hamilton-Jacobi structure}

\label{Hailton-Jacobi-subsection}

Eq.~(\ref{W-evolution-compact}) has the form of a functional
Hamilton-Jacobi equation. Changing the notation
\begin{equation}
g\rightarrow q\,,\quad W(g,t)\rightarrow S(q,t)\,\,,
\end{equation}
we easily see that we deal with the equation
\begin{equation}
\frac{\partial S}{\partial t}=-H\left( \frac{\partial S}{\partial q},q\right)\,,
\label{Hamilton-Jacobi}
\end{equation}
where the Hamiltonian (written in terms of discrete variables $p_{n},q_{n}$)
\begin{equation}
H(p,q)=\frac{1}{2}\sum\limits_{ijmn}K_{ijmn}q_{i}q_{j}p_{m}p_{n}
\end{equation}
is quadratic both in coordinates $q_{n}$ and in momenta
\begin{equation}
p_{n}=\frac{\partial S}{\partial q_{n}}\,.
\end{equation}

\subsection{Evolution equation in the next-to-leading order}

Inserting ansatz (\ref{W-def-1}) into the evolution equation (\ref{evolution-4}),
we can also derive an equation for the pre-exponential factor
$A(g,t)$
\begin{align}
\frac{\partial }{\partial t}\ln A(g,t)& =-\left\{ \left( g\otimes g\right)
\cdot K\cdot \left[ \frac{\delta W(g,t)}{\delta g}\otimes \frac{\delta \ln
A(g,t)}{\delta g}\right] \right\}  \notag \\
& +b(t)\left\{ \left( g\otimes g\right) \cdot K\cdot \left[ \frac{\delta
W(g,t)}{\delta g}\otimes \frac{\delta W(g,t)}{\delta g}\right] \right\} 
\notag \\
& -\frac{1}{2}\left\{ \left(g\otimes g\right) \cdot K\cdot \left[ \left( 
\frac{\delta }{\delta g}\otimes \frac{\delta }{\delta g}\right) W(g,t)\right]
\right\}\,,  \label{A-subleading-evolution}
\end{align}
where
\begin{equation}
b(t)=\frac{1}{2}\lim_{N_{c}\rightarrow \infty }\left\{ N_{c}\left[ 1-\frac{
N_{c}+1}{4\pi }\frac{\alpha _{s}(\mu )}{a(\mu )}\right] \right\} \,.
\end{equation}
Taking the difference of two equations (\ref{A-subleading-evolution})
for baryons $B_1$, $B_2$ described by the pre-exponential factors
$A_{B_1}(g,t)$ and $A_{B_2}(g,t)$, we obtain the following equation for
the evolution of $A_{B_1}(g,t)/A_{B_2}(g,t)$
\begin{equation}
\frac{\partial }{\partial t} \frac{A_{B_1}(g,t)}{A_{B_2}(g,t)}=-\left\{
\left( g\otimes g\right) \cdot K\cdot \left[ \frac{\delta W(g,t)}{\delta g}
\otimes  \frac{\delta }{\delta g} \frac{A_{B_1}(g,t)}{A_{B_2}(g,t)}
 \right] \right\} \,.  \label{A12-evolution}
\end{equation}
This is a linear homogeneous equation for $A_{B_1}(g,t)/A_{B_2}(g,t)$. The
structure of this equation agrees with the statement about the factorization
of $A_{B}(g,t)$ into ``elementary functionals'' $\xi _{i}(g,t)$ [cf. Eq. (\ref{A-factorization})]:
\begin{equation}
A_{B}(g,t)=A_{0}(g,t)\prod_{i}\left[ \xi _{i}(g,t)\right] ^{n_{i}}\,.
\label{A-decomposition-LC}
\end{equation}
Functionals $\xi _{i}(g,t)$ correspond to the elementary excitations
 $\omega _{i}$ in the mass formula (\ref{M-harmonic}).
The factorization (\ref{A-decomposition-LC}) is compatible with the
evolution equation (\ref{A12-evolution}) if functionals
$\xi _{i}(g,t)$ obey Eq.~(\ref{A12-evolution}):
\begin{equation}
\frac{\partial }{\partial t}\xi _{i}(g,t)=-\left\{ \left( g\otimes g\right)
\cdot K\cdot \left[ \frac{\delta W(g,t)}{\delta g}\otimes \frac{\delta \xi
_{i}(g,t)}{\delta g}\right] \right\} \,.
\end{equation}
Strictly speaking, expression (\ref{A-decomposition-LC}) must be modified
by the zero mode factors which are discussed in Sec.~\ref{Zero-modes-section}.

\section{Asymptotic limit}

\label{Asymptotic-limit-section}

\subsection{Double limit $N_{c}\rightarrow \infty $, 
$t \rightarrow \infty $}

At finite $N_{c}$ the asymptotic large-$\mu $ behavior of the baryon
distribution amplitude is well known [see Eq. (\ref{Psi-B-C-B})]. In terms of the variable $t$ (\ref{dt-d-mu}) the limit of large scales $\mu$
corresponds to $t\to\infty$. Now we want to study the large-$t$
behavior of the functional $W(g,t) $. Since the functional $W(g,t)$ is defined
via the large-$N_{c}$
limit, we face the problem of the double limit $N_{c}\rightarrow \infty $,
$t\rightarrow \infty $. As we shall see, the order in which these two limits
are taken is important. One of the results of our analysis of the
double limit $N_{c}\rightarrow \infty $, $t\rightarrow \infty $ will be a
general expression for the anomalous dimensions of the leading-twist baryon
operators in the large $N_{c}$ limit. Later (in Sec. \ref{Anomalous-dimensions-section})
this expression will be used for the calculation of these anomalous dimensions.

\subsection{Linearized theory of the asymptotic limit}

Now we turn to the investigation of the asymptotic limit of $W(g,t)$ at 
$t$ $\rightarrow \infty $. In this section the analysis will be based on
the evolution equation (\ref{W-evolution-compact}). A complete systematic
analysis of the nonlinear evolution equation (\ref{W-evolution-compact}) in
the asymptotic regime is beyond the scope of this paper. Our aim is more
pragmatic: we want to find the physical solution. On the way to this
physical solution we shall use various assumptions about the structure of the
solution. The justification of these intermediate assumptions comes from the
final form of the solution and from the independent analysis of the
asymptotic limit in Sec. \ref{Asymptotic-solution-method-2}.

We expect the following structure of the large-$t$ behavior of the
functional $W(g,t)$
\begin{equation}
W(g,t)\overset{t\rightarrow \infty }{\rightarrow }W_{\mathrm{as}}(g,t)\equiv
W_{0}(g)-\sigma (t)\,,  \label{W-0-def}
\end{equation}
where the function $\sigma (t)$ is $g$ independent and the functional
$W_{0}(g)$ does not depend on $t$.

Inserting the ansatz (\ref{W-0-def}) into the evolution equation
(\ref{W-evolution-compact}), we arrive at the system of equations
\begin{equation}
\frac{1}{2}\left\{\left( g\otimes g\right) \cdot K\cdot\left[ \frac{\delta W_{0}(g)
}{\delta g}\otimes\frac{\delta W_{0}(g)}{\delta g}\right]\right\} =E\,,
\label{W0-c-eq}
\end{equation}
\begin{equation}
\sigma(t)=Et\,,  \label{sigma-eq}
\end{equation}
where $E$ is some constant.

According to Eqs. (\ref{W-scaling}) and (\ref{W-0-def}) the functional
$W_{0}(g) $ must obey the condition
\begin{equation}
W_{0}(\lambda g)=W_{0}(g)+\ln \lambda \,.  \label{W-0-lambda}
\end{equation}

Note that in terms of the Hamilton-Jacobi interpretation 
(\ref{Hamilton-Jacobi}) of the evolution equation (\ref{W-evolution-compact}) the
asymptotic equations (\ref{W0-c-eq}) and (\ref{sigma-eq}) correspond to
fixing the energy $E$ in the Hamilton-Jacobi equation:
\begin{equation}
H\left( \frac{\partial S}{\partial q},q\right) =E\,,\quad\frac{\partial S}{
\partial t}=-E\,.  \label{Hamiltoni-Jacobi-static}
\end{equation}

This ``energy'' interpretation of the asymptotic regime is quite similar to
the standard analysis of the asymptotic distribution amplitudes of hadrons
in terms of the lowest eigenstates of the effective Hamiltonian describing
the evolution. However, our current large-$N_{c}$ evolution equation is
nonlinear and this leads to certain complications.

The true physical functional $W(g,t)$ is different from the asymptotic
solution $W_{\mathrm{as}}(g,t)$ (\ref{W-0-def}):
\begin{equation}
W(g,t)=W_{\mathrm{as}}(g,t)+\Delta W(g,t)\,.  \label{W-near-as}
\end{equation}
Both $W(g,t)$ and $W_{\mathrm{as}}(g,t)$ satisfy condition 
(\ref{W-scaling}). Therefore we must have
\begin{equation}
\Delta W(\lambda g,t)=\Delta W(g,t)\,.  \label{Delta-W-scale-inv}
\end{equation}
At large $t$ the $\Delta W(g,t)$ becomes small and we can linearize the
evolution equation (\ref{W-evolution-compact}) in $\Delta W(g,t)$:
\begin{equation}
\frac{\partial }{\partial t}\Delta W(g,t)=-\left\{ \left( g\otimes g\right)
\cdot K\cdot \left[ \frac{\delta W_{\mathrm{as}}(g,t)}{\delta g}\otimes 
\frac{\delta \Delta W(g,t)}{\delta g}\right] \right\} \,.
\end{equation}
According to Eq. (\ref{W-0-def}) we have
\begin{equation}
\frac{\delta W_{\mathrm{as}}(g,t)}{\delta g}=\frac{\delta W_{0}(g)}{\delta g}
\,.
\end{equation}
Therefore
\begin{equation}
\frac{\partial }{\partial t}\Delta W(g,t)=-\left\{ \left( g\otimes g\right)
\cdot K\cdot \left[ \frac{\delta W_{0}(g)}{\delta g}\otimes \frac{\delta
\Delta W(g,t)}{\delta g}\right] \right\} \,.  \label{Delta-W-evolution}
\end{equation}
In order to solve this equation, let us first consider the spectral problem
\begin{equation}
\left( g\otimes g\right) \cdot K\cdot \left[ \frac{\delta W_{0}(g)}{\delta g}
\otimes \frac{\delta \xi _{k}(g)}{\delta g}\right] =\Omega _{k}\xi _{k}(g)
\label{chi-n-eq-1}
\end{equation}
in the class of functionals $\xi _{k}(g)$ obeying condition (\ref{Delta-W-scale-inv})
\begin{equation}
\xi _{k}(\lambda g)=\xi _{k}(g)\,.  \label{chi-W-scale-inv}
\end{equation}

The solution of Eq.~(\ref{Delta-W-evolution}) can be represented in the
form
\begin{equation}
\Delta W(g,t)=\sum\limits_{k}a_{k}e^{-\Omega_{k}t}\xi_{k}(g)\,.
\label{Delta-W-decomposition}
\end{equation}
Now we insert Eqs. (\ref{W-0-def}), (\ref{sigma-eq}), and
(\ref{Delta-W-decomposition}) into Eq. (\ref{W-near-as}):
\begin{equation}
W(g,t)\overset{t\rightarrow\infty}{=}W_{0}(g)-Et+\sum\limits_{k}a_{k}e^{-
\Omega_{k}t}\xi_{k}(g)\,.  \label{W-spectral-decomposition}
\end{equation}
This equation is derived in the double limit $N_{c}\rightarrow\infty$, 
$t\rightarrow\infty$. Note that the limit $N_{c}\rightarrow\infty$ is taken
first [remember that the functional $W(g,t)$ is defined by Eq. (\ref{W-def-1})
via the large-$N_{c}$ limit].

\subsection{Asymptotic behavior in terms of anomalous dimensions}

Formally Eq.~(\ref{W-spectral-decomposition}) solves the problem of the
asymptotic limit $t\rightarrow\infty$. Nevertheless
one can wonder how this solution is connected with the standard
representation of the $t$ dependence in terms of the decomposition in
operators with given anomalous dimensions. For the generating functional 
$\Phi(g,t)$ (\ref{Phi-def-1}) this decomposition has the form

\begin{equation}
\Phi(g,t)\overset{\,t\rightarrow\infty}{\rightarrow}\sum\limits_{\alpha}
\Phi_{\alpha}(g)\exp\left( -\Gamma_{\alpha}t\right) \,.
\label{Phi-expansion-2}
\end{equation}
Here $\Gamma_{\alpha}$ are anomalous dimensions of operators diagonalizing
the evolution. The expansion (\ref{Phi-expansion-2}) is written for finite 
$N_{c}$. If we take the large-$N_{c}$ limit in Eq. (\ref{Phi-expansion-2}),
then we must be careful about the order of the two limits 
$N_{c}\rightarrow\infty$ and $t\rightarrow\infty$. The approach based on
Eq.~(\ref{Phi-expansion-2}) assumes that the limit of large $t$ is
taken before the limit $N_{c}\rightarrow\infty$. On the contrary, Eq. 
(\ref{W-spectral-decomposition}) deals with the opposite order of the two limits
(first $N_{c}\rightarrow \infty$ and then $t\rightarrow\infty$).

At large $N_{c}$ we expect the following structure of the anomalous dimensions 
$\Gamma_{\alpha}$
\begin{equation}
\Gamma_{\alpha}=c_{1}N_{c}+c_{0}(\alpha)+c_{-1}
(\alpha)N_{c}^{-1}+O(N_{c}^{-2})\,.  \label{gamma-N-expansion}
\end{equation}
Note that the leading order coefficient $c_{1}$ is independent of $\alpha$.
This independence will be explained later.

If we insert decomposition (\ref{gamma-N-expansion}) into
Eq.~(\ref{Phi-expansion-2}), then the 
$1/N_{c}$ expansion for $\Gamma_{\alpha}$ will be exponentiated:
\begin{equation}
\Phi(g,t)\overset{\,t\rightarrow\infty,N_{c}\rightarrow\infty}{\rightarrow }
\sum\limits_{\alpha}\Phi_{\alpha}(g)\exp\left\{ -t\left[ c_{1}N_{c}+c_{0}
(\alpha)+c_{-1}(\alpha)N_{c}^{-1}+\ldots\right] \right\} \,.
\label{Phi-expansion-3}
\end{equation}
This exponentiation of the $1/N_{c}$ expansion for $\Gamma_{\alpha}$ means
that we have a rather nontrivial sensitivity to the order of the limits 
$t\rightarrow\infty$ and $N_{c}\rightarrow\infty$.

If one is interested in the other order of limits (first $N_{c}\rightarrow
\infty$ and then $t\rightarrow\infty$), then one should start from
Eq. (\ref{W-spectral-decomposition}). Inserting the decomposition 
(\ref{W-spectral-decomposition}) into Eq. (\ref{W-def-1}), we find
\begin{equation}
\Phi(g,t)\overset{N_{c}\rightarrow\infty,\,t\rightarrow\infty}{\rightarrow }
A(g,t)\exp\left\{ N_{c}\left[ W_{0}(g)-Et+\sum\limits_{k}a_{k}e^{-
\Omega_{k}t}\xi_{k}(g)\right] \right\} \,.  \label{Phi-expansion-1}
\end{equation}

Now we can compare the two asymptotic expressions (\ref{Phi-expansion-3}), 
(\ref{Phi-expansion-1}) derived for functional $\Phi(g,t)$ in the double
limit $N_{c}\rightarrow\infty,\,t\rightarrow\infty$ assuming different
orders of these two limits. Separating the factors containing the exponentiated
product $N_{c}t$, we conclude from the comparison of Eqs. (\ref{Phi-expansion-3}),
(\ref{Phi-expansion-1}) that
\begin{equation}
e^{-c_{1}tN_{c}}=e^{-EtN_{c}}\,.
\end{equation}
We see that
\begin{equation}
c_{1}=E\,.  \label{c1-E}
\end{equation}
This result explains why the coefficient $c_{1}$ appearing in large-$N_{c}$
expansion (\ref{gamma-N-expansion}) for $\Gamma_{\alpha}$ is independent of 
$\alpha$.

One can also match the two asymptotic representations
(\ref{Phi-expansion-3})
and
(\ref{Phi-expansion-1})
in the next order of the $1/N_{c}$ expansion. To this aim we expand the
RHS of Eq. (\ref{Phi-expansion-1}) in powers of 
$a_{k}e^{-\Omega_{k}t}\xi_{k}(g)$:
\begin{align}
& \Phi(g,t)\overset{N_{c}\rightarrow\infty,\,t\rightarrow\infty}{\rightarrow 
}A(g,t)\exp\left\{ N_{c}W_{0}(g)\right\}  \notag \\
& \times\sum_{\{n_{k}\}}\exp\left[ -\left(
N_{c}E+\sum\limits_{k}n_{k}\Omega_{k}\right) t\right] \prod_{k}\frac{1}
{n_{k}!}\left[ N_{c}a_{k}\xi_{k}(g)\right] ^{n_{k}}\,.
\end{align}
Comparing this decomposition with Eq. (\ref{Phi-expansion-3}), we see that
\begin{equation}
c_{0}(\alpha)=\Delta
E+\sum\limits_{k}n_{k}\Omega_{k}\quad(n_{k}=0,1,2,\ldots)\,,
\end{equation}
where the $O(N_{c}^{0})$ contribution $\Delta E$ comes from exponential part
of the large-$t$ asymptotic behavior of $A(g,t)\sim e^{-t\Delta E}$.
Inserting this expression for $c_{0}(\alpha)$ into 
Eq. (\ref{gamma-N-expansion}) and using the expression (\ref{c1-E}) for $c_{1}$, we
find
\begin{equation}
\Gamma_{\{n_{k}\}}=N_{c}E+\left( \Delta E+\sum\limits_{k}n_{k}\Omega
_{k}\right) +c_{-1}(n_{1},n_{2},\ldots)N_{c}^{-1}+O(N_{c}^{-2})\,.
\label{gamma-n-expansion}
\end{equation}
Here we label the anomalous dimensions $\Gamma_{\alpha}$ with sets of
integer numbers $\{n_{k}\}$:
\begin{equation}
\alpha=\{n_{k}\}\quad(n_{k}=0,1,2,\ldots)\,.
\end{equation}
At finite $N_{c}$ the spectrum of the anomalous dimensions comes from the
diagonalization of the $N_{c}$-particle ``effective Hamiltonian'' and the
anomalous dimensions $\Gamma_{\alpha}$ are labeled by $N_{c}-1$ ``quantum
numbers'' $\alpha_{1},\alpha_{2},\ldots\alpha_{N_{c}-1}$ (one degree of
freedom is eliminated by the momentum conservation). In the limit 
$N_{c}\rightarrow\infty$ we arrive at the ``oscillator spectrum'' 
(\ref{gamma-n-expansion}) and the anomalous dimensions are parametrized by a set
of integer excitation numbers $n_{k}$.

The dependence of the anomalous dimensions on some quantum numbers
(e.g. helicity) appears only in the order $O(N_{c}^{-1})$.
Denoting these quantum numbers by $\rho$, we can write the correct
generalization of Eq. (\ref{gamma-n-expansion})
\begin{equation}
\Gamma_{\{n_{k},\rho\}}=N_{c}E+\left( \Delta
E+\sum\limits_{k}n_{k}\Omega_{k}\right)
+c_{-1}(n_{1},n_{2},\ldots;\rho)N_{c}^{-1}+O(N_{c}^{-2})\,.
\label{gamma-n-expansion-beta}
\end{equation}
This structure of the spectrum corresponding to an effective harmonic
oscillator in the order $N_{c}^{0}$ (with anharmonic corrections appearing
in the next orders of the $1/N_{c}$ expansion) is typical for semiclassical
systems. The semiclassical nature of the $1/N_{c}$ expansion is known since
long ago. Depending on the methods used for the construction of the $1/N_{c}$
expansion, the semiclassical features can arise via the mean field
approximation, saddle point approximation in path integrals, WKB expansion
etc.

Our derivation of the expansion (\ref{gamma-n-expansion-beta}) was based on
the asymptotic linearized analysis of the Hamilton-Jacobi equation 
(\ref{Delta-W-evolution}) in terms of the large-$t$ decomposition 
(\ref{W-spectral-decomposition}). In this approach the spectrum of $\Omega _{k}$
is given by the eigenvalues of the spectral problem (\ref{chi-n-eq-1})
for the functionals $\xi_{k}(g) $.

In the rest of the paper we compute parameters $E,\Delta E$ and $\Omega _{k}$
appearing in the expansion (\ref{gamma-n-expansion-beta}). The results are given
by Eq. (\ref{c-gamma-Nf-1}) for $E$ and Eq. (\ref{Delta-E-res}) for $\Delta
E $. The parameters $\Omega _{k}$ are computed in Eqs. (\ref{Omega-minus}), 
(\ref{Omega-plus}) in the case of one quark flavor $N_{f}=1$. The role
of parameter $\rho$ is played by the helicity $J_3$ in this case.
The $O(N_c^{-1})$
term $c_{-1}(0,0,\ldots;J_3)N_{c}^{-1}$ is computed for the lowest
anomalous dimension corresponding to the asymptotic wave function of
baryons with given helicity $J_3$ [see Eq. (\ref{gamma-lowest-exact})].

\section{Calculation of the asymptotic functional $W_{0}(g)$}

\label{W0-calculation-section}

\subsection{Methods}

We want to compute the asymptotic functional $W_{0}(g)$ which determines the
asymptotic behavior (\ref{W-0-def}) of $W(g,t)$ at large $t$. Two methods
can be used for the calculation:

1) We can solve Eq.~(\ref{W0-c-eq}).

2) We can use the traditional approach to the determination of the large-$t$
asymptotic baryon distribution amplitude at finite $N_{c}$. Using this
finite-$N_{c}$ asymptotic distribution amplitude we can construct the
corresponding generating functional by analogy with Eq. (\ref{Phi-def-1}).
The large-$N_{c}$ asymptotic behavior of this functional will be dominated
by the exponential term containing information on $W_{0}(g)$.

The first method is described in Sec. \ref{W0-solution-section}. 
In Sec. \ref{Asymptotic-solution-method-2} we show how the second method works.
In Sec. \ref{Two-methods-comparison-section} we compare these two methods. The
computed functional $W_{0}(g)$ depends on the number of quark flavors $N_{f}$.
The nature of this dependence is studied in
Sec. \ref{Nf-dependence-section}. In Sec. \ref{Analytical-properties-section} we
comment on the analytical properties of the functional $W(g,t)$ and
illustrate the general statements using our result for the asymptotic
functional $W_{0}(g)$.

\subsection{ Solution of the equation for $W_{0}(g)$}

\label{W0-solution-section}

According to Eq. (\ref{W-0-def}) the main quantity characterizing the
asymptotic limit is the functional $W_{0}(g)$. This functional can be found
by solving Eq.~(\ref{W0-c-eq}).

Our approach to Eq.~(\ref{W0-c-eq}) will be rather heuristic. We simply
use an ansatz which allows us to solve Eq.~(\ref{W0-c-eq}) for some
special value of $E$. The relevance of this value of $E$ for the description
of the asymptotic behavior will be seen from the analysis of 
Sec. \ref{Asymptotic-solution-method-2}. The form of the solution $W_{0}(g)$ of
Eq. (\ref{W0-c-eq}) depends on the number of quark flavors $N_{f}$. We start from
the simplest case $N_{f}=1$. The result found for $N_{f}=1$ is easily
generalized to the more interesting case $N_{f}=2$.

\subsubsection{Case $N_{f}=1$}

In order to solve Eq.~(\ref{W0-c-eq}) we use the property of the evolution
kernel $\tilde{K}^{s_{1}s_{2}}$
\begin{equation}
\int dy_{1}'\int dy_{2}'
\tilde{K}^{s_{1}s_{2}}(y_{1},y_{2};y_{1}',y_{2}')y_{1}'y_{2}'
e^{-X(g)(y_{1}'+y_{2}')}=\frac{1}{2}\delta^{s_{1}s_{2}}y_{1}y_{2}e^{-X(g)(y_{1}+y_{2})}
\label{K-tilde-x-exp}
\end{equation}
which is valid for any functional $X(g)$.
This identity follows from Eq. (\ref{K-tilde-x-exp-0}) derived in
Appendix~\ref{Evolution-kernel-Appendix}. Keeping in mind this identity,
we make the following ansatz
\begin{equation}
\frac{\delta W_{0}(g)}{\delta g_{s}(y)}=ye^{-X(g)y}U_{s}(g)\,,
\label{delta-W0-ansatz-Nf-1}
\end{equation}
where $X(g)$ and $U_{s}(g)$ are some functionals of $g_{s}$. Inserting this
ansatz into Eq. (\ref{W0-c-eq}) and using Eq. (\ref{K-tilde-x-exp}), we find
\begin{equation}
\sum\limits_{s}\left[ G_{s}(g)U_{s}(g)\right] ^{2}=4E\,,
\label{GU-E}
\end{equation}
where 
\begin{equation}
G_{s}(g)\equiv\int_{0}^{\infty}g_{s}(y)ye^{-X(g)y}\,.
\label{G-Nf-1}
\end{equation}
Combining this expression for $G_{s}(g)$ with Eq. (\ref{delta-W0-ansatz-Nf-1}),
we find
\begin{equation}
\int_{0}^{\infty}dyg_{s}(y)\frac{\delta W_{0}(g)}{\delta g_{s}(y)}
=G_{s}(g)U_{s}(g)\,.
\end{equation}
According to Eq.~(\ref{W-scaling-2}) we have
\begin{equation}
\sum\limits_{s}\int_{0}^{\infty}dyg_{s}(y)\frac{\delta W_{0}(g)}{
\delta g_{s}(y)}=1\,.
\end{equation}
Combining the last two equations, we conclude that
\begin{equation}
\sum\limits_{s}G_{s}(g)U_{s}(g)=1\,.  \label{GG-UU-eq-1}
\end{equation}
We can satisfy both Eq. (\ref{GU-E}) and Eq. (\ref{GG-UU-eq-1}) by taking
\begin{equation}
G_{s}(g)U_{s}(g)=\frac{1}{2}\,,  \label{GU-half}
\end{equation}
which corresponds to
\begin{equation}
E=\frac{1}{8}\,.  \label{c-gamma-Nf-1}
\end{equation}
Now it follows from Eqs. (\ref{delta-W0-ansatz-Nf-1}), (\ref{G-Nf-1}),
and (\ref{GU-half})
\begin{equation}
\frac{\delta W_{0}(g)}{\delta g_{s}(y)}=\frac{1}{2}ye^{-X(g)y}\left[
\int_{0}^{\infty}dy^{\prime}g_{s}(y^{\prime})y^{\prime}e^{-X(g)y^{
\prime}}\right] ^{-1}\,.  \label{var-W-eq}
\end{equation}

We can look for the solution of this equation in the form
\begin{equation}
W_{0}(g)=F\left[ X(g)\right] +\frac{1}{2}\sum\limits_{s}\ln\left[
\int_{0}^{\infty}dyg_{s}(y)ye^{-X(g)y}\right] \,,
\label{W-via-F-Nf-1}
\end{equation}
where $F(X)$ is an arbitrary function and the functional $X(g)$ is
implicitly determined by the equation
\begin{equation}
\frac{\partial}{\partial X}\left\{ F(X)+\frac{1}{2}\sum\limits_{s}\ln\left[
\int_{0}^{\infty}dyg_{s}(y)ye^{-Xy}\right] \right\} =0\quad
\Longrightarrow\quad X=X(g)\,.  \label{dZ-eq-Nf-1}
\end{equation}
Thus we have found an infinite set of solutions $W_{0}(g)$ of
Eq.~(\ref{W0-c-eq}) corresponding to the same constant $E$ (\ref{c-gamma-Nf-1}).
These solutions are parametrized by arbitrary functions $F(X)$. The physical
solution is fixed by the condition (\ref{W-momentum}) which leads to
the following choice of $F(X)$:
\begin{equation}
F(X)=X\,.  \label{F-physical-Nf-1}
\end{equation}
Indeed, taking this function $F(X)$ and computing the LHS of
Eq.~(\ref{dZ-eq-Nf-1}) for functions $g^{(y_{0},\beta)}$
(\ref{g-delta}), we obtain
\begin{equation}
\frac{\partial}{\partial X}\left\{ F(X)+\frac{1}{2}\sum\limits_{s}\ln\left[
\int_{0}^{\infty}dyg_{s}^{(y_{0},\beta)}(y)ye^{-Xy}\right] \right\}
=1-y_{0}\,.  \label{W-naive-Nf-1}
\end{equation}
Therefore Eq. (\ref{dZ-eq-Nf-1}) has no solutions at $y_{0}\neq1$ in
agreement with the condition (\ref{W-momentum}).

We can combine Eqs. (\ref{W-via-F-Nf-1}), (\ref{dZ-eq-Nf-1}), 
and (\ref{F-physical-Nf-1}) into the final representation for $W_{0}(g)$ 
\begin{equation}
W_{0}(g)=\underset{X}{\mathrm{extremum}}\left\{ X+\frac{1}{2}\sum
\limits_{s}\ln\left[ \int_{0}^{\infty}dyg_{s}(y)ye^{-Xy}\right]
\right\} \,.  \label{W0-res-Nf-1}
\end{equation}
The word \emph{extremum} must be understood in the sense of the equation
\begin{equation}
\frac{\partial}{\partial X}\left\{ X+\frac{1}{2}\sum\limits_{s}\ln\left[
\int_{0}^{\infty}dyg_{s}(y)ye^{-Xy}\right] \right\} =0\,.
\label{X-g-equation-Nf-1}
\end{equation}
In the general case both $g_{s}$ and $X$ can be complex so that
we cannot speak about a maximum or a minimum. Equation 
(\ref{X-g-equation-Nf-1}) determines the functional $X(g)$ and this functional
should be used on the RHS of Eq. (\ref{W0-res-Nf-1}) for the calculation of
$W_0(g)$.

\subsubsection{Case $N_{f}=2$}

The generalization of the above results to the case $N_{f}=2$ is
straightforward. By analogy with Eq. (\ref{delta-W0-ansatz-Nf-1}) we make
the ansatz
\begin{equation}
\frac{\delta W_{0}(g)}{\delta g_{fs}(y)}=ye^{-X(g)y}U_{fs}(g)\,.
\label{delta-W0-ansatz}
\end{equation}
Introducing the notation
\begin{equation}
G_{fs}(g)\equiv\int_{0}^{\infty}dyg_{fs}(y)ye^{-X(g)y}\,,
\label{G-fs-def}
\end{equation}
we find the $N_{f}=2$ analogs of Eqs. (\ref{GU-E}) and (\ref{GU-half})
\begin{equation}
\sum
\limits_{f_{1}f_{2}s}G_{f_{1}s}(g)G_{f_{2}s}(g)U_{f_{1}s}(g)U_{f_{2}s}(g)=4E
\,,  \label{GG-UU-eq}
\end{equation}
\begin{equation}
\sum\limits_{fs}G_{fs}(g)U_{fs}(g)=1\,.
\end{equation}
One can satisfy both equations by taking
\begin{equation}
\sum\limits_{f}G_{fs_{1}}(g)U_{fs_{2}}(g)=\frac{1}{2}\delta_{s_{1}s_{2}}\,,
\label{G-U-inv}
\end{equation}
which leads to the same value
\begin{equation}
E=\frac{1}{8}  \label{c-gamma}
\end{equation}
as in the $N_{f}=1$ case (\ref{c-gamma-Nf-1}).

According to Eqs. (\ref{G-fs-def}) and (\ref{G-U-inv}) we have
\begin{equation}
U_{fs}(g)=\frac{1}{2}\left[ G(g)\right] _{sf}^{-1}=\frac{1}{2}\left\{ \left[
\int_{0}^{\infty }dyg(y)ye^{-X(g)y}\right] ^{-1}\right\} _{sf}\,.
\end{equation}
Inserting this result into Eq. (\ref{delta-W0-ansatz}), we find
\begin{equation}
\frac{\delta W_{0}(g)}{\delta g_{fs}(y)}=\frac{1}{2}ye^{-X(g)y}\left\{ \left[
\int_{0}^{\infty }dy^{\prime }g(y^{\prime })ye^{-X(g)y^{\prime }}
\right] ^{-1}\right\} _{sf}\,.  \label{delta-W-result}
\end{equation}
The general solution of this equation compatible with the momentum
conservation constraint (\ref{W-momentum}) is
\begin{equation}
W_{0}(g)=\underset{X}{\mathrm{extremum}}\left\{ X+\frac{1}{2}\ln \det_{fs}
\left[ \int_{0}^{\infty }dyg_{fs}(y)ye^{-Xy}\right] \right\}\,.
\label{W0-g-res}
\end{equation}
Similarly to the $N_f=1$ case, this compact representation should be understood
in the sense of the (generally speaking complex) ``extremum equation''
\begin{equation}
\frac{\partial }{\partial X}\left\{ X+\frac{1}{2}\ln \det_{fs}\left[
\int_{0}^{\infty }dyg_{fs}(y)ye^{-Xy}\right] \right\} =0\,,
\label{X-saddle-point}
\end{equation}
which determines the functional $X(g)$.
Eq. (\ref{X-saddle-point}) can be rewritten in the form
\begin{equation}
\mathrm{Tr}\left\{ \left[ \int_{0}^{\infty }dyg(y)y^{2}e^{-Xy}\right] 
\left[ \int_{0}^{\infty }dy^{\prime }g(y^{\prime })y^{\prime
}e^{-Xy^{\prime }}\right] ^{-1}\right\} =2\,.  \label{X-saddle-point-0}
\end{equation}

Inserting Eq. (\ref{W0-g-res}) into Eq. (\ref{W-0-def}), we find the final
expression for the asymptotic functional:
\begin{equation}
W_{\mathrm{as}}(g,t)=\underset{X}{\mathrm{extremum}}\left\{ X+\frac{1}{2}\ln
\det_{fs}\left[ \int_{0}^{\infty }dyg_{fs}(y)ye^{-Xy}\right] \right\}
-Et\,.  \label{W-mu-as-res}
\end{equation}

\subsection{Direct construction of the asymptotic functional}

\label{Asymptotic-solution-method-2}

Let us consider the case of $N_{f}=2$ flavors. The asymptotic distribution
amplitude of baryons with helicity 
\begin{equation}
|J_{3}|\leq T
\label{J3-T-condition}
\end{equation}
is
\begin{align}
&
\Psi_{(f_{1}s_{1})(f_{2}s_{2})\ldots(f_{N_{c}}s_{N_{c}})}^{(B)}(x_{1},x_{2},
\ldots,x_{N_{c}})  \notag \\
&
=C_{(f_{1}s_{1})(f_{2}s_{2})\ldots(f_{N_{c}}s_{N_{c}})}^{(B)}x_{1}x_{2}
\ldots x_{N_{c}}\delta(x_{1}+x_{2}+\ldots+x_{N_{c}}-1)\,.  
\label{Psi-B-C-B}
\end{align}
The helicity-flavor part $C_{(f_{1}s_{1})(f_{2}s_{2})
\ldots(f_{N_{c}}s_{N_{c}})}^{(B)}$ coincides with the wave function of the
toy quark model (\ref{psi-NRQM}) up to a normalization factor:
\begin{equation}
C_{(f_{1}s_{1})(f_{2}s_{2})\ldots(f_{N_{c}}s_{N_{c}})}^{(B)}=\tilde{a}
^{(B)}N_{c}^{3N_{c}/2}\psi_{(f_{1}s_{1})(f_{2}s_{2})
\ldots(f_{N_{c}}s_{N_{c}})}^{TT_{3}J_{3}}\,.
\label{C-a-psi}
\end{equation}
The factor of $N_{c}^{3N_{c}/2}$ is inserted in order to simplify the
subsequent expressions (the question about the large-$N_{c}$ behavior of the
coefficient $\tilde{a}^{(B)}$ requires a separate analysis).

Combining Eqs. (\ref{psi-NRQM}), (\ref{Psi-B-C-B}), and (\ref{C-a-psi}), we
find
\begin{align}
& \Psi _{(f_{1}s_{1})(f_{2}s_{2})\ldots
(f_{N_{c}}s_{N_{c}})}^{(B)}(x_{1},x_{2},\ldots ,x_{N_{c}})=\tilde{a}
^{(B)}c_{TN_{c}}N_{c}^{3N_{c}/2}  \notag \\
& \times \left[ \int
dRD_{J_{3}T_{3}}^{T}(R^{-1})\prod\limits_{k=1}^{N_{c}}R_{f_{k}s_{k}}\right]
x_{1}x_{2}\ldots x_{N_{c}}\delta (x_{1}+x_{2}+\ldots +x_{N_{c}}-1)\,.
\label{Psi-as-1}
\end{align}

It is easy to see that this wave function diagonalizes the RHS of the
evolution equation (\ref{evolution-2}). Indeed, using identity 
(\ref{K-tilde-delta}), we find
\begin{equation}
K_{ij}\Psi_{(f_{1}s_{1})(f_{2}s_{2})\ldots(f_{N_{c}}s_{N_{c}})}^{(B)}=\frac{1
}{2}\delta_{s_{i}s_{j}}\Psi_{(f_{1}s_{1})(f_{2}s_{2})
\ldots(f_{N_{c}}s_{N_{c}})}^{(B)}\,.
\end{equation}
The helicity-flavor wave function $C_{(f_{1}s_{1})(f_{2}s_{2})
\ldots(f_{N_{c}}s_{N_{c}})}^{(B)}$ of the baryon with helicity $J_{3}$ has 
$\frac{N_{c}}{2}+J_{3}$ indices $s_{k}=+1/2$ and $\frac{N_{c}}{2}-J_{3}$
indices $s_{k}=-1/2$. Therefore
\begin{align}
& \sum\limits_{1\leq i<j\leq
N_{c}}K_{ij}\Psi_{(f_{1}s_{1})(f_{2}s_{2})\ldots(f_{N_{c}}s_{N_{c}})}^{(B)}=
\sum\limits_{1\leq i<j\leq N_{c}}\frac{1}{2}\delta_{s_{i}s_{j}}
\Psi_{(f_{1}s_{1})(f_{2}s_{2})\ldots(f_{N_{c}}s_{N_{c}})}^{(B)}  \notag \\
& =\frac{1}{2}\left[ \frac{\left( \frac{N_{c}}{2}+J_{3}\right) \left( \frac{
N_{c}}{2}+J_{3}-1\right) }{2}+\frac{\left( \frac{N_{c}}{2}-J_{3}\right)
\left( \frac{N_{c}}{2}-J_{3}-1\right) }{2}\right]
\Psi_{(f_{1}s_{1})(f_{2}s_{2})\ldots(f_{N_{c}}s_{N_{c}})}^{(B)}  \notag \\
& =\frac{1}{2}\left[ \frac{N_{c}}{2}\left( \frac{N_{c}}{2}-1\right) +\left(
J_{3}\right) ^{2}\right] \Psi_{(f_{1}s_{1})(f_{2}s_{2})
\ldots(f_{N_{c}}s_{N_{c}})}^{(B)}\,.  \label{sum-K-as}
\end{align}

Therefore the evolution equation (\ref{evolution-2}) for the wave function
(\ref{Psi-B-C-B}) becomes
\begin{equation}
\mu\frac{\partial}{\partial\mu}\Psi^{\mu}(x_{1},\ldots,x_{N_{c}})=-\frac{
N_{c}+1}{2N_{c}}\frac{\alpha_{s}(\mu)}{2\pi}\left[ \frac{N_{c}}{2}\left( 
\frac{N_{c}}{2}-1\right) +\left( J_{3}\right) ^{2}\right] \Psi^{\mu}(x_{1},
\ldots,x_{N_{c}}).
\label{evolution-asymptotic-exact}
\end{equation}
Taking the limit of large $N_{c}$ and assuming that $J_{3}=O(N_{c}^0)$ in this
limit, we find
\begin{equation}
\mu\frac{\partial}{\partial\mu}\Psi^{\mu}(x_{1},\ldots,x_{N_{c}})=-\frac{
N_{c}^{2}}{16\pi}\alpha_{s}(\mu)\Psi^{\mu}(x_{1},\ldots,x_{N_{c}})\,.
\end{equation}
This corresponds to the $\mu$ dependence
\begin{equation}
\Psi^{\mu}(x_{1},\ldots,x_{N_{c}})=\Psi^{\mu_{0}}(x_{1},\ldots,x_{N_{c}})
\exp\left\{ N_{c}\left[ \sigma(\mu_{0})-\sigma(\mu)\right] \right\}\,,
\label{Psi-mu-as-evolution}
\end{equation}
where function $\sigma(\mu)$ obeys the equation
\begin{equation}
\mu\frac{\partial\sigma(\mu)}{\partial\mu}=\frac{N_{c}}{16\pi}
\alpha_{s}(\mu)\,.  \label{sigma-evolution}
\end{equation}

Changing from $\mu$ to the variable $t$ (\ref{dt-d-mu}) we find
\begin{equation}
\sigma=\frac{1}{8}t\,.  \label{sigma-t-again}
\end{equation}
This agrees with the above results (\ref{sigma-eq}) and (\ref{c-gamma}).

Thus the evolution of the asymptotic distribution amplitude
(\ref{Psi-as-1}) is described by the $\mu$ dependent factor 
\begin{equation}
\tilde{a}^{(B)}(\mu )=a^{(B)}\exp \left[ -N_{c}\sigma (\mu )\right] \,,
\end{equation}
so that
\begin{align}
& \Psi _{(f_{1}s_{1})(f_{2}s_{2})\ldots (f_{N_{c}}s_{N_{c}})}^{(B,\mu
)}(x_{1},x_{2},\ldots ,x_{N_{c}})=a^{(B)}c_{TN_{c}}N_{c}^{3N_{c}/2}\exp 
\left[ -N_{c}\sigma (\mu )\right]  \notag \\
& \times \left[ \int
dRD_{J_{3}T_{3}}^{T}(R^{-1})\prod\limits_{k=1}^{N_{c}}R_{f_{k}s_{k}}\right]
x_{1}x_{2}\ldots x_{N_{c}}\delta (x_{1}+x_{2}+\ldots +x_{N_{c}}-1)  \notag \\
& =a^{(B)}c_{TN_{c}}N_{c}^{(3N_{c}/2)+1}\exp \left[ -N_{c}\sigma (\mu )
\right] \int dRD_{J_{3}T_{3}}^{T}(R^{-1})\int \frac{dZ}{2\pi }
e^{-iZN_{c}}\prod\limits_{k=1}^{N_{c}}\left(
R_{f_{k}s_{k}}x_{k}e^{iZx_{k}N_{c}}\right) \,.
\end{align}

Next we compute the generating functional (\ref{Phi-def-1}) for this wave
function
\begin{align}
& \Phi_{\mu}(g)\overset{\mu\rightarrow\infty}{=}\frac{a^{(B)}c_{TN_{c}}N_{c}
}{2\pi}\exp\left[ -N_{c}\sigma(\mu)\right] \int dR\int
dZD_{J_{3}T_{3}}^{T}(R^{-1})  \notag \\
& \times\exp\left\{ N_{c}\left\{ -iZ+\ln\left[ R_{fs}\int_{0}^{
\infty}dyg_{fs}(y)ye^{iZy}\right] \right\} \right\} \,.  \label{Phi-as-RZ-int}
\end{align}
The integrals over $R$ and $Z$ can be taken using the saddle point method.
Let us first perform the integration over $R$ at fixed $Z$. This can be done
by repeating the calculation of Sec. \ref{Saddle-point-NRQM-section} for the
quark model. We must simply replace
\begin{equation}
g\rightarrow\int_{0}^{\infty}dyg(y)ye^{iZy}
\end{equation}
in Eqs. (\ref{Phi-start-0}) and (\ref{R-via-g-saddle-point}). As a result, we
find the saddle point for the $R$ integral in Eq. (\ref{Phi-as-RZ-int}):
\begin{equation}
R^{-1}=\pm\left[ \int_{0}^{\infty}dyg(y)ye^{iZy}\right] ^{\mathrm{tr}
}\left\{ \det\left[ \int_{0}^{\infty}dyg(y)ye^{iZy}\right] \right\}
^{-1/2}\,.  \label{R-sp-method-2}
\end{equation}
Inserting this result into the exponent of (\ref{Phi-as-RZ-int}), we find
with the exponential accuracy (neglecting factors independent of $g$ and
$\mu $)
\begin{align}
\Phi_{\mu}^{\mathrm{as}}(g) & \sim\exp\left[ -N_{c}\sigma(\mu)\right]  \notag
\\
& \times\int dZ\exp\left\{ N_{c}\left\{ -iZ+\frac{1}{2}\ln\det\left[
\int_{0}^{\infty}dyg(y)ye^{iZy}\right] \right\} \right\} \,.
\label{Phi-as-Z-int}
\end{align}

The saddle point equation for the $Z$ integral is
\begin{equation}
\frac{\partial}{\partial Z}\left\{ -iZ+\frac{1}{2}\ln\det\left[
\int_{0}^{\infty}dyg(y)ye^{iZy}\right] \right\} =0\,.
\label{Z-saddle-point}
\end{equation}
This equation coincides with Eq.~(\ref{X-saddle-point}) if we set
\begin{equation}
X=-iZ\,.
\label{X-via-Z}
\end{equation}
Now we see that the saddle point integration in Eq. (\ref{Phi-as-Z-int})
reproduces our old results (\ref{W0-g-res}) and (\ref{W-mu-as-res}).

\subsection{Comparison of the two methods}

\label{Two-methods-comparison-section}

Thus the two methods described in Sections \ref{W0-solution-section}
and \ref{Asymptotic-solution-method-2} have led us to the same result 
(\ref{W0-g-res}) for the functional $W_0(g)$. Note that in the second method we assumed that the baryon
obeys the condition $|J_{3}|\leq T$ which was needed in order to make use of
the nonrelativistic spin-flavor wave functions (\ref{C-a-psi}). In principle
the condition $|J_{3}|\leq T$ is automatically satisfied for the lowest
$O(N_{c}^{-1})$ excited baryons (\ref{M-T-J}) with $J=T$. But the discussion
of the universality of the functional $W(g)$ in 
Sec. \ref{Universality-W-subsection} shows that the same functional $W(g)$ must
describe not only the lowest $J=T$ baryons (\ref{M-T-J}) but also higher 
$O(N_{c}^{0})$ excitations (\ref{M-harmonic}) and 
baryon-meson scattering states. Note that the first method based on the
asymptotic analysis of the evolution equation for $W(g)$ is completely
compatible with the universality of $W(g)$. Actually the condition 
$|J_{3}|\leq T$ used in the second method can be omitted. Indeed, the origin
of this condition is the totally symmetric form of the $x_{k}$ dependent part of
the asymptotic wave function (\ref{Psi-B-C-B}). In the case of the 
baryons with $|J_{3}|> T$ we cannot keep this form of the $x_{k}$ dependence.
But one can construct the corresponding asymptotic wave function and show
that it is ``almost symmetric'' in the sense that in the large-$N_{c}$ limit
this wave function leads to the generating functional $\Phi (g)$ which has
the same $W(g)$ exponent, and all effects of the asymmetry of the wave
function are localized in the pre-exponential factor $A(g)$.

One of the advantages of the second method considered in 
Sec. \ref{Asymptotic-solution-method-2} is that it allows us to trace the connection
with the traditional analysis of the asymptotic baryon distribution
amplitude. Another good feature of this method is that it clarifies the
saddle point origin (\ref{R-sp-method-2}), (\ref{Z-saddle-point}) of
Eqs. (\ref{delta-W-result}), (\ref{X-saddle-point}) which appear rather formally
in the first method. Concerning this saddle point interpretation, we must
make an important comment. In principle the large-$N_{c}$ limit justifies
the applicability of the saddle point method to the calculation of the
integral (\ref{Phi-as-RZ-int}). However, the large-$N_{c}$ limit \emph{does
not guarantee} that

1) the saddle point equation (\ref{Z-saddle-point}) has one and only one
solution,

2) the $R$ and $Z$ ``integration contours'' can be properly deformed so that
the solution of the saddle point equation gives the dominant contribution.

Generally speaking, the validity of the conditions needed for the applicability
of the saddle point approximation depends on the function $g$ for which we
compute the asymptotic functional $W_{0}(g)$. It is easy to construct
examples of functions $g$ for which the saddle point method works and
counter-examples illustrating the violation of the saddle point method.

To summarize, our representation for the asymptotic functional 
(\ref{W-mu-as-res}) should be considered as a formal expression that carries
information about the saddle point equation but does not fix the relevant
solution of this equation.

\subsection{$N_{f}$ dependence}

\label{Nf-dependence-section}

The results of our analysis of the cases $N_{f}=1$ and $N_{f}=2$
are represented by expressions (\ref{W0-res-Nf-1}) and (\ref{W0-g-res})
for the functional $W_0(g)$. We stress that
the $N_f=1$ and $N_f=2$ results are different. Let us use the notation
$W_0^{(N_f)}(g^{(N_f)})$ in order to mark the $N_f$ dependence.
Starting from some $g_{s}^{(1)}$, let us define
\begin{equation}
g_{fs}^{(2)}=\left\{ 
\begin{array}{cc}
g_{s}^{(1)} & \,\mathrm{if}\,f=1, \\ 
0 & \mathrm{if}\,f=2 \,.
\end{array}
\right.   
\label{g-Nf2-special}
\end{equation}
One could naively expect that
this choice of $g_{fs}^{(2)}$ selects only the contribution of the quarks with 
$f=1 $ so that the values $W_{0}^{(2)}(g^{(2)})$ and 
$W_{0}^{(1)}(g^{(1)})$ must coincide. But this does not happen.
Moreover, the functional $W_{0}^{(2)}(g^{(2)})$ (\ref{W0-g-res}) has a
singularity for the functions (\ref{g-Nf2-special}) because of the vanishing
determinant
\begin{equation}
\det_{fs}\left[ \int_{0}^{\infty}dyg_{fs}^{(2)}(y)ye^{-Xy}
\right] =0\,,
\end{equation}
\begin{equation}
\mathrm{Re}\,W_{0}^{(2)}(g^{(2)})=-\infty\,.  \label{W-Nf2-sing}
\end{equation}
What stands behind this difference of the $N_{f}=1$ and $N_{f}=2$ cases?
Obviously the $N_{f}$ dependence of the nonperturbative dynamics of QCD has
nothing to do with this problem because the evolution equation does not know
anything about this nonperturbative dynamics. This difference has another
origin.

The large-$N_{c}$ baryons which we studied in the $N_{f}=1$ case consist of
quarks with the same flavor $u$. If we ``embed'' this $(uu\ldots u)$ baryon
into the $N_{f}=2$ theory, then it will be classified as a state with
isospin $T_3=N_{c}/2$. Thus our large-$N_{c}$ work in the $N_{f}=1$ sector
should be interpreted in the $N_f=2$ terms as the limit
\begin{equation}
N_{c}\rightarrow\infty\,,\quad T_3=\frac{N_{c}}{2}\rightarrow\infty\,.
\end{equation}
However, the functional $W_{0}^{(2)}(g^{(2)})$ was introduced for another limit:
\begin{equation}
N_{c}\rightarrow\infty\,,\quad T=O(N_{c}^{0})\,.
\end{equation}
Although we did not emphasize the condition $T=O(N_{c}^{0})$, it was really
essential in our $N_f=2$ analysis.
For example, in the saddle point calculation of the $R$ integral 
(\ref{Phi-as-RZ-int}) we did not include the function 
$D_{J_{3}T_{3}}^{T}(R^{-1}) $ into the saddle point equation, which is
allowed only if $T$ is kept fixed in the limit $N_{c}\rightarrow\infty$.

Now we understand the meaning of the singularity (\ref{W-Nf2-sing}). After
the exponentiation this singularity leads to zero:
\begin{equation}
\Phi^{(2)}(g^{(2)})=A^{(2)}(g^{(2)})\exp\left[
N_{c}\,W_{0}^{(2)}(g^{(2)})\right] =0\,.
\end{equation}
The origin of this zero is obvious: function $g_{fs}^{(2)}$ 
(\ref{g-Nf2-special}) selects the states with $T_3=N_{c}/2$ whereas the baryon has
a finite isospin which is kept fixed at large $N_{c}$. This discrepancy
leads to the vanishing functional $\Phi^{(2)}(g^{(2)})$.

Thus our analysis of the two cases $N_{f}=1$ and $N_{f}=2$ used different
assumptions about the behavior of the baryon isospin at large $N_{c}$.

Now we can turn to the case of $N_{f}=3$ flavors. Our experience with the
isospin shows that one has to distinguish between two cases:

1) fixed strangeness $S=O(N_{c}^{0})$,

2) growing strangeness $S=O(N_{c})$.

The case of fixed strangeness can be trivially reduced to the $N_{f}=2$
case. This follows from the universality of the generating functional $W(g)$.
Indeed, in the large-$N_c$ world with $N_f=3$, the nonstrange baryons and baryons with
$O(N_c^0)$ strangeness are described by the same functional
$W^{(3)}(g^{(3)})$. For the nonstrange baryons the functional
$\Phi^{(3)}(g^{(3)})$ (\ref{Phi-def-1}) obviously does not depend on the
strange components of $g_{fs}^{(3)}$ with $f=3$.
Now we apply Eq. (\ref{W-def-1}) to these $g_{3s}^{(3)}$-independent
functionals $\Phi^{(3)}(g^{(3)})$ and conclude that
the functional $W^{(3)}(g^{(3)})$ is also $g_{3s}^{(3)}$-independent.
Due to the universality this property of $W^{(3)}(g^{(3)})$ can be extended
to the case of baryons with $O(N_c^0)$ strangeness.
Thus in the $N_f=3$ world, the baryons with $O(N_c^0)$ strangeness are described by
the functional $W^{(3)}(g^{(3)})$ which is independent of the components
$g_{3s}^{(3)}$. The dependence on $g_{3s}^{(3)}$ appears only in the
pre-exponential functionals $A_B^{(3)}(g^{(3)})$.

\subsection{Analytical properties of $W(g)$}

\label{Analytical-properties-section}

According to Eq. (\ref{Phi-def-1}) $\Phi (g)$ is a homogeneous polynomial
functional of $g$. In particular, $\Phi (g)$ is a
holomorphic functional of $g$. Does this mean that the functional $W(g)$
defined by Eq. (\ref{W-def-1}) is also holomorphic in $g$? The situation is
rather subtle. The expression for $W(g)$ via $\Phi (g)$ following from 
Eq. (\ref{W-def-1}),
\begin{equation}
W(g)=\lim_{N_{c}\rightarrow \infty }\frac{1}{N_{c}}\ln \Phi (g)\,,
\end{equation}
contains a logarithm which can lead to a violation of the analyticity. In
principle, the appearance of this logarithm is a mere notational effect:
instead of Eq. (\ref{W-def-1}) we could work with the large-$N_{c}$
asymptotic representation 
\begin{equation}
\Phi _{B}(g,N_{c})=N_{c}^{\nu }A_{B}(g)\left[ E(g)\right] ^{N_{c}}\left[
1+O(N_{c}^{-1})\right] \,.
\label{Phi-via-Psi}
\end{equation}
Obviously
\begin{equation}
E(g)\equiv e^{W(g)}\,.
\label{Phi-via-Psi-0}
\end{equation}
Using Eq. (\ref{Phi-via-Psi}), we find
\begin{equation}
\left[ E(g)\right] ^{2}=\lim_{N_{c}\rightarrow \infty }\frac{\Phi
_{B}(g,N_{c}+2)}{\Phi _{B}(g,N_{c})}\,.
\label{E2-repr}
\end{equation}
Here $N_{c}$ is shifted by two in the numerator of the RHS. This is
necessary for the cancellation of the factors $A_{B}(g)$. Indeed, according
to Eq. (\ref{TS-sequence}) the quantum numbers of baryons $B$ are different
in theories with odd and even $N_{c}$. Eq.~(\ref{E2-repr}) shows that the
functional $\left[ E(g)\right] ^{2}$ is well defined and has no ambiguities
in contrast to the functional $W(g)$.

In order to illustrate these problems let us consider the toy quark model
described in Sec. \ref{NRQM-section}. Our result (\ref{W-NRQM-res}) for the
toy analog $w(g)$ of the functional $W(g)$ contains a logarithmic
singularity for degenerate matrices  $g$ with $\det g=0$. According to 
Eq. (\ref{W-NRQM-res}) in the toy model we have the analog $\exp w(g)$ of the
functional (\ref{Phi-via-Psi-0})
\begin{equation}
e(g)\equiv \exp w(g)=\sqrt{2\det g}\,.
\end{equation}
We see that the analyticity of $e(g)$ is broken by the square root
singularity that appears for degenerate matrices $g$. As was mentioned 
in Sec. \ref{NRQM-section}, the sign uncertainties of this square root are insignificant
for $\left[ e(g)\right] ^{N_{c}}$ at even $N_{c}$. For odd $N_{c}$ this
sign ambiguity is compensated by a similar root ambiguity of the
pre-exponential $D$ function in Eq. (\ref{A-NRQM-res}).

Now let us turn to the analytical properties of the functional $W_{0}(g)$
describing the asymptotic behavior of the baryon distribution amplitude at high
normalization points $\mu\rightarrow\infty$. To be specific, let us consider
the case $N_{f}=2$ when $W_{0}(g)$ is given by Eq. (\ref{W0-g-res}). It is
easy to see that this expression for $W_{0}(g)$ has two ambiguities.

1) Expression (\ref{W0-g-res}) contains the logarithmic term
\begin{equation}
\frac{1}{2}\ln \det_{fs}\left[ \int_{0}^{\infty }dyg_{fs}(y)ye^{-Xy}
\right]
\end{equation}
which has an additive $\pi in$ ambiguity. This logarithm is accompanied by
the coefficient $1/2$ so that the uncertainties of $\exp \left[ N_{c}W_{0}(g)
\right] $ reduce to $(\pm 1)^{N_{c}}$ and disappear in $\Phi_B(g)$ in the same way like
in the case of the toy quark model.

2) Parameter $X$ in Eq. (\ref{W0-g-res}) is defined implicitly via the
extremum equation (\ref{X-saddle-point}). The ambiguities of the
determination of $X$ from this extremum equation were discussed 
in Sec. \ref{Two-methods-comparison-section}. These ambiguities can also lead to a
violation of the analyticity of the functional $W_{0}(g)$.

Note that the appearance of singularities in $W_{0}(g)$ and the violation of
analyticity are natural from the point of view of the WKB method used in our
large-$N_{c}$ analysis. Such well-known ``singular'' features of the WKB
method like caustics, Stokes lines, etc. have their analogs in our problem of
the baryon wave function at large $N_{c}$.

\section{Diagonalization of anomalous dimensions}

\label{Anomalous-dimensions-section}

\subsection{Anomalous dimensions of baryon operators at large $N_{c}$}

We already know that at large $N_{c}$ the anomalous dimensions $\Gamma
_{\{n_{k}\}}$ of baryonic operators are given by the relation 
(\ref{gamma-n-expansion-beta}). Parameter $E$ appearing in this equation
is determined by Eq.~(\ref{c-gamma-Nf-1}) so that
\begin{equation}
\Gamma_{\{n_{k},J_{3}\}}=\frac{1}{8}N_{c}+\left( \Delta
E+\sum\limits_{k}n_{k}\Omega_{k}\right)
+c_{-1}(n_{1},n_{2},\ldots;J_{3})N_{c}^{-1}+O(N_{c}^{-2})\,.
\label{gamma-n-J3}
\end{equation}
Here $J_{3}$ is the helicity of the corresponding operator. In the case of one
quark flavor $N_{f}=1$, the helicity $J_{3}$ is the only quantum number
which does not contribute in the order $O(N_{c}^{0})$.

The lowest anomalous dimension corresponds to the case when all numbers $n_{k}$
vanish. We know the corresponding anomalous dimension exactly
from the evolution equation for the exact asymptotic wave function
(\ref{evolution-asymptotic-exact}):
\begin{equation}
\Gamma _{\{n_{k}=0\}}=\frac{1}{4}\left( \frac{N_{c}}{2}-1\right) +\frac{
\left( J_{3}\right) ^{2}}{2N_{c}}\,.  \label{gamma-lowest-exact}
\end{equation}
Note that we do not include the factor
\begin{equation}
\frac{N_{c}+1}{N_{c}}\frac{\alpha (\mu )}{2\pi }
\end{equation}
of the evolution equation (\ref{evolution-4}) into the anomalous dimensions 
(\ref{gamma-n-J3}), (\ref{gamma-lowest-exact}) and treat this factor separately. The
exact result for the anomalous dimension $\Gamma _{\{n_{k}=0\}}$ 
(\ref{gamma-lowest-exact}) agrees with the general structure of the $1/N_{c}$
expansion (\ref{gamma-n-J3}). Comparing the exact result
(\ref{gamma-lowest-exact}) with the $1/N_{c}$ expansion 
(\ref{gamma-n-J3}), we can find the parameter
\begin{equation}
\Delta E=-\frac{1}{4}\,.  \label{Delta-E-res}
\end{equation}

\subsection{Traditional approach to the calculation of anomalous dimensions}

The ``excitation energies'' $\Omega_{k}$ appearing in the large-$N_{c}$
representation (\ref{gamma-n-J3}) for the anomalous dimensions of the baryon
operators can be computed by solving Eq.~(\ref{chi-n-eq-1}). Our
derivation of this equation was based on the linearized analysis of the
asymptotic behavior (\ref{W-spectral-decomposition}) of the solution $W(g,t)$
to the evolution equation (\ref{W-evolution-compact}). Let
us show how the same result comes from the traditional approach to the
calculation of anomalous dimensions.

In the traditional approach one has to diagonalize the effective Hamiltonian
corresponding to the evolution equation (\ref{evolution-2})
\begin{equation}
H\tilde{\Psi}_{\alpha}(x_{1},\ldots, x_{N_{c}})=\Gamma_{\alpha}\tilde{\Psi }
_{\alpha}(x_{1},\ldots, x_{N_{c}})\,,  \label{H-Psi-Gamma-Psi}
\end{equation}
\begin{equation}
H=\sum\limits_{1\leq i<j\leq N_{c}}K_{ij}\,.
\end{equation}
Let us show that the eigenvalues $\Gamma_{\alpha}$ appearing here coincide
with the expression (\ref{gamma-n-J3}) at large $N_{c}$. First we rewrite
Eq.~(\ref{H-Psi-Gamma-Psi}) in terms of the generating functional for 
the function $\tilde{\Psi}_{\alpha}(x_{1},\ldots, x_{N_{c}})$. By analogy with Eq. 
(\ref{Phi-def-1}) we define
\begin{align}
\tilde{\Phi}_{\alpha}(g) &
=N_{c}^{-N_{c}/2}\int_{0}^{\infty}dy_{1}\int_{0}^{
\infty}dy_{2}\ldots\int_{0}^{
\infty}dy_{N_{c}}g_{f_{1}s_{1}}(y_{1})g_{f_{2}s_{2}}(y_{2})\ldots
g_{f_{N_{c}}s_{N_{c}}}(y_{N_{c}})  \notag \\
& \times\tilde{\Psi}_{\alpha,(f_{1}s_{1})(f_{2}s_{2})
\ldots(f_{N_{c}}s_{N_{c}})}\left( \frac{y_{1}}{N_{c}},\frac{y_{2}}{N_{c}}
,\ldots ,\frac{y_{N_{c}}}{N_{c}}\right) \,.
\end{align}
Repeating the same steps that were used in the derivation of 
Eq. (\ref{evolution-4}), we find
\begin{equation}
\frac{1}{2}\left( g\otimes g\right) \cdot K\cdot\left[ \left( \frac{\delta}{
\delta g}\otimes\frac{\delta}{\delta g}\right) \tilde{\Phi }_{\alpha}(g)
\right] =\Gamma_{\alpha}\tilde{\Phi}_{\alpha}(g)\,.
\label{Phi-alpha-Schroedinger}
\end{equation}
Next we write the large-$N_{c}$ ansatz (\ref{W-def-1}) 
\begin{equation}
\tilde{\Phi}_{\alpha}(g)=N_{c}^{\nu_{\alpha}}\tilde{A}_{\alpha}(g)\exp\left[
N_{c}W_{0}(g)\right] \left[ 1+O\left( N_{c}^{-1}\right) \right]
\label{Pfi-A-alpha}
\end{equation}
and expand
\begin{equation}
\Gamma_{\alpha}=N_{c}E+\Delta\Gamma_{\alpha}+O\left( N_{c}^{-1}\right) \,.
\label{Gamma-alpha-Nc}
\end{equation}

We use the tilded notation $\tilde{\Psi}_{\alpha}$, $\tilde{\Phi}_{\alpha}(g)$, 
$\tilde{A}_{\alpha}(g)$ in order to distinguish the wave functions and
functionals associated with the diagonalization problem 
(\ref{H-Psi-Gamma-Psi}) from the wave functions and functionals
$\Psi$, $\Phi(g)$, $A(g)$ corresponding to the physical baryons states. However, note that
the functionals $W_{0}(g)$, $\xi_{k}(g)$ appearing in this section are the
same as before.

Inserting ansatz (\ref{Pfi-A-alpha}) into Eq. (\ref{Phi-alpha-Schroedinger}),
we reproduce Eq. (\ref{W0-c-eq}) in the leading order of the $1/N_{c}$
expansion. In the next order we obtain the equation
\begin{gather}
\left( g\otimes g\right) \cdot K\cdot\left[  \frac{\delta W_{0}(g)}{
\delta g}\otimes\frac{\delta\ln\tilde{A}_{\alpha}(g)}{\delta g}
\right]  \notag \\
+\frac{1}{2}\left( g\otimes g\right) \cdot K\cdot\left[ \left( \frac{\delta}{
\delta g}\otimes\frac{\delta}{\delta g}\right) W_{0}(g)\right]
=\Delta\Gamma_{\alpha}\,.  \label{A-alpha-eq}
\end{gather}

In addition to this equation the functionals $\tilde{A}_{\alpha}(g)$ must be
homogeneous,
\begin{equation}
\tilde{A}_{\alpha}(\lambda g)=\tilde{A}_{\alpha}(g)\,,
\label{A-alpha-homogeneous}
\end{equation}
for any number $\lambda$. This condition follows from the property
\begin{equation}
\tilde{\Phi}_{\alpha}(\lambda g)=\lambda^{N_{c}}\tilde{\Phi}_{\alpha}(g)
\end{equation}
and from Eq. (\ref{W-0-lambda}).

Let us use the value $\alpha=0$ for the ground state. Taking the difference
of the equations (\ref{A-alpha-eq}) with $\alpha\neq 0$ and with $\alpha=0$,
we find
\begin{equation}
\left( g\otimes g\right) \cdot K\cdot\left[  \frac{\delta W_{0}(g)}{
\delta g}\otimes\frac{\delta}{\delta g}\ln\frac{\tilde{A}_{\alpha}(g)}{
\tilde{A}_{0}(g)} \right] =\Delta\Gamma_{\alpha}-\Delta \Gamma_{0}\,.
\end{equation}
Now let us make the ansatz
\begin{equation}
\tilde{A}_{\alpha}(g)=\tilde{A}_{0}(g)\prod\limits_{k}\left[ \xi _{k}(g)
\right] ^{n_{k}}\,,  \label{A-chi-decomposition}
\end{equation}
where $\xi_{k}(g)$ are solutions of Eq. (\ref{chi-n-eq-1}). Then
\begin{equation}
\left( g\otimes g\right) \cdot K\cdot\left[  \frac{\delta W_{0}(g)}{
\delta g}\otimes\frac{\delta}{\delta g}\sum\limits_{k}n_{k}\ln\xi
_{k}(g) \right] =\Delta\Gamma_{\alpha}-\Delta\Gamma_{0}\,.
\label{Delta-E-alpha-2}
\end{equation}
According to Eq. (\ref{chi-n-eq-1}) we have
\begin{equation}
\left( g\otimes g\right) \cdot K\cdot\left[ \frac{\delta W_{0}(g)}{\delta g}
\otimes\frac{\delta}{\delta g}\ln\xi_{k}(g)\right] =\Omega_{k}\,.
\label{ln-chi-eq}
\end{equation}
Inserting Eq. (\ref{ln-chi-eq}) into Eq. (\ref{Delta-E-alpha-2}), we find
\begin{equation}
\Delta\Gamma_{\alpha}-\Delta\Gamma_{0}=\sum\limits_{k}n_{k}\Omega_{k}\,.
\end{equation}
Now we combine this with Eq. (\ref{Gamma-alpha-Nc}):
\begin{equation}
\Gamma_{\alpha}-\Gamma_{0}=\sum\limits_{k}n_{k}\Omega_{k}+O(N_{c}^{-1})\,.
\end{equation}
Taking $\alpha=\{n_{k}\}$, we see that this result agrees with 
Eq. (\ref{gamma-n-J3}) and
\begin{equation}
\Delta\Gamma_{0}=\Delta E\,.  \label{Delta-Gamma-0-Delta-E}
\end{equation}
Thus at large $N_{c}$ the diagonalization of the effective Hamiltonian 
(\ref{H-Psi-Gamma-Psi}) reproduces our old expression for the anomalous
dimensions of baryon operators (\ref{gamma-n-J3}).

\subsection{Calculation of excitation energies $\Omega_{k}$}

\label{Calculation-Omega-section}

Now we want to find the excitation energies $\Omega_{k}$. This can be done
by solving Eq. (\ref{chi-n-eq-1}). In this paper we solve this equation in
the simplest case of one flavor $N_{f}=1$. In order to construct this
solution we need auxiliary functionals
\begin{equation}
M_{ks}(g)=\int_{0}^{\infty}dyg_{s}(y)y^{k+1}e^{-X(g)y}\,,
\label{M-def-new}
\end{equation}
\begin{equation}
T_{ks}(g)=\frac{M_{ks}(g)}{M_{0s}(g)}\,.  \label{T-def-new}
\end{equation}

We shall see that the solutions $\xi_{m}(g)$ of Eq.~(\ref{chi-n-eq-1})
can be constructed as polynomials
\begin{equation}
\xi_{m}(g)=P_{m}[T(g)]\,.  \label{chi-via-T}
\end{equation}
Obviously we have for any constant $\lambda$
\begin{equation}
T_{ks}(\lambda g)=T_{ks}(g)\,.
\end{equation}
Therefore the functionals $\xi_{m}(g)$ constructed as polynomials of $T_{ks}$
(\ref{chi-via-T}) will obey condition (\ref{chi-W-scale-inv}).

Assuming the structure (\ref{chi-via-T}) of the solution, we find
\[
\frac{\delta\xi_{m}(g)}{\delta g_{s_{2}}(y)}=\sum\limits_{ks}\frac{\partial
P_{m}(T)}{\partial T_{ks}}\frac{\delta T_{ks}(g)}{\delta g_{s_{2}}(y)}\,,
\]
so that Eq.~(\ref{chi-n-eq-1}) can be rewritten in the form
\begin{equation}
\sum\limits_{ks}\frac{\partial P_{m}(T)}{\partial T_{ks}}\left\{ \left(
g\otimes g\right) \cdot K\cdot\left[ \frac{\delta W_{0}(g)}{\delta g}\otimes
\frac{\delta T_{ks}(g)}{\delta g}\right] \right\} =\Omega_{m}P_{m}[T(g)]\,.
\label{chi-n-eq-F}
\end{equation}
Below we establish an important property of functionals $T_{ks}(g)$:
\begin{align}
& (g\otimes g)\cdot K\cdot\left[ \frac{\delta W_{0}(g)}{\delta g}\otimes\,
\frac{\delta T_{ks}(g)}{\delta g}\right]  \notag \\
& =\frac{1}{4}\left[ -T_{ks}(g)+\sum\limits_{s_{1}}\sum
\limits_{n=0}^{k}B_{kn}^{s_{1}s}T_{ns_{1}}(g)T_{k-n,s}(g)\right] \,,
\label{K-T-action}
\end{align}
where $B_{kn}^{s_{1}s}$ are some coefficients. The derivation of the
equality (\ref{K-T-action}) and the calculation of the coefficients 
$B_{kn}^{s_{1}s}$ requires some work 
[see Eq.~(\ref{K-T-action-0}) in Appendix~\ref{T-properties-appendix}]. But once this work is done, the solution of
Eq.~(\ref{chi-n-eq-F}) becomes simple. Inserting Eq. (\ref{K-T-action})
into Eq. (\ref{chi-n-eq-F}), we find
\begin{equation}
\frac{1}{4}\sum\limits_{ks}\frac{\partial P_{m}(T)}{\partial T_{ks}}\left[
-T_{ks}+\sum\limits_{s_{1}}\sum
\limits_{n=0}^{k}B_{kn}^{s_{1}s}T_{ns_{1}}T_{k-n,s}\right]
=\Omega_{m}P_{m}(T)\,.  \label{T-algebraic}
\end{equation}
We look for the polynomial solutions (\ref{chi-via-T})
\begin{equation}
P_{m}(T)=\sum\limits_{\{k_{i},s_{i}\}}a_{\{k_{i}s_{i}
\}}^{(m)}T_{k_{1}s_{1}}T_{k_{2}s_{2}}\ldots T_{k_{n}s_{n}}\,.
\end{equation}
Let us assign the ``weight'' $k$ to the functional $T_{ks}$. Then the
product $T_{k_{1}s_{1}}T_{k_{2}s_{2}}\ldots T_{k_{n}s_{n}}$ will have the
weight
\begin{equation}
m=k_{1}+k_{2}+\ldots +k_{n}\,.  \label{m-weight}
\end{equation}
Obviously Eq.~(\ref{T-algebraic}) is closed with respect to
polynomials $P_{m}(T)$ of any given weight $m$. Therefore we can classify
the solutions according to their weight (\ref{m-weight}). This is the reason
why we identify the notation $m$ for weight (\ref{m-weight}) and the index 
$m $ of the solution $P_{m}(T)$. Let us concentrate on the linear terms 
$T_{ms}$ in $P_{m}(T)$:
\begin{equation}
P_{m}(T)=\sum\limits_{s}a_{ms}T_{ms}+\mathrm{(nonlinear\,terms)}\,.
\label{F-T-str}
\end{equation}
Separating these linear terms in Eq.~(\ref{T-algebraic}), we find
\begin{equation}
\frac{1}{4}\sum\limits_{s}a_{ms}\left[ -T_{ms}+\sum\limits_{s_{1}}\left(
B_{mm}^{s_{1}s}T_{ms_{1}}+B_{m0}^{s_{1}s}T_{ms}\right) \right] =\Omega
_{m}\sum\limits_{s}a_{ms}T_{ms}\,.  \label{a-T-Omega}
\end{equation}
We have taken into account the property
\begin{equation}
T_{0s}(g)=1  \label{T-0-one}
\end{equation}
which is an obvious consequence of Eq. (\ref{T-def-new}). The coefficients 
$B_{mn}^{s_{1}s_{2}}$ depend only on the relative orientation of the
helicities $s_{1}$, $s_{2}$ so that we can use the notation
\begin{equation}
B_{mn}^{ss}=B_{mn}^{+}\,,\quad B_{mn}^{-s,s}=B_{mn}^{-}\,.
\end{equation}
Then
\begin{equation}
\sum\limits_{s_{1}}\left(
B_{mm}^{s_{1}s}T_{ms_{1}}+B_{m0}^{s_{1}s}T_{ms}\right) =\left(
B_{mm}^{+}+B_{m0}^{+}\right) T_{ms}+\left(
B_{mm}^{-}T_{m,-s}+B_{m0}^{-}T_{ms}\right)
\end{equation}
and Eq. (\ref{a-T-Omega}) takes the form
\begin{equation}
\frac{1}{4}\sum\limits_{s}a_{ms}\left[ \left(
B_{mm}^{+}+B_{m0}^{+}+B_{m0}^{-}-1\right) T_{ms}+B_{mm}^{-}T_{m,-s}\right]
=\Omega _{m}\sum\limits_{s}a_{ms}T_{ms}\,.  \label{a-T-Omega-2}
\end{equation}
Obviously we have two solutions $a_{ms}^{\pm }$ for a given weight $m$. One
solution is even in $s$
\begin{equation}
a_{ms}^{+}=a_{m,-s}^{+}\,,
\end{equation}
and the other is odd
\begin{equation}
a_{ms}^{-}=-a_{m,-s}^{-}\,.
\end{equation}
According to Eqs. (\ref{chi-via-T}) and (\ref{F-T-str}) these two possibilities
lead to the solutions
\begin{equation}
\xi _{m}^{\pm }(g)\equiv P_{m}^{\pm }[T(g)]=\left[ T_{m+}(g)\pm T_{m-}(g)
\right] +\mathrm{[terms\,nonlinear\,in\,}T_{ks}(g)\mathrm{]}\,.
\end{equation}
Parameters $\Omega _{m}^{\pm }$ associated with these solutions can be found
from Eq. (\ref{a-T-Omega-2}):
\begin{equation}
\Omega _{m}^{(\pm )}=\frac{1}{4}\left( B_{mm}^{+}\pm
B_{mm}^{-}+B_{m0}^{+}+B_{m0}^{-}-1\right) \,.  \label{Omega-via-B}
\end{equation}
The explicit expressions for $B_{mm}^{\pm }$ and $B_{m0}^{\pm }$ can be
found in Appendix \ref{T-properties-appendix} [see Eqs. (\ref{B-kk}) -- 
(\ref{B-k0})]. As a result, we find
\begin{equation}
\Omega _{m}^{-}=\frac{1}{2}\left[ \frac{1}{2}-\frac{1}{m+2}+\frac{1}{
(m+1)(m+2)}\right] +\sum\limits_{j=2}^{m+1}\frac{1}{j}\,,
\label{Omega-minus}
\end{equation}
\begin{equation}
\Omega _{m}^{+}=\Omega _{m}^{-}-\frac{2}{m(m+2)}\,.  \label{Omega-plus}
\end{equation}
The corresponding solutions of Eq.~(\ref{chi-n-eq-F}) will be
denoted $\xi _{m}^{\pm }(g)$:
\begin{equation}
\left( g\otimes g\right) \cdot K\cdot \left[ \frac{\delta W_{0}(g)}{\delta g}
\otimes \frac{\delta \xi _{m}^{\pm }(g)}{\delta g}\right] =\Omega _{m}^{\pm
}\xi _{m}^{\pm }(g)\,.  \label{Omega-m-zeta}
\end{equation}

In the above derivation we had $m\geq 1$. However, one should keep in mind
that
\begin{equation}
\Omega _{1}^{+}=0\,.
\end{equation}
The solution of Eq.~(\ref{chi-n-eq-F}) for this zero mode is
\begin{equation}
\xi _{1}^{+}(g)=T_{1,+}(g)+T_{1,-}(g)\,.
\end{equation}
Computing the derivative with respect to $X$ in Eq. (\ref{X-g-equation-Nf-1}),
we find
\begin{equation}
\sum\limits_{s}T_{1s}(g)=2\,.  \label{T1-sum}
\end{equation}
This means that
\begin{equation}
\xi _{1}^{+}(g)=2
\end{equation}
so that Eq.~(\ref{Omega-m-zeta}) for $\xi _{1}^{+}(g)$
\begin{equation}
\left( g\otimes g\right) \cdot K\cdot \left[ \frac{\delta W_{0}(g)}{\delta g}
\otimes \frac{\delta \xi _{1}^{+}(g)}{\delta g}\right] =0
\end{equation}
is satisfied in a trivial way.

In Appendix \ref{T-properties-appendix} we find another zero mode of
Eq.~(\ref{chi-n-eq-1}). According to Eq. (\ref{M1-pm-zero-mode}) we have
\begin{equation}
(g\otimes g)\cdot K\cdot\left[ \frac{\delta W_{0}(g)}{\delta g}\otimes\,
\frac{\delta\xi_{\mathrm{rot}}(g)}{\delta g}\right] =0\,,  \label{chi-rot-eq}
\end{equation}
where
\begin{equation}
\xi_{\mathrm{rot}}(g)=\frac{M_{0+}(g)}{M_{0-}(g)}\,.  \label{J-zero-mode}
\end{equation}

This zero mode corresponds to the spin rotations around the space component
of the light-cone vector $n^{\mu}$ used in the definition of the
distribution amplitude (\ref{chi-psi}), (\ref{BDA-def}). The role of this
zero mode will be discussed in Sec. \ref{Helicity-section}.

Let us summarize. The spectrum of the anomalous dimensions of baryon
operators is described by Eq.~(\ref{gamma-n-J3}),
\begin{equation}
\Gamma _{\{n_{m}^{\pm },J_{3}\}}=\frac{1}{8}N_{c}+\left( \Delta
E+\sum\limits_{m\geq 2}n_{m}^{+}\Omega _{m}^{+}+\sum\limits_{m\geq
1}n_{m}^{-}\Omega _{m}^{-}\right) +O(N_{c}^{-1})\,,  \label{gamma-final}
\end{equation}
with the parameter $\Delta E$ (\ref{Delta-E-res}) and with the excitation
energies $\Omega _{k}^{\pm }$ (\ref{Omega-minus}), (\ref{Omega-plus}).

\subsection{Helicity}

\label{Helicity-section}

Our result for the anomalous dimensions (\ref{gamma-final}) is degenerate in
helicity $J_{3}$. This degeneracy is lifted in the order $O(N_{c}^{-1})$. It
can be seen from the exact (in $N_{c}$) result (\ref{gamma-lowest-exact})
for the operators corresponding to $n_{m}^{\pm}=0$. In fact, the conservation of
helicity allows us to diagonalize the evolution in terms of generating
functionals $\tilde{A}_{\alpha }(g)$ (\ref{A-chi-decomposition}) without
computing the $O(N_{c}^{-1})$ corrections to the anomalous dimensions.

Under the axial rotations corresponding to the helicity
\begin{equation}
g_{s}^{\prime }=g_{s}e^{is\phi }\,,  \label{helicity-rotation}
\end{equation}
the functionals $M_{ks}(g)$ (\ref{M-def-new}) and $T_{ks}(g)$ 
(\ref{T-def-new}) transform as follows
\begin{align}
M_{ks}(g^{\prime })& =e^{i\phi s}M_{ks}(g)\,,  \label{M-helicity-rotation} \\
T_{ks}(g^{\prime })& =T_{ks}(g)\,.
\end{align}
The solutions $\xi _{m}^{\pm }(g)$ are constructed as functions of $T_{ks}(g)
$ [see Eq. (\ref{chi-via-T})]. Therefore these solutions are invariant under
rotations (\ref{helicity-rotation})
\begin{equation}
\xi _{m}^{\pm }(g^{\prime })=\xi _{m}^{\pm }(g)\,.  \label{chi-rotation}
\end{equation}
Using Eq. (\ref{M-helicity-rotation}), we find for the zero mode 
(\ref{J-zero-mode})
\begin{equation}
\xi _{\mathrm{rot}}(g^{\prime })=e^{i\phi }\xi _{\mathrm{rot}}(g)\,.
\label{chi-rot-rotation}
\end{equation}

In the previous section we showed that the anomalous dimensions of baryon
operators diagonalizing the evolution can be parametrized by
\begin{equation}
\alpha =(n_{m}^{\pm },J_{3})\,,
\end{equation}
where $n_{m}^{\pm }$ are integer numbers and $J_{3}$ is the helicity of the
baryon operator. Assuming this meaning of the subscript $\alpha $ in the
generating functional $\tilde{A}_{\alpha }$ (\ref{Pfi-A-alpha})
\begin{equation}
\tilde{A}_{\alpha }(g)\equiv \tilde{A}_{\{n_{m}^{\pm },J_{3}\}}(g)\,,
\end{equation}
we can write for the rotations (\ref{helicity-rotation})
\begin{equation}
\tilde{A}_{\{n_{m}^{\pm },J_{3}\}}(g^{\prime })=e^{iJ_{3}\phi }\tilde{A}
_{\{n_{m}^{\pm },J_{3}\}}(g)\,.  \label{A-rotation}
\end{equation}

Now let us rewrite the decomposition (\ref{A-chi-decomposition}) in the form
\begin{equation}
\tilde{A}_{\{n_{m}^{\pm},J_{3}\}}(g)=\tilde{A}_{0}(g)\left\{ \prod
\limits_{m\geq2}\left[ \xi_{m}^{+}(g)\right] ^{n_{m}^{+}}\right\} \left\{
\prod\limits_{m\geq1}\left[ \xi_{m}^{-}(g)\right] ^{n_{m}^{-}}\right\} \left[
\xi_{\mathrm{rot}}(g)\right] ^{J_{3}}\,.  \label{A-decomposition-2}
\end{equation}
Here we have separated the contribution of the zero mode $\xi_{\mathrm{rot}}(g)$
(\ref{J-zero-mode}). Applying transformation (\ref{helicity-rotation})
to the decomposition (\ref{A-decomposition-2}) and using 
Eqs. (\ref{chi-rotation}), (\ref{chi-rot-rotation}), and (\ref{A-rotation}), we find
\begin{equation}
\tilde{A}_{0}(g^{\prime})=\tilde{A}_{0}(g)\,.  \label{A0-rotation}
\end{equation}

\subsection{Functional $\tilde{A}_{0}(g)$}

Now we want to compute the functional $\tilde{A}_{0}(g)$ appearing in the
decomposition (\ref{A-decomposition-2}). First let us derive an equation for
this functional. Taking Eq. (\ref{A-alpha-eq}) for the ground state $\alpha
=0$ and using relations (\ref{Delta-E-res}), (\ref{Delta-Gamma-0-Delta-E}),
we obtain
\begin{gather}
\left( g\otimes g\right) \cdot K\cdot \left[  \frac{\delta W_{0}(g)}{
\delta g}\otimes \frac{\delta \ln \tilde{A}_{0}(g)}{\delta g} \right]
\notag \\
+\frac{1}{2}\left( g\otimes g\right) \cdot K\cdot \left[ \left( \frac{\delta 
}{\delta g}\otimes \frac{\delta }{\delta g}\right) W_{0}(g)\right] =-\frac{1
}{4}\,.  \label{A-0-eq}
\end{gather}
The solution of this equation can be written in terms of functionals $T_{ks}$
(\ref{T-def-new}):
\begin{equation}
\tilde{A}_{0}(g)=\left\{ \sum\limits_{s}\left\{ T_{2s}(g)
-\left[T_{1s}(g)\right] ^{2}\right\} \right\} ^{-1/2}\,.  \label{A0-res}
\end{equation}
A straightforward but tedious calculation [using relations (\ref{K-T-action}) and
(\ref{delta-W-M1})] shows that this functional obeys equation 
(\ref{A-0-eq}). Note that Eq.~(\ref{A-0-eq}) does not fix the solution 
$\tilde{A}_{0}(g)$ completely. For example, due to Eq. (\ref{chi-rot-eq}) we
have the freedom of transformations
\begin{equation}
\tilde{A}_{0}(g)\rightarrow \tilde{A}_{0}(g)\left[ \xi _{\mathrm{rot}}(g)
\right] ^{n}\,.
\end{equation}
The ambiguity is fixed by the additional condition (\ref{A0-rotation}).
Obviously solution (\ref{A0-res}) obeys constraint (\ref{A0-rotation}) as
well as the constraint (\ref{A-alpha-homogeneous}).

\section{Functional $A_{B}(g)$ in the case $N_{f}=2$}

\label{A-B-Nf-2-main-section}

\subsection{Functional $A_{B}(g,t)$ in the asymptotic regime $t\rightarrow
\infty $}

\label{A-B-Nf-2-section}

In Sec. \ref{Two-methods-comparison-section} we showed how the saddle point
method can be used for the calculation of the functional $W_{0}(g)$
(\ref{W0-g-res}). The same saddle point method also allows us to compute the
pre-exponential factor $A_{B}(g)$. To this aim we return to 
Eq. (\ref{Phi-as-RZ-int}) and perform the integration over $R$ more carefully, using
Eq. (\ref{R-int}). As a result, we find
\begin{align}
& \Phi _{\mu }(g)\overset{\mu \rightarrow \infty }{=}\frac{
a^{(B)}c_{TN_{c}}N_{c}}{2\pi }\frac{2^{N_{c}+2}}{\sqrt{2\pi N_{c}^{3}}}\exp 
\left[ -N_{c}\sigma (\mu )\right]  \notag \\
& \times \int dZD_{J_{3}T_{3}}^{T}\left( R_{0}^{-1}\right) \exp \left\{
N_{c}\left\{ -iZ+\frac{1}{2}\ln \det \left[ \int_{0}^{\infty
}dyg(y)ye^{iZy}\right] \right\} \right\}\,,
\end{align}
where
\begin{equation}
R_{0}^{-1}=\frac{g_{\mathrm{eff}}^{\mathrm{tr}}}{\sqrt{\det g_{\mathrm{eff}}}
}\,,  \label{R0-res}
\end{equation}
\begin{equation}
g_{\mathrm{eff}}=\int_{0}^{\infty }dyg(y)ye^{iZy}\,.
\end{equation}
Next we perform the saddle point integration over $Z$ which yields
\begin{align}
& \Phi _{\mu }(g)\overset{\mu \rightarrow \infty }{=}\frac{2^{N_{c}+1}\sqrt{2
}a^{(B)}c_{TN_{c}}}{\pi N_{c}}\exp \left[ -N_{c}\sigma (\mu )\right]
D_{J_{3}T_{3}}^{T}(R_{0}^{-1})  \notag \\
& \times \left\{ -\frac{\partial ^{2}}{\partial Z^{2}}\ln \det \left[
\int_{0}^{\infty }dyg(y)ye^{iZy}\right] \right\} ^{-1/2}\exp \left\{
N_{c}\left\{ -iZ+\frac{1}{2}\ln \det \left[ \int_{0}^{\infty
}dyg(y)ye^{iZy}\right] \right\} \right\}\,,
\label{Phi-Nf2-calc-2}
\end{align}
where $Z$ is determined by the saddle point equation (\ref{Z-saddle-point}).
It is convenient to use the variable $X=-iZ$ (\ref{X-via-Z}) instead of $Z$.
Introducing the compact notation
\begin{equation}
L_{k}=\int_{0}^{\infty }dyg(y)y^{k+1}e^{-Xy}\,,  \label{L-k-def}
\end{equation}
we can rewrite Eq. (\ref{R0-res}) in the form
\begin{equation}
R_{0}^{-1}=\frac{L_{0}^{\mathrm{tr}}}{\sqrt{\det L_{0}}}\,.
\label{R0-via-L0}
\end{equation}
Using notation (\ref{L-k-def}), we find
\begin{equation}
\frac{\partial }{\partial Z}\left\{ -iZ+\frac{1}{2}\ln \det \left[
\int_{0}^{\infty }dyg(y)ye^{iZy}\right] \right\} =i\left[ -1+\frac{1}{
2}\mathrm{Tr}\left( L_{0}^{-1}L_{1}\right) \right]\,.  \label{dZ-via-L}
\end{equation}
Applying the identity
\begin{equation}
\frac{\partial }{\partial Z}\mathrm{Tr}\left( L_{0}^{-1}L_{1}\right) =i
\mathrm{Tr}\left[ L_{0}^{-1}L_{2}-\left( L_{0}^{-1}L_{1}\right) ^{2}\right]
\,,  \label{dZ-L0-L1}
\end{equation}
we derive from Eq. (\ref{dZ-via-L})
\begin{equation}
\frac{\partial ^{2}}{\partial Z^{2}}\left\{ \ln \det \left[
\int_{0}^{\infty }dyg(y)ye^{iZy}\right] \right\} =-\mathrm{Tr}\left[
L_{0}^{-1}L_{2}-\left( L_{0}^{-1}L_{1}\right) ^{2}\right] \,.
\label{Nf-2-d2Z}
\end{equation}
Inserting Eq. (\ref{dZ-via-L}) into the saddle point equation
(\ref{Z-saddle-point}), we find
\begin{equation}
\mathrm{Tr}\left( L_{0}^{-1}L_{1}\right) =2\,.  \label{L0-L1-id}
\end{equation}
This saddle point equation implicitly defines the dependence of $X$ on $g$
\begin{equation}
X=X(g)\,.
\end{equation}
Inserting this dependence $X(g)$ into Eq. (\ref{L-k-def}) we see that $L_{k}$
also become functionals of $g$ only
\begin{equation}
L_{k}=L_{k}(g)\,.
\end{equation}

Using Eqs. (\ref{L-k-def}), (\ref{R0-via-L0}), and (\ref{Nf-2-d2Z}), we find from
Eq. (\ref{Phi-Nf2-calc-2})
\begin{align}
& \Phi _{\mu }(g)\overset{\mu \rightarrow \infty }{=}\frac{2^{N_{c}+1}\sqrt{2
}a^{(B)}c_{TN_{c}}}{\pi N_{c}}\exp \left[ -N_{c}\sigma (\mu )\right]
D_{J_{3}T_{3}}^{T}\left( \frac{L_{0}^{\mathrm{tr}}}{\sqrt{\det L_{0}}}\right)
\notag \\
& \times \left\{ \mathrm{Tr}\left[ L_{0}^{-1}L_{2}-\left(
L_{0}^{-1}L_{1}\right) ^{2}\right] \right\} ^{-1/2}\exp \left[ N_{c}\left(
-iZ+\frac{1}{2}\ln \det L_{0}\right) \right] \,.
\end{align}
Now we insert expression (\ref{CT-Nc-Large-Nc}) for $c_{TN_{c}}$ and apply
the property of Wigner functions (\ref{D-R-transposed}):
\begin{align}
& \Phi _{\mu }(g)\overset{\mu \rightarrow \infty }{=}a^{(B)}2^{N_{c}/2}
\left( \frac{8}{\pi ^{3}N_{c}}\right) ^{1/4}\exp \left[ -N_{c}\sigma (\mu )
\right] \sqrt{2T+1}D_{T_{3}J_{3}}^{T}\left( \frac{L_{0}}{\sqrt{\det L_{0}}}
\right)   \notag \\
& \times \left\{ \mathrm{Tr}\left[ L_{0}^{-1}L_{2}-\left(
L_{0}^{-1}L_{1}\right) ^{2}\right] \right\} ^{-1/2}\exp \left[ N_{c}\left(
-iZ+\frac{1}{2}\ln \det L_{0}\right) \right] \,.
\end{align}
In the leading order of the $1/N_{c}$ expansion, $\sigma (\mu )$ is given by
expression (\ref{sigma-t-again}). We must also take into account the
$O(N_{c}^{0})$ correction $\Delta E$ to the energy $N_cE$. Then
\begin{align}
& \Phi (g,t)\overset{t\rightarrow \infty }{=}a^{(B)}N_{c}^{-1}2^{N_{c}/2}
\left( \frac{8}{\pi ^{3}N_{c}}\right) ^{1/4}\exp \left[ -\left(
N_{c}E+\Delta E\right) t\right] \sqrt{2T+1}D_{T_{3}J_{3}}^{T}\left( \frac{
L_{0}}{\sqrt{\det L_{0}}}\right)   \notag \\
& \times \left\{ \mathrm{Tr}\left[ L_{0}^{-1}L_{2}-\left(
L_{0}^{-1}L_{1}\right) ^{2}\right] \right\} ^{-1/2}\exp \left\{ N_{c}\left[
X(g)+\frac{1}{2}\ln \det L_{0}\right] \right\} \,.
\end{align}
This leads to the old result (\ref{W-mu-as-res}) for the functional $W_{
\mathrm{as}}(g,t)$
\begin{equation}
W_{\mathrm{as}}(g,t)=W_{0}(g)-Et\,,
\end{equation}
\begin{equation}
W_{0}(g)=X(g)+\frac{1}{2}\ln \left[ \det L_{0}(g)\right]   \label{W0-L0-Nf2}
\end{equation}
but now we also know the functional $A_{B}(g,t)$
\begin{equation}
A_{B}^{T,T_{3}J_{3}}(g,t)=c_{B}\sqrt{2T+1}D_{T_{3}J_{3}}^{T}\left[ Q_{0}(g)
\right] A_{0}(g,t)\,  \label{A-B-via-L-direct}\,,
\end{equation}
where $c_{B}$ is a constant independent of $g$ and
\begin{equation}
Q_{0}(g)=\frac{L_{0}(g)}{\sqrt{\det L_{0}(g)}}\,,  \label{Q-L0-Nf-2}
\end{equation}
\begin{equation}
A_{0}(g,t)=\exp \left( -t\Delta E\right) \left\{ \mathrm{Tr}\left\{
L_{0}^{-1}(g)L_{2}(g)-\left[ L_{0}^{-1}(g)L_{1}(g)\right] ^{2}\right\}
\right\} ^{-1/2}\,.  \label{A-0-res-Nf-2}
\end{equation}
Note that the functional $A_{B}(g,t)$ depends on the type of the state $B$.
In the asymptotic regime $t\rightarrow \infty $ this dependence appears in
Eq. (\ref{A-B-via-L-direct}) only via the quantum numbers $TJ_{3}T_{3}$ and
via the constant $c_{B}$. The origin of this simple dependence is obvious:
the asymptotic distribution amplitude is controlled by the single operator with the
lowest anomalous dimension corresponding to given $TJ_{3}T_{3}$. The result
(\ref{A-B-via-L-direct}) can be used only for baryons obeying condition
$|J_{3}|\leq T$ (\ref{J3-T-condition}), because the underlying expression
(\ref{Psi-B-C-B}) for the asymptotic distribution amplitude was valid only for
this case. For the lowest $O(N_{c}^{-1})$ baryon excitations with $J=T$
(\ref{TS-sequence}), the condition $|J_{3}|\leq T$ is satisfied automatically.

\subsection{Comments on the noncommutativity of limits $N_{c}\rightarrow
\infty $ and $t\rightarrow \infty $}

In the previous section we have computed the functional $A_{B}(g,t)$ in the
asymptotic regime. This calculation was based on the analysis of the double
limit $N_{c}\rightarrow \infty $ and $t\rightarrow \infty $. As was
discussed in Sec. \ref{Asymptotic-limit-section}, these two limits do not
commute. Therefore our results (\ref{A-B-via-L-direct}) -- 
(\ref{A-0-res-Nf-2}) must be taken with a certain care. The calculation of
$A_{B}(g,t)$ made in Sec. \ref{A-B-Nf-2-section} relied on the results of
Sec. \ref{Asymptotic-solution-method-2}, which were based on the explicit
expression for the asymptotic distribution amplitude. Therefore the work of
Sec. \ref{A-B-Nf-2-section} corresponded to the case when the asymptotic
limit $t\rightarrow \infty $ precedes the large-$N_{c}$ limit: the
asymptotic limit selects the contribution corresponding to the lowest
anomalous dimension (depending on the quantum numbers $JJ_{3}TT_{3}$), and
only after that we use the large-$N_{c}$ asymptotic distribution amplitude
corresponding to these chosen quantum numbers $JJ_{3}TT_{3}$. Strictly
speaking, the calculation of Sec. \ref{A-B-Nf-2-section} should be
interpreted as a calculation of the $N_{f}=2$ analog of the functional
$\tilde{A}_{\alpha }(g)$ (\ref{Pfi-A-alpha}), which appears in the problem of
the diagonalization of the anomalous dimensions at large $N_{c}$.

Although the \emph{method} standing behind the calculation in
Sec. \ref{A-B-Nf-2-section} is based on the limit
\begin{equation}
\mathrm{first}\,t\rightarrow \infty, \,\mathrm{then\,}N_{c}\rightarrow \infty
\,,  \label{t-first-Nc-next}
\end{equation}
the \emph{form} of presentation used in Sec. \ref{A-B-Nf-2-section}
corresponds to the limit
\begin{equation}
\mathrm{first}\,N_{c}\rightarrow \infty, \,\mathrm{then\,}t\rightarrow \infty
\,.  \label{Nc-first-t-next}
\end{equation}
Indeed, the functional $A_{B}(g,t)$ is defined only in the limit
$N_{c}\rightarrow \infty $. Before investigating the large-$t$ behavior of
$A_{B}(g,t)$ we must take the large-$N_{c}$ limit in order to define this
functional.

To summarize, the results of the previous section were derived for the
regime (\ref{t-first-Nc-next}). The possibility of the extrapolation of
these results to the region (\ref{Nc-first-t-next}) requires an additional
investigation.

If we stay in the ``safe'' regime (\ref{t-first-Nc-next}) corresponding to
the problem of the diagonalization of the anomalous dimensions, then we can
rewrite Eq. (\ref{A-B-via-L-direct}) in the form similar to the $N_{f}=1$
equation (\ref{A-decomposition-2}):
\begin{equation}
\tilde{A}_{TT_{3}J_{3}}(g)=\sqrt{2T+1}D_{T_{3}J_{3}}^{T}\left[ Q_{0}(g)
\right] \tilde{A}_{0}(g)\,,
\label{A-tilde-TT3J}
\end{equation}
where
\begin{equation}
\tilde{A}_{0}(g)=\left\{ \mathrm{Tr}\left\{ L_{0}^{-1}(g)L_{2}(g)-\left[
L_{0}^{-1}(g)L_{1}(g)\right] ^{2}\right\} \right\} ^{-1/2}\,.
\end{equation}
The zero mode factor $\sqrt{2T+1}D_{T_{3}J_{3}}^{T}\left[ Q_{0}(g)\right] $
in Eq. (\ref{A-tilde-TT3J}) plays the same role as the $N_{f}=1$ zero mode
factor $\left[ \xi _{\mathrm{rot}}(g)\right] ^{J_{3}}$ in
Eq. (\ref{A-decomposition-2}). Equation (\ref{A-tilde-TT3J}) contains no analogs of
the nonzero mode factors $\xi _{m}^{\pm }(g)$ of
Eq. (\ref{A-decomposition-2}) because the representation (\ref{A-tilde-TT3J}) was derived for the
asymptotic wave function corresponding to the lowest anomalous dimension
(for given $TT_{3}J_3$).

The analogy between Eq. (\ref{A-tilde-TT3J}) and the $N_{f}=1$ equations of
Sec. \ref{Calculation-Omega-section} can be continued. For example, the
$N_{f}=2$ zero mode $Q_{0}(g)$ obeys the same equation (\ref{chi-rot-eq}) as
the $N_{f}=1$ zero mode $\xi _{\mathrm{rot}}(g)$:
\begin{equation}
(g\otimes g)\cdot K\cdot \left[ \frac{\delta W_{0}(g)}{\delta g}
\otimes \,\frac{\delta Q_{0}(g)}{\delta g}\right] =0\,.
\end{equation}
The check of this equation is straightforward but tedious. One has to use
expressions (\ref{W0-L0-Nf2}), (\ref{Q-L0-Nf-2}) for the functionals
$W_{0}(g)$, $Q_{0}(g)$ via functionals $L_{k}(g)$ and the identity 
(\ref{L0-L1-id}) for $L_{k}(g)$.

\subsection{Zero modes and the structure of the functional $A_{B}(g)$}
\label{Zero-modes-section}

In Sec. \ref{A-factorization-section} we have suggested the factorized form
(\ref{A-factorization}) for the functional $A_{B}(g)$. However, this naive
factorized form is modified by zero modes. The precise form of this
modification depends on the involved symmetries. For example, in the problem
of the asymptotic distribution amplitude for $N_{f}=1$ flavor we have only
one nontrivial zero mode $\xi _{\mathrm{rot}}(g)$ (\ref{J-zero-mode}). This
zero mode corresponds to the Abelian symmetry associated with the helicity$J_3$.
Therefore the zero mode factor $\left[ \xi _{\mathrm{rot}}(g)\right] ^{J_{3}}$
appearing in Eq. (\ref{A-decomposition-2}) does not violate the naive
factorized form (\ref{A-factorization}) of the functional $A_{B}(g)$.

The case of $N_{f}=2$ quark flavors is more complicated. The results
obtained in this paper give only indirect information about the functional
$A_{B}(g)$. We know the functional $A_{B}(g)$

1) in the toy quark model where it is given by expression (\ref{A-NRQM-res})
for the baryons with $T=J$,

2) in the asymptotic $t\rightarrow \infty $ case described by 
Eq. (\ref{A-tilde-TT3J}), which was derived for baryons with the helicity $J_{3}$
constrained by the condition $|J_{3}|\leq T$.

Comparing Eqs. (\ref{A-NRQM-res}) and (\ref{A-tilde-TT3J}), we see that in
both cases the functional $A_{B}(g)$ has the form
\begin{equation}
A_{T=J,T_{3}J_{3}}(g)=\mathrm{const\,}D_{T_{3}J_{3}}^{T=J}\left[ Q(g)\right]
A_{0}(g)\,.  \label{A-D-Q-general}
\end{equation}
In the toy quark model we have
\begin{equation}
Q(g)=\frac{g}{\sqrt{\det g}}\,,\quad A_{0}(g)=1\quad (\mathrm{
toy\,quark\,model})\,,  \label{Q-A0-NRQM}
\end{equation}
whereas in the case of the asymptotic wave function ($t\rightarrow \infty $)
\begin{equation}
Q(g)=\frac{L_{0}(g)}{\sqrt{\det L_{0}(g)}}\,,\quad A_{0}(g)=\left\{
\mathrm{Tr}\left\{ L_{0}^{-1}(g)L_{2}(g)-\left[ L_{0}^{-1}(g)L_{1}(g)\right]
^{2}\right\} \right\} ^{-1/2}\quad (\mathrm{asymptotic\,regime})\,.
\label{Q-A0-asymptotic}
\end{equation}
In both cases $Q(g)$ is an $SL(2,C)$ matrix.

Comparing the results obtained in the naive quark model (\ref{Q-A0-NRQM})
and in the asymptotic limit (\ref{Q-A0-asymptotic}), we can make the
conjecture that in the nonasymptotic regime the true QCD functional
$A_{TT_{3}J_{3}}(g,t)$ for the lowest $O(N_{c}^{-1})$ excited baryons with
$J=T$ will have the same structure as Eq. (\ref{A-D-Q-general})
\begin{equation}
A_{T=J,T_{3}J_{3}}(g,t)=\mathrm{const}D_{T_{3}J_{3}}^{T=J}
\left[ Q(g,t)\right] A_{0}(g,t)\,,  \label{A-Q-conjecture}
\end{equation}
\begin{equation}
\det Q(g,t)=1\,,  \label{det-Q-g-t-one}
\end{equation}
where $Q(g,t)$ is some unknown functional of $g$ with values in $SL(2,C)$
which is determined by the dynamics of large-$N_{c}$ QCD.

Let us show that the conjecture (\ref{A-Q-conjecture}) and the $SL(2,C)$
constraint (\ref{det-Q-g-t-one}) are compatible with the evolution equation.
Applying the general evolution equation (\ref{A12-evolution}) to the
expression (\ref{A-Q-conjecture}), we see that it takes the form
\begin{equation}
\frac{\partial }{\partial t} \frac{D_{T_{3}^{\prime }J_{3}^{\prime
}}^{T^{\prime }=J^{\prime }}\left[ Q(g,t)\right] }{D_{T_{3}J_{3}}^{T=J}\left[
Q(g,t)\right] }
=-\left\{ \left( g\otimes g\right) \cdot K\cdot\left[  \frac{
\delta W(g,t)}{\delta g}\otimes  \frac{\delta }{\delta g} \frac{
D_{T_{3}^{\prime }J_{3}^{\prime }}^{T^{\prime }=J^{\prime }}\left[ Q(g,t)
\right] }{D_{T_{3}J_{3}}^{T=J}\left[ Q(g,t)\right] }\right] \right\}
\,.  \label{DD-evoluiton}
\end{equation}
Let us show that this equation will hold automatically for any
$T=J,T_{3},J_{3}$ and $T^{\prime }=J^{\prime },T_{3}^{\prime },J_{3}^{\prime
} $ if the functional $Q(g,t)$ obeys the evolution equation
\begin{equation}
\frac{\partial }{\partial t}Q_{fs}(g,t)=-\left\{ \left( g\otimes g\right)
\cdot K\cdot \left[ \frac{\delta W(g,t)}{\delta g}\otimes \frac{\delta
Q_{fs}(g,t)}{\delta g}\right] \right\} \,.  \label{Q-evolution}
\end{equation}
Indeed, Eq. (\ref{Q-evolution}) has an obvious property: If $Q(g,t)$ obeys
this equation, then any function $F(Q)$ depending on the matrix elements
$Q_{fs}$ will obey the same equation:
\begin{equation}
\frac{\partial }{\partial t}F\left[ Q(g,t)\right] =-\left\{ \left( g\otimes
g\right) \cdot K\cdot \left[ \frac{\delta W(g,t)}{\delta g}\otimes \frac{
\delta F\left[ Q(g,t)\right] }{\delta g}\right] \right\} \,.
\label{F-Q-evolution}
\end{equation}
Applying this property to the function
\begin{equation}
F(Q)=\frac{D_{T_{3}^{\prime }J_{3}^{\prime }}^{T^{\prime }=J^{\prime }}(Q)}{
D_{T_{3}J_{3}}^{T=J}(Q)}\,,
\end{equation}
we derive equation (\ref{DD-evoluiton}). If we take
\begin{equation}
F(Q)=\det Q\,,
\end{equation}
then we find from Eq. (\ref{F-Q-evolution})
\begin{equation}
\frac{\partial }{\partial t}\det Q(g,t)=-\left\{ \left( g\otimes g\right)
\cdot K\cdot \left[ \frac{\delta W(g,t)}{\delta g}\otimes \frac{\delta \det
Q(g,t)}{\delta g}\right] \right\} \,.
\end{equation}
Obviously this equation is consistent with the $SL(2,C)$ constraint 
(\ref{det-Q-g-t-one}).

Certainly the consistency of the conjecture (\ref{A-Q-conjecture}), 
(\ref{det-Q-g-t-one}) with the evolution equation and with the asymptotic case
cannot replace the proof of this conjecture. In Ref. \cite{Pobylitsa-05} we
present additional arguments in favor of Eq. (\ref{A-Q-conjecture}) based on
the large-$N_{c}$ spin-flavor symmetry
\cite{Bardakci-84,GS-84,DM-93,Jenkins-93,DJM-94}
and on the soft-pion theorem for the baryon distribution amplitude
\cite{PPS-01}. 
In Ref. \cite{Pobylitsa-05} we also derive the correct version of the naive factorized
ansatz (\ref{A-factorization}) for the $O(N_{c}^{0})$ excited states, taking
into account the modification of Eq. (\ref{A-factorization}) due to the zero
modes.

\section{Conclusions}

This paper shows how the systematic $1/N_{c}$ expansion can be applied to the
baryon wave function. The main idea of the method is to introduce a
generating functional for the baryon wave function and to construct the 
$1/N_{c}$ expansion for this generating functional rather than for the baryon
wave function itself.

A specific feature of this functional is its exponential dependence on 
$N_{c} $. The exponential part of this functional is universal for all
low-lying baryons [including the $O(N_{c}^{-1})$ and $O(N_{c}^{0})$ excited
resonances and meson-baryon scattering states]. The pre-exponential
factors depend on the baryon (baryon-meson) state but these factors have
a simple representation in terms of functionals associated with elementary
excitations (or single meson scattering states).

The developed formalism was applied to the light-cone baryon wave function
(distribution amplitude). It is shown that the structure of the $1/N_{c}$
expansion is compatible with the evolution equation. A nonlinear evolution
equation is derived for the leading (exponential) term of the generating
functional for the baryon distribution amplitude. The nonlinearity of this
equation can be understood in terms of the well-known analogy between the
large-$N_{c}$ limit and the semiclassical approximation. The usual linear
evolution equation is an analog of the quantum nonstationary Schr\"{o}dinger
equation, whereas the nonlinear evolution equation appearing in the
large-$N_{c}$ limit corresponds to the classical Hamilton-Jacobi equation.

The nonlinear evolution equation derived in this paper was successfully used
for the diagonalization of the anomalous dimensions of baryon operators of
the leading twist. This problem was
reduced to the linearized analysis of perturbations of the asymptotic
solution of the Hamilton-Jacobi equation controlling the evolution in the
large $N_{c}$ limit. Although the nonlinear evolution equation has a rather
nontrivial functional form, this problem was solved analytically, and simple
expressions were found for the anomalous dimensions of baryon operators in
two orders of the $1/N_{c}$ expansion.

\acknowledgments
I am grateful to Ya.~I.~Azimov, G. Kirilin,
N.~Kivel, A.~Manashov, V.~Yu.~ Petrov, M.~V.~Polyakov and N.~Stefanis for
interesting discussions. This work was supported by DFG and BMBF.

\appendix

\section{Baryon wave function in the toy quark model}

\label{NRQM-appendix}

\subsection{Wigner $D$ functions}

The quark wave function (\ref{psi-NRQM}) is represented as an integral over
the $SU(2)$ group containing Wigner functions $D_{m_{1}m_{2}}^{j}(R)$ with
the well known properties 
\begin{equation}
D_{m_{1}m_{2}}^{j}(R_{1}R_{2})=\sum
\limits_{m=-j}^{j}D_{m_{1}m}^{j}(R_{1})D_{mm_{2}}^{j}(R_{2}),
\label{D-property-1}
\end{equation}
\begin{equation}
\int dR\left[ D_{m_{1}m_{2}}^{j}(R)\right] ^{\ast }D_{m_{1}^{\prime
}m_{2}^{\prime }}^{j^{\prime }}(R)=\frac{1}{2j+1}\delta _{jj^{\prime
}}\delta _{m_{1}m_{1}^{\prime }}\delta _{m_{2}m_{2}^{\prime }}\,.
\label{D-property-2}
\end{equation}
For the transposed matrix $R^{\mathrm{tr}}$ we have
\begin{equation}
D_{m_{1}m_{2}}^{j}(R^{\mathrm{tr}})=D_{m_{2}m_{1}}^{j}(R)\,.
\label{D-R-transposed}
\end{equation}

The characters of the irreducible $SU(2)$ representations  
\begin{equation}
\chi _{j}(R)=\sum\limits_{m=-j}^{j}D_{mm}^{j}(R)\,  \label{character-1}
\end{equation}
have the properties
\begin{equation}
\chi _{j}(R)=\chi _{j}^{\ast }(R)=\chi _{j}(R^{-1})\,,  \label{character-2}
\end{equation}
\begin{equation}
\chi _{1/2}(R)=\mathrm{Tr}R\,,  \label{character-3}
\end{equation}
\begin{equation}
\int dR\chi _{j_{1}}^{\ast }(R)\chi _{j_{2}}(R)=\delta _{j_{1}j_{2}}\,.
\end{equation}

\subsection{Normalization integral}

\label{Normalization-appendix}

Let us compute the constant $c_{TN_{c}}$ appearing in Eq. (\ref{psi-NRQM}).
This constant $c_{TN_{c}}$ is fixed by the normalization condition 
(\ref{psi-normalization}):
\begin{equation}
\left| c_{TN_{c}}\right| ^{2}\int dR^{\prime }\int
dRD_{J_{3}T_{3}}^{T}(R^{-1})\left[ D_{J_{3}T_{3}}^{T}(R^{\prime -1})\right]
^{\ast }\left[ \mathrm{Tr}\left( RR^{\prime +}\right) \right] ^{N_{c}}=1\,.
\label{c-T-condition}
\end{equation}
Changing the integration variable $R\rightarrow RR^{\prime }$ and using the
properties of the $D$ functions (\ref{D-property-1}) -- (\ref{D-property-2}),
we can simplify this integral:
\begin{align}
& \int dR^{\prime }\int dRD_{J_{3}T_{3}}^{T}(R^{-1})\left[
D_{J_{3}T_{3}}^{T}(R^{\prime -1})\right] ^{\ast }\left[ \mathrm{Tr}\left(
RR^{\prime +}\right) \right] ^{N_{c}}  \notag \\
& =\frac{1}{(2T+1)^{2}}\int dR\sum\limits_{m}D_{mm}^{T}(R^{-1})\left( 
\mathrm{Tr}R\right) ^{N_{c}}\,.
\end{align}
According to Eqs. (\ref{character-1}) -- (\ref{character-3}) we have
\begin{equation}
\sum\limits_{m}D_{mm}^{T}(R^{-1})\left( \mathrm{Tr}R\right) ^{N_{c}}=\chi
_{T}^{\ast }(R)\left[ \chi _{1/2}(R)\right] ^{N_{c}}\,.  \label{D-Tr-chi-chi}
\end{equation}
Now we derive from Eqs. (\ref{c-T-condition}) -- (\ref{D-Tr-chi-chi})
\begin{equation}
c_{TN_{c}}=\frac{2T+1}{\sqrt{I_{N_{c},T}}}\,,  \label{C-T-via-K}
\end{equation}
where
\begin{equation}
I_{N_{c},j}\equiv \int dR\chi _{j}^{\ast }(R)\left( \mathrm{Tr}R\right)
^{N_{c}}=\int dR\chi _{j}^{\ast }(R)\left[ \chi _{1/2}(R)\right] ^{N_{c}}\,.
\label{K-int-def}
\end{equation}
This integral has a simple meaning. It counts how many times the $j$
representation appears in the decomposition of the tensor product of $N_{c}$
spin-$\frac{1}{2}$ representations. Combining this group interpretation of
the $I_{N_{c},j}$ with the standard rules of the spin addition, we arrive at
the recursion relation
\begin{equation}
I_{j,N_{c}+1}=I_{j+\frac{1}{2},N_{c}}+I_{j-\frac{1}{2},N_{c}}
\label{I-recursion}
\end{equation}
with the initial conditions
\begin{equation}
I_{\frac{1}{2},1}=1,\quad I_{-\frac{1}{2},N_{c}}=0\,.  \label{I-initial}
\end{equation}
It is easy to check that the solution of Eqs. (\ref{I-recursion}), 
(\ref{I-initial}) is
\begin{equation}
I_{j,N_{c}}=\frac{N_{c}!(2j+1)}{\left( \frac{N_{c}}{2}+j+1\right) !\left( 
\frac{N_{c}}{2}-j\right) !}\,.  \label{K-jN-res}
\end{equation}
Inserting this result into Eq. (\ref{C-T-via-K}), we find
\begin{equation}
c_{TN_{c}}=\sqrt{\frac{2T+1}{N_{c}!}\left( \frac{N_{c}}{2}+T+1\right)
!\left( \frac{N_{c}}{2}-T\right) !}\,.  \label{c-T-res-0}
\end{equation}

\subsection{Function $\protect\phi_{TT_{3}J_{3}}(g)$}

According to Eq. (\ref{Phi-start-0})
\begin{equation}
\phi _{TT_{3}J_{3}}(g)=c_{TN_{c}}\int dRD_{J_{3}T_{3}}^{T}(R^{-1})\left[ 
\mathrm{Tr}\left( Rg^{\mathrm{tr}}\right) \right] ^{N_{c}}\,.
\end{equation}
In the case when $g$ is an $SU(2)$ matrix we can compute this integral
changing the variable $R\rightarrow R\left( g^{\mathrm{tr}}\right) ^{-1}$
\begin{align}
\phi _{TT_{3}J_{3}}(g)& =c_{TN_{c}}\int dRD_{J_{3}T_{3}}^{T}(g^{\mathrm{tr}
}R^{-1})\left( \mathrm{Tr} R \right) ^{N_{c}}  \notag \\
& =c_{TN_{c}}\sum\limits_{T_{3}^{\prime }=-T}^{T}\int
dRD_{J_{3}T_{3}^{\prime }}^{T}(g^{\mathrm{tr}})D_{T_{3}^{\prime
}T_{3}}^{T}(R^{-1})\left( \mathrm{Tr} R \right) ^{N_{c}}\,.
\end{align}
Using the characters (\ref{character-1}) -- (\ref{character-3}), we can
express this integral in terms of $I_{j,N_{c}}$ (\ref{K-int-def})
\begin{gather}
\int dRD_{T_{3}^{\prime }T_{3}}^{T}(R^{-1})\left( \mathrm{Tr} R
\right) ^{N_{c}}=\frac{\delta _{T_{3}^{\prime }T_{3}}}{2T+1}\int
dR\sum\limits_{m=-T}^{T}D_{mm}^{T}(R^{-1})\left( \mathrm{Tr} R
\right) ^{N_{c}}  \notag \\
=\frac{\delta _{T_{3}^{\prime }T_{3}}}{2T+1}\int dR\chi _{T}(R)\left( 
\mathrm{Tr} R \right) ^{N_{c}}=\frac{\delta _{T_{3}^{\prime
}T_{3}}}{2T+1}I_{T,N_{c}}\,.
\end{gather}

Thus
\begin{equation}
\phi _{TT_{3}J_{3}}(g)=D_{J_{3}T_{3}}^{T}(g^{\mathrm{tr}})\frac{
c_{TN_{c}}I_{T,N_{c}}}{2T+1}\,.
\end{equation}
Using the above result for $c_{TN_{c}}$ (\ref{C-T-via-K}) and the property
of Wigner functions (\ref{D-R-transposed}), we find 
\begin{equation}
\phi _{TT_{3}J_{3}}(g)=\sqrt{I_{T,N_{c}}}D_{T_{3}J_{3}}^{T}(g)\,.
\label{Phi-D-SU2}
\end{equation}
This result is derived for $SU(2)$ matrices $g$. By definition 
(\ref{Phi-start-0}), $\phi _{TT_{3}J_{3}}(g)$ is an analytical function of $g$.
Therefore we can ``analytically continue'' the equality (\ref{Phi-D-SU2}) to
arbitrary $SL(2,C)$ matrices $g$ [assuming the $SL(2,C)$ version of the
Wigner function $D_{T_{3}J_{3}}^{T}(g)$].

Now let us consider arbitrary matrices $g$. According to the definition of 
$\phi _{TT_{3}J_{3}}(g)$ (\ref{Phi-start-0}) we have
\begin{equation}
\phi _{TT_{3}J_{3}}(\lambda g)=\lambda ^{N_{c}}\phi _{TT_{3}J_{3}}(g)\,.
\end{equation}
Therefore
\begin{equation}
\phi _{TT_{3}J_{3}}(g)=\left( \det g\right) ^{N_{c}/2}\phi _{TT_{3}J_{3}} 
\left[ g\left( \det g\right) ^{-1/2}\right] \,.  \label{Phi-g-scaled}
\end{equation}
Since $g\left( \det g\right) ^{-1/2}$ is an $SL(2,C)$ matrix, we can use the
result (\ref{Phi-D-SU2})
\begin{equation}
\phi _{TT_{3}J_{3}}\left[ g\left( \det g\right) ^{-1/2}\right] =\sqrt{
I_{T,N_{c}}}D_{T_{3}J_{3}}^{T}\left[ g\left( \det g\right) ^{-1/2}\right] \,.
\label{Phi-D}
\end{equation}
Combining Eqs. (\ref{Phi-g-scaled}), (\ref{Phi-D}) and using the expression 
(\ref{C-T-via-K}), we find
\begin{equation}
\phi _{TT_{3}J_{3}}(g)=\frac{2T+1}{c_{TN_{c}}}\left( \det g\right)
^{N_{c}/2}D_{T_{3}J_{3}}^{T}\left[ g\left( \det g\right) ^{-1/2}\right] \,.
\label{Phi-NRQM-res-0}
\end{equation}

\section{Evolution kernel}

\label{Evolution-kernel-Appendix}

A detailed discussion of the evolution equation, kernels and anomalous
dimensions for the baryon distribution amplitude can be found in the
original paper \cite{LB-80} and in later publications,
e.g. \cite{BFKL-85,BB-89}. Here we list only those relations
which are used in this paper.

\subsection{$N_{c}$ dependence}

The leading-order evolution equations for the quark distribution amplitudes
of mesons
\begin{equation}
\mu\frac{\partial}{\partial\mu}\Psi_{\mathrm{meson}}^{\mu}(x_{1},x_{2})=-
\frac{N_{c}^{2}-1}{2N_{c}}\frac{\alpha_{s}(\mu)}{\pi}K_{12}\Psi_{\mathrm{
meson}}^{\mu}(x_{1},x_{2})  \label{meson-evolution}
\end{equation}
and baryons
\begin{equation}
\mu\frac{\partial}{\partial\mu}\Psi_{\mathrm{baryon}}^{\mu}(x_{1},\ldots,
x_{N_{c}})=-\frac{N_{c}+1}{2N_{c}}\frac{\alpha_{s}(\mu)}{\pi}\sum
\limits_{1\leq i<j\leq N_{c}}K_{ij}\Psi_{\mathrm{baryon}}^{\mu}(x_{1},\ldots,
x_{N_{c}})  \label{baryon-evolution}
\end{equation}
contain the same ``pair interactions'' $K_{ij}$. The operator $K_{ij}$ acts
on the variables $x_{i},x_{j}$.
The $N_{c}$-dependent factors in Eqs. (\ref{meson-evolution}) and
(\ref{baryon-evolution}) originate from the contraction of the color structure of
the one-gluon exchange
\begin{equation}
\sum\limits_{a=1}^{N_{c}^{2}-1}t_{c_{1}c_{1}^{\prime
}}^{a}t_{c_{2}c_{2}^{\prime }}^{a}=\frac{1}{2}\left( \delta
_{c_{1}c_{2}^{\prime }}\delta _{c_{2}c_{1}^{\prime }}-\frac{1}{N_{c}}\delta
_{c_{1}c_{1}^{\prime }}\delta _{c_{2}c_{2}^{\prime }}\right) \,,
\end{equation}
\begin{equation}
\mathrm{Sp}\left( t^{a}t^{b}\right) =\frac{1}{2}\delta ^{ab}\,
\end{equation}
with the color singlet projectors corresponding to the hadron states.
This gives for mesons
\begin{equation}
\left(
\sum\limits_{a=1}^{N_{c}^{2}-1}t_{c_{1}c_{1}^{\prime}}^{a}t_{c_{2}c_{2}^{
\prime}}^{a}\right) \delta_{c_{1}c_{2}^{\prime}}=\frac{N_{c}^{2}-1}{2N_{c}}
\delta_{c_{1}^{\prime}c_{2}}\,
\end{equation}
and for baryons
\begin{equation}
\left(
\sum\limits_{a=1}^{N_{c}^{2}-1}t_{c_{1}c_{1}^{\prime}}^{a}t_{c_{2}c_{2}^{
\prime}}^{a}\right) \varepsilon_{c_{1}^{\prime}c_{2}^{\prime }c_{3}\ldots
c_{N_{c}}}=-\frac{N_{c}+1}{2N_{c}}\varepsilon_{c_{1}c_{2}c_{3}\ldots
c_{N_{c}}}\,.
\end{equation}

\subsection{Properties of the evolution kernel}

The evolution equations (\ref{meson-evolution}) and (\ref{baryon-evolution})
are written in a compact form in terms of operators $K_{ij}$. The action of 
$K_{12}$ is described by the kernel $\tilde{K}
^{s_{1}s_{2}}(x_{1},x_{2};y_{1},y_{2})$ introduced in Eq. (\ref{K-tilde-def}).
The spin structure of this kernel is
\begin{equation}
\tilde{K}^{s_{1}s_{2}}(x_{1},x_{2};y_{1},y_{2})=\delta
^{s_{1}s_{2}}K^{+}(x_{1},x_{2};y_{1},y_{2})+\delta
^{s_{1},-s_{2}}K^{-}(x_{1},x_{2};y_{1},y_{2})\,.
\end{equation}
We assume that the momentum conserving delta function is included into
the definition of the evolution kernel:
\begin{equation}
\tilde{K}^{s_{1}s_{2}}(x_{1},x_{2};y_{1},y_{2})
\sim\delta(x_1+x_2-y_1-y_2)\,.
\label{K-ss-delta}
\end{equation}

One can introduce homogeneous polynomials \cite{BFKL-85,Ohrndorf-82}
\begin{equation}
\,R_{n+1}(x_{1},x_{2})\,=\,\sum_{k=0}^{n}\frac{n!(n+2)!(-1)^{k}}{
k!(k+1)!(n-k)!(n-k+1)!}\,x_{1}^{k}x_{2}^{n-k}\,,
\label{R-explicit}
\end{equation}
which can be expressed in terms of Gegenbauer polynomials $C_{n}^{3/2}$:
\begin{equation}
R_{n+1}(x_{1},x_{2})=\frac{2}{n+1}(x_{1}+x_{2})^{n}C_{n}^{3/2}\left( \frac{
x_{2}-x_{1}}{x_{1}+x_{2}}\right) \,,  \label{R-n-via-Cn}
\end{equation}
The polynomials $R_{n+1}(x_{1},x_{2})$ diagonalize the evolution kernel:
\begin{equation}
\int_0^\infty dy_{1}\int_0^\infty dy_{2}\tilde{K}
^{s_{1}s_{2}}(x_{1},x_{2};y_{1},y_{2})R_{n+1}(y_{1},y_{2})y_{1}y_{2}=\frac{1
}{2}\gamma_{n}^{s_{1}s_{2}}R_{n+1}(x_{1},x_{2})x_{1}x_{2}\,.
\end{equation}
The anomalous dimensions $\gamma_{n}^{s_{1}s_{2}}$ are
\begin{equation}
\gamma_{n}^{s_{1}s_{2}}=1+4\sum\limits_{k=2}^{n+1}\frac{1}{k}-\frac{2\delta
_{s_{1},-s_{2}}}{(n+1)(n+2)}\,.  \label{gamma-n-my}
\end{equation}
We have
\begin{equation}
\int_0^\infty dy_{1}\int_0^\infty
dy_{2}K^{\pm}(x_{1},x_{2};y_{1},y_{2})R_{n+1}(y_{1},y_{2})y_{1}y_{2}=\frac{1
}{2}\gamma_{n}^{\pm}R_{n+1}(x_{1},x_{2})x_{1}x_{2}\,.
\end{equation}
Here
\begin{equation}
\gamma_{n}^{s_{1}s_{2}}=\delta^{s_{1}s_{2}}\gamma_{n}^{+}+
\delta^{s_{1},-s_{2}}\gamma_{n}^{-},  \label{gamma-s-gamma-pm}
\end{equation}
\begin{equation}
\gamma_{n}^{+}=1+4\sum\limits_{k=2}^{n+1}\frac{1}{k}\,,
\end{equation}
\begin{equation}
\gamma_{n}^{-}=1+4\sum\limits_{k=2}^{n+1}\frac{1}{k}-\frac{2}{(n+1)(n+2)}\,.
\end{equation}

Since the kernel $K^{\pm}(x_{1},x_{2};y_{1},y_{2})$ contains
the delta function (\ref{K-ss-delta}), we immediately conclude that
\begin{equation}
\int_0^\infty dy_{1}\int_0^\infty
dy_{2}K^{
\pm}(x_{1},x_{2};y_{1},y_{2})R_{n+1}(y_{1},y_{2})y_{1}y_{2}f(y_{1}+y_{2})=
\frac{1}{2}\gamma_{n}^{\pm}R_{n+1}(x_{1},x_{2})x_{1}x_{2}f(x_{1}+x_{2})
\label{K-R-f}
\end{equation}
for any function $f$.

In particular,
\begin{align}
& \int_0^\infty dy_{1}\int_0^\infty
dy_{2}K^{\pm}(x_{1},x_{2};y_{1},y_{2})R_{n+1}(y_{1},y_{2})y_{1}y_{2}
\delta(y_{1}+y_{2}-\lambda)  \notag \\
& =\frac{1}{2}\gamma_{n}^{\pm}R_{n+1}(x_{1},x_{2})x_{1}x_{2}\delta
(x_{1}+x_{2}-\lambda)\,,  \label{K-R-action}
\end{align}
\begin{equation}
\int_0^\infty dy_{1}\int_0^\infty
dy_{2}K^{
\pm}(x_{1},x_{2};y_{1},y_{2})R_{n+1}(y_{1},y_{2})y_{1}y_{2}e^{-a(y_{1}+y_{2})}=
\frac{1}{2}\gamma_{n}^{\pm}R_{n+1}(x_{1},x_{2})x_{1}x_{2}e^{-a(x_{1}+x_{2})}
\,.
\end{equation}
Taking $n=0$ in these equations, we obtain
\begin{equation}
\int_0^\infty dy_{1}\int_0^\infty dy_{2}\tilde{K}
^{s_{1}s_{2}}(x_{1},x_{2};y_{1},y_{2})y_{1}y_{2}\delta(y_{1}+y_{2}-\lambda)=
\frac{1}{2}\delta^{s_{1}s_{2}}x_{1}x_{2}\delta(x_{1}+x_{2}-\lambda)\,,
\label{K-tilde-delta}
\end{equation}
\begin{equation}
\int_0^\infty dy_{1}\int_0^\infty dy_{2}\tilde{K}
^{s_{1}s_{2}}(x_{1},x_{2};y_{1},y_{2})y_{1}y_{2}e^{-a(y_{1}+y_{2})}=\frac{1}{
2}\delta^{s_{1}s_{2}}x_{1}x_{2}e^{-a(x_{1}+x_{2})}\,.
\label{K-tilde-x-exp-0}
\end{equation}

The evolution kernel can be decomposed in Gegenbauer polynomials
\begin{align}
4 & \sum\limits_{n=0}^{\infty}\frac{\gamma_{n}^{\varepsilon}}{h_{n}}
C_{n}^{3/2}(2y-1)C_{n}^{3/2}\left( 2x-1\right)  \notag \\
& =-\left( \frac{1}{|x-y|}+\delta_{\varepsilon,-}\right) \left[ \frac{
\theta(y-x)}{(1-x)y}+\frac{\theta(x-y)}{(1-y)x}\right] \quad(x\neq y)\,,
\label{CC-kernel}
\end{align}
where
\begin{equation}
h_{n}=\frac{(n+2)(n+1)}{n+\frac{3}{2}}
\end{equation}
is the normalization constant for Gegenbauer polynomials:
\begin{equation}
\int_{-1}^{1}dz(1-z^{2})C_{m}^{3/2}\left( z\right) C_{n}^{3/2}\left(
z\right) =h_{m}\delta_{mn}\,.  \label{CC-ort}
\end{equation}

Special care is needed for the singularities appearing at $x=y$ in Eq. 
(\ref{CC-kernel}). The precise integral form of this equality is
\begin{align}
4 & \sum\limits_{n=0}^{\infty}\frac{\gamma_{n}^{\varepsilon}}{h_{n}}
C_{n}^{3/2}(2y-1)\int_{0}^{1}dx\,C_{n}^{3/2}\left( 2x-1\right)
x(1-x)\phi(x)  \notag \\
& =-\int_{0}^{1}dx \left[ \delta_{\varepsilon,-}\phi(x)+\frac{
\phi(x)-\phi(y)}{|x-y|}\right] \left[ \frac{x}{y}\theta(y-x)+\frac{1-x}{1-y}
\theta(x-y)\right]  +\frac{1}{2}\phi(y)\,,  \label{CC-sum-phi-action}
\end{align}
where $\phi(x)$ is an arbitrary function.

\section{Functionals $T_{ks}(g)$}

\label{T-properties-appendix}

In this appendix we study the properties of functionals $T_{ks}(g)$. These
functionals are defined by Eq. (\ref{T-def-new}). They play an important
role in Sec. \ref{Calculation-Omega-section} where we solve
Eq.~(\ref{chi-n-eq-1}).

\subsection{Variational derivatives}

Let us start from the variational derivative of the functional $M_{ks}(g)$
(\ref{M-def-new}):

\begin{equation}
\frac{\delta }{\delta g_{s}(y)}M_{ks^{\prime }}(g)=\delta _{ss^{\prime
}}y^{k+1}e^{-X(g)y}-M_{k+1,s^{\prime }}(g)\frac{\delta X(g)}{\delta g_{s}(y)}
\,.  \label{delta-M-delta-X}
\end{equation}
Combining this result with Eq. (\ref{T-def-new}), we find 
\begin{equation}
\frac{\delta T_{ks^{\prime }}(g)}{\delta g_{s}(y)}=\frac{\delta _{ss^{\prime
}}}{M_{0s}(g)}\left[ y^{k+1}-yT_{ks}(g)\right] e^{-X(g)y}-\left[
T_{k+1,s^{\prime }}(g)-T_{ks^{\prime }}(g)T_{1s^{\prime }}(g)\right] \frac{
\delta X(g)}{\delta g_{s}(y)}\,.  \label{delta-T-delta-g}
\end{equation}
The variational derivative of the identity (\ref{T1-sum}) with respect to
$\delta g_{s}(y)$ yields 
\begin{equation}
\sum\limits_{s^{\prime }}\frac{\delta T_{1s^{\prime }}(g)}{\delta g_{s}(y)}
=0\,.  \label{delta-T-delta-g-2}
\end{equation}
We insert Eq. (\ref{delta-T-delta-g}) into Eq. (\ref{delta-T-delta-g-2}): 
\begin{equation}
\sum\limits_{s^{\prime }}\left\{ \frac{\delta _{ss^{\prime }}}{M_{0s}(g)} 
\left[ y^{2}-yT_{1s}(g)\right] e^{-X(g)y}-\left\{ T_{2s^{\prime }}(g)-\left[
T_{1s^{\prime }}(g)\right] ^{2}\right\} \frac{\delta X(g)}{\delta g_{s}(y)}
\right\} =0\,.
\end{equation}
Then 
\begin{equation}
\frac{\delta X(g)}{\delta g_{s}(y)}=\frac{N(g)}{M_{0s}(g)}\left[ y-T_{1s}(g) 
\right] ye^{-X(g)y}\,,  \label{delta-X-dela-g}
\end{equation}
where 
\begin{equation}
N(g)=\left\{ \sum\limits_{s^{\prime }}\left\{ T_{2s^{\prime }}(g)-\left[
T_{1s^{\prime }}(g)\right] ^{2}\right\} \right\} ^{-1}\,.  \label{N-g-def}
\end{equation}

\subsection{Identities for the action of the evolution kernel $K$}

\subsubsection{Action of $K$ on $\protect\delta W_{0}/\protect\delta g\otimes
\protect\delta X/\protect\delta g$}

Now we want to prove the following identity

\begin{equation}
(g\otimes g)\cdot K\cdot \left[ \frac{\delta W_{0}(g)}{\delta g}\otimes \,
\frac{\delta X(g)}{\delta g}\right] =0\,.  \label{gg-K-W-X}
\end{equation}
According to Eqs. (\ref{var-W-eq}) and  (\ref{M-def-new}) we have 
\begin{equation}
\frac{\delta W_{0}(g)}{\delta g_{s}(y)}=\frac{ye^{-X(g)y}}{2M_{0s}(g)}\,.
\label{delta-W-M1}
\end{equation}
Using Eqs. (\ref{delta-X-dela-g}) and (\ref{delta-W-M1}), we find 
\begin{gather}
(g\otimes g)\cdot K\cdot \left[ \frac{\delta W_{0}(g)}{\delta g}\otimes \,
\frac{\delta X(g)}{\delta g}\right]   \notag \\
=\sum\limits_{s_{1}s_{2}}(g_{s_{1}}\otimes g_{s_{2}})\cdot \tilde{K}
^{s_{1}s_{2}}\left\{ \left[ \frac{y_{1}e^{-X(g)y_{1}}}{2M_{0s_{1}}(g)}
\right] \,\left\{ y_{2}e^{-X(g)y_{2}}\frac{N(g)}{M_{0s_{2}}(g)}\left[
y_{2}-T_{1s_{2}}(g)\right] \right\} \right\}   \notag \\
=\frac{N(g)}{2}\sum\limits_{s_{1}s_{2}}\frac{1}{M_{0s_{1}}(g)M_{0s_{2}}(g)}
\left\{ (g_{s_{1}}\otimes g_{s_{2}})\cdot \tilde{K}^{s_{1}s_{2}}\left\{ 
\left[ y_{1}y_{2}e^{-X(g)\left( y_{1}+y_{2}\right) }\right] \,\left[
y_{2}-T_{1s_{2}}(g)\right] \right\} \right\} \,.  \label{K-delta-X-calc-1a}
\end{gather}
In order to compute the action of the operator $K$ we use Eq. (\ref{K-R-f}) 
\begin{align}
& \tilde{K}^{s_{1}s_{2}}\left\{ \left[ y_{1}y_{2}e^{-X(g)\left(
y_{1}+y_{2}\right) }\right] \,\left[ y_{2}-T_{1s_{2}}(g)\right] \right\}  
\notag \\
& =\tilde{K}^{s_{1}s_{2}}\left\{ \left[ y_{1}y_{2}e^{-X(g)\left(
y_{1}+y_{2}\right) }\right] \,\left[ -\,T_{1s_{2}}(g)+\frac{1}{2}\left(
y_{1}+y_{2}\right) +\frac{1}{2}(y_{2}-y_{1})\right] \right\}   \notag \\
& =\frac{1}{2}\left\{ \left[ y_{1}y_{2}e^{-X(g)\left( y_{1}+y_{2}\right) }
\right] \,\left\{ \,\gamma _{0}^{s_{1}s_{2}}\left[ -T_{1s_{2}}(g)+\frac{1}{2}
\left( y_{1}+y_{2}\right) \right] +\frac{1}{2}\gamma
_{1}^{s_{1}s_{2}}(y_{2}-y_{1})\right\} \right\} \,.
\end{align}
Inserting this into Eq. (\ref{K-delta-X-calc-1a}), we obtain 
\begin{align}
& (g\otimes g)\cdot K\cdot \left[ \frac{\delta W_{0}(g)}{\delta g}\otimes \,
\frac{\delta X(g)}{\delta g}\right]   \notag \\
& =\,\frac{N(g)}{4}\sum\limits_{s_{1}s_{2}}\left\{ \gamma
_{0}^{s_{1}s_{2}}\left\{ -T_{1s_{2}}(g)+\frac{1}{2}\left[
T_{1s_{1}}(g)+T_{1s_{2}}(g)\right] \right\} -\frac{1}{2}\gamma
_{1}^{s_{1}s_{2}}\left[ T_{1s_{1}}(g)-T_{1s_{2}}(g)\right] \right\}  
\notag \\
& =\,\frac{N(g)}{8}\sum\limits_{s_{1}s_{2}}\left( \gamma
_{0}^{s_{1}s_{2}}-\gamma _{1}^{s_{1}s_{2}}\right) \left[
T_{1s_{1}}(g)-T_{1s_{2}}(g)\right]
\label{K-delta-X-calc-2a}
\end{align}
Here 
\begin{equation}
\gamma _{0}^{s_{1}s_{2}}-\gamma _{1}^{s_{1}s_{2}}=\gamma
_{0}^{s_{2}s_{1}}-\gamma _{1}^{s_{2}s_{1}}
\end{equation}
is symmetric in $s_{1}$, $s_{2}$ whereas 
\begin{equation}
T_{1s_{1}}(g)-T_{1s_{2}}(g)
\end{equation}
is antisymmetric. Therefore 
\begin{equation}
\sum\limits_{s_{1}s_{2}}\left( \gamma _{0}^{s_{1}s_{2}}-\gamma
_{1}^{s_{1}s_{2}}\right) \left[ T_{1s_{1}}(g)-T_{1s_{2}}(g)\right] =0\,.
\end{equation}
Thus the RHS of Eq. (\ref{K-delta-X-calc-2a}) vanishes. This proves the
identity (\ref{gg-K-W-X}).

\subsubsection{Action of $K$ on $\protect\delta W_{0}/\protect\delta 
g\otimes \protect\delta M_{0s}/\protect\delta g$}

Taking $k=0$ in Eq. (\ref{delta-M-delta-X}) and using Eq. (\ref{gg-K-W-X}),
we find 
\begin{align}
& (g\otimes g)\cdot K\cdot \left[ \frac{\delta W_{0}(g)}{\delta g}\otimes \,
\frac{\delta M_{0s}(g)}{\delta g}\right]   \notag \\
& =\sum\limits_{s_{1}s_{2}}\left\{ (g_{s_{1}}\otimes g_{s_{2}})\cdot \tilde{K
}^{s_{1}s_{2}}\left\{ \frac{\delta W_{0}(g)}{\delta g_{s_{1}}(y_{1})}\,\left[
\delta _{ss_{2}}y_{2}e^{-X(g)y_{2}}\right] \right\} \right\} \,.
\end{align}
Inserting Eq. (\ref{delta-W-M1}), we obtain
\begin{align}
& (g\otimes g)\cdot K\cdot \left[ \frac{\delta W_{0}(g)}{\delta g}\otimes \,
\frac{\delta M_{0s}(g)}{\delta g}\right]   \notag \\
& =\sum\limits_{s_{1}}\frac{1}{2M_{0s_{1}}}(g_{s_{1}}\otimes g_{s})\cdot 
\tilde{K}^{s_{1}s}\left[ y_{1}y_{2}e^{-X(g)\left( y_{1}+y_{2}\right) }\right]
\,.
\end{align}
Here according to Eq. (\ref{K-tilde-x-exp}) 
\begin{equation}
\tilde{K}^{s_{1}s}\left[ y_{1}y_{2}e^{-X(g)\left( y_{1}+y_{2}\right) }\right]
=\frac{1}{2}\delta _{s_{1}s}\left[ y_{1}y_{2}e^{-X(g)\left(
y_{1}+y_{2}\right) }\right] \,.
\end{equation}
Therefore 
\begin{equation}
(g_{s_{1}}\otimes g_{s})\cdot \tilde{K}^{s_{1}s}\left[ y_{1}y_{2}e^{-X(g)
\left( y_{1}+y_{2}\right) }\right] =\frac{1}{2}\delta
_{s_{1}s}M_{0s_{1}}M_{0s}\,.
\end{equation}
Thus 
\begin{equation}
(g\otimes g)\cdot K\cdot \left[ \frac{\delta W_{0}(g)}{\delta g}\otimes \,
\frac{\delta M_{0s}(g)}{\delta g}\right] =\frac{1}{4}M_{0s}(g)\,.
\end{equation}
As a consequence, we obtain the identity 
\begin{equation}
(g\otimes g)\cdot K\cdot \left[ \frac{\delta W_{0}(g)}{\delta g}\otimes \,
\frac{\delta }{\delta g}\frac{M_{0+}(g)}{M_{0-}(g)}\right] =0\,.
\label{M1-pm-zero-mode}
\end{equation}

\subsubsection{Action of $K$ on $\protect\delta W_{0}/\protect\delta 
g\otimes \protect\delta T_{ks}/\protect\delta g$}

Now we want to derive relation (\ref{K-T-action}). According to
Eqs. (\ref{delta-T-delta-g}) and (\ref{gg-K-W-X}) we have 
\begin{align}
& (g\otimes g)\cdot K\cdot \left[ \frac{\delta W_{0}(g)}{\delta g}\otimes \,
\frac{\delta T_{ks}(g)}{\delta g}\right]   \notag \\
& =\sum\limits_{s_{1}s_{2}}(g_{s_{1}}\otimes g_{s_{2}})\cdot \tilde{K}
^{s_{1}s_{2}}\left\{ \frac{\delta W_{0}(g)}{\delta g_{s_{1}}(y_{1})}
\,\left\{ \frac{\delta _{ss_{2}}}{M_{0s}(g)}\left[ y_{2}^{k+1}-y_{2}T_{ks}(g)
\right] e^{-X(g)y_{2}}\right\} \right\} \,.
\end{align}
Inserting Eq. (\ref{delta-W-M1}), we find 
\begin{align}
& (g\otimes g)\cdot K\cdot \left[ \frac{\delta W_{0}(g)}{\delta g}\otimes \,
\frac{\delta T_{ks}(g)}{\delta g}\right]   \notag \\
& =\sum\limits_{s_{1}}\frac{1}{2M_{0s_{1}}(g)M_{0s}(g)}(g_{s_{1}}\otimes
g_{s})\cdot \tilde{K}^{s_{1}s}\left\{ \left[ y_{2}^{k}-T_{ks}(g)\right]
y_{1}y_{2}e^{-X(g)\left( y_{1}+y_{2}\right) }\right\} \,.
\label{K-T-ks-action}
\end{align}

Let us decompose $u^{k}$ in Gegenbauer polynomials $C_{n}^{3/2}\left(
1-2u\right) $. 
\begin{equation}
u^{k}=\sum\limits_{n=0}^{k}\alpha _{nk}C_{n}^{3/2}\left( 1-2u\right) \,.
\label{a-nk-def}
\end{equation}
Taking here 
\begin{equation}
u=\frac{y_{2}}{y_{1}+y_{2}}\,,
\end{equation}
we find 
\begin{equation}
y_{2}^{k}=\sum\limits_{n=0}^{k}\alpha
_{nk}(y_{1}+y_{2})^{k}C_{n}^{3/2}\left( \frac{y_{1}-y_{2}}{y_{1}+y_{2}}
\right) \,.
\label{y2-k-alpha-C}
\end{equation}
In order to compute the action of the operator $K$ in
Eq.~(\ref{K-T-ks-action}) we use Eqs. (\ref{R-n-via-Cn}), (\ref{K-R-f}), and
(\ref{y2-k-alpha-C}): 
\begin{align}
& \tilde{K}^{s_{1}s}\left\{ y_{2}^{k}\left[ \left( y_{1}y_{2}\right)
e^{-X(g)\left( y_{1}+y_{2}\right) }\right] \right\}   \notag \\
& =\frac{1}{2}\left[ \left( y_{1}y_{2}\right) e^{-X(g)\left(
y_{1}+y_{2}\right) }\right] \sum\limits_{n=0}^{k}\alpha _{nk}\gamma
_{n}^{s_{1}s}(y_{1}+y_{2})^{k}C_{n}^{3/2}\left( \frac{y_{1}-y_{2}}{
y_{1}+y_{2}}\right) \,.
\end{align}
Here we deal with a homogeneous polynomial in $y_{1},y_{2}$: 
\begin{equation}
(y_{1}+y_{2})^{k}\sum\limits_{n=0}^{k}\alpha _{nk}\gamma
_{n}^{s_{1}s_{2}}C_{n}^{3/2}\left( \frac{y_{1}-y_{2}}{y_{1}+y_{2}}\right)
=\sum\limits_{m=0}^{k}B_{km}^{s_{1}s_{2}}y_{1}^{m}y_{2}^{k-m}\,.
\label{B-coeff-def}
\end{equation}
We consider this equation as a definition of coefficients $B_{km}^{s_{1}s_{2}}$.
Now we have 
\begin{equation}
\tilde{K}^{s_{1}s}\left\{ y_{2}^{k}\left[ \left( y_{1}y_{2}\right)
e^{-X(g)\left( y_{1}+y_{2}\right) }\right] \right\} =\frac{1}{2}\left[
\left( y_{1}y_{2}\right) e^{-X(g)\left( y_{1}+y_{2}\right) }\right]
\sum\limits_{m=0}^{k}B_{km}^{s_{1}s}y_{1}^{m}y_{2}^{k-m}\,.
\label{K-y2-k}
\end{equation}
According to Eq. (\ref{K-tilde-x-exp}) 
\begin{equation}
\tilde{K}^{s_{1}s}\left[ \left( y_{1}y_{2}\right) e^{-X(g)\left(
y_{1}+y_{2}\right) }\right] =\frac{1}{2}\delta _{s_{1}s}\left[ \left(
y_{1}y_{2}\right) e^{-X(g)\left( y_{1}+y_{2}\right) }\right] \,.
\label{K-y2-0}
\end{equation}
Now we take the difference of Eqs. (\ref{K-y2-k}) and (\ref{K-y2-0})
\begin{align}
& \tilde{K}^{s_{1}s}\left\{ \left[ y_{2}^{k}-T_{ks}(g)\right] \left[ \left(
y_{1}y_{2}\right) e^{-X(g)\left( y_{1}+y_{2}\right) }\right] \right\}  
\notag \\
& =\frac{1}{2}\left[ \left( y_{1}y_{2}\right) e^{-X(g)\left(
y_{1}+y_{2}\right) }\right] \left[ \sum
\limits_{m=0}^{k}B_{km}^{s_{1}s}y_{1}^{m}y_{2}^{k-m}-\delta
_{s_{1}s}T_{ks}(g)\right] \,.
\end{align}
We insert this expression into Eq. (\ref{K-T-ks-action})
\begin{align}
& (g\otimes g)\cdot K\cdot \left[ \frac{\delta W_{0}(g)}{\delta g}\otimes \,
\frac{\delta T_{ks}(g)}{\delta g}\right]   \notag \\
& =\sum\limits_{s_{1}}\frac{1}{4M_{0s_{1}}(g)M_{0s}(g)}(g_{s_{1}}\otimes
g_{s})\cdot \left\{ \left[ \left( y_{1}y_{2}\right) e^{-X(g)\left(
y_{1}+y_{2}\right) }\right] \left[ \sum
\limits_{m=0}^{k}B_{km}^{s_{1}s}y_{1}^{m}y_{2}^{k-m}-\delta
_{s_{1}s}T_{ks}(g)\right] \right\}   \notag \\
& =\sum\limits_{s_{1}}\frac{1}{4M_{0s_{1}}(g)M_{0s}(g)}\left[
\sum\limits_{m=0}^{k}B_{km}^{s_{1}s}M_{ms_{1}}(g)M_{k-m,s}(g)-\delta
_{s_{1}s}T_{ks}(g)M_{0s_{1}}(g)M_{0s}(g)\right]   \notag \\
& =\frac{1}{4}\left[ -T_{ks}(g)+\sum\limits_{s_{1}}\sum
\limits_{n=0}^{k}B_{kn}^{s_{1}s}T_{ns_{1}}(g)T_{k-n,s}(g)\right] \,.
\label{K-T-action-0}
\end{align}
Thus relation (\ref{K-T-action}) is proved.

\subsection{Coefficients $\protect\alpha_{nk}$ and $B_{kn}^{s_{1}s_{2}}$}

\subsubsection{Properties of $\protect\alpha_{nk}$}

Coefficients $\alpha _{nk}$ are defined by Eq. (\ref{a-nk-def}) which can be
rewritten in the form 
\begin{equation}
\left( \frac{1-z}{2}\right) ^{k}=\sum\limits_{n=0}^{k}\alpha
_{nk}C_{n}^{3/2}\left( z\right) \,.  \label{1-z-C-n}
\end{equation}
Using the orthogonality of Gegenbauer polynomials (\ref{CC-ort}), we find 
\begin{equation}
\alpha _{nk}=\frac{1}{h_{n}}\int_{-1}^{1}dz(1-z^{2})\left( \frac{1-z}{
2}\right) ^{k}C_{n}^{3/2}(z)\,.  \label{alpha-explicit-1}
\end{equation}

\subsubsection{Properties of $B_{kn}^{s_{1}s_{2}}$}

Coefficients $B_{kn}^{s_{1}s_{2}}$ are defined by Eq. (\ref{B-coeff-def}).
According to Eq. (\ref{gamma-s-gamma-pm}) the anomalous dimensions $\gamma
_{n}^{s_{1}s_{2}}$ depend only on relative helicities: 
\begin{equation}
\gamma _{n}^{ss}\equiv \gamma _{n}^{+}\,,\quad \gamma _{n}^{s,-s}=\gamma
_{n}^{-}\,.
\end{equation}
The same holds for the coefficients $B_{kn}^{s_{1}s_{2}}$ (\ref{B-coeff-def})
so that we can introduce $B_{kn}^{\pm}$:
\begin{equation}
B_{kn}^{ss}=B_{kn}^{+}\,,\quad B_{kn}^{s,-s}=B_{kn}^{-}\,.
\end{equation}
In Sec. \ref{Calculation-Omega-section} we need the values for
$B_{kk}^{\varepsilon }$ and $B_{k0}^{\varepsilon }$, 
\begin{equation}
B_{kk}^{\varepsilon }=-\frac{2\delta _{\varepsilon ,-}}{(k+1)(k+2)}-\frac{2}{
k(k+1)}\quad (k>0)\,,  \label{B-kk}
\end{equation}
\begin{equation}
B_{k0}^{\varepsilon }=1-\frac{2}{k+2}\delta _{\varepsilon
,-}+2\sum\limits_{j=2}^{k+1}\frac{1}{j}\quad (k\geq 0)\,,  \label{B-k0}
\end{equation}
which will be computed below. At $k=0$ we have 
\begin{equation}
B_{00}^{\varepsilon }=\delta _{\varepsilon ,+}\,.
\end{equation}

In order to derive relations (\ref{B-kk}) and (\ref{B-k0}) we rewrite
Eq. (\ref{B-coeff-def}) in the form 
\begin{equation}
(y_{1}+y_{2})^{k}\sum\limits_{n=0}^{k}\alpha_{nk}\gamma_{n}^{
\varepsilon}C_{n}^{3/2}\left( \frac{y_{1}-y_{2}}{y_{1}+y_{2}}\right) =\sum
\limits_{m=0}^{k}B_{km}^{\varepsilon}y_{1}^{m}y_{2}^{k-m}\,.
\end{equation}
Inserting Eq. (\ref{alpha-explicit-1}), we find 
\begin{gather}
(y_{1}+y_{2})^{k}\sum\limits_{n=0}^{k}\frac{1}{h_{n}}\gamma_{n}^{\varepsilon
}C_{n}^{3/2}\left( \frac{y_{1}-y_{2}}{y_{1}+y_{2}}\right)  \notag \\
\times\int_{-1}^{1}dz(1-z^{2})\left( \frac{1-z}{2}\right)
^{k}C_{n}^{3/2}(z)=\sum\limits_{m=0}^{k}B_{km}^{
\varepsilon}y_{1}^{m}y_{2}^{k-m}\,.
\end{gather}
The summation over $n$ on the LHS can be extended to infinity since
$C_{n}^{3/2}(z)$ is orthogonal to polynomials of degree smaller than $n$: 
\begin{gather}
(y_{1}+y_{2})^{k}\int_{-1}^{1}dz\left( \frac{1-z}{2}\right)
^{k}(1-z^{2})  \notag \\
\times\left[ \sum\limits_{n=0}^{\infty}\frac{1}{h_{n}}\gamma_{n}^{
\varepsilon}C_{n}^{3/2}\left( \frac{y_{1}-y_{2}}{y_{1}+y_{2}}\right)
C_{n}^{3/2}(z)\right] =\sum\limits_{m=0}^{k}B_{km}^{
\varepsilon}y_{1}^{m}y_{2}^{k-m}\,.
\end{gather}
Changing the integration variable 
\begin{equation}
z=2x-1\,,
\end{equation}
we obtain 
\begin{gather}
8(y_{1}+y_{2})^{k}\int_{0}^{1}dx\,x\left( 1-x\right) ^{k+1}  \notag \\
\times\left[ \sum\limits_{n=0}^{\infty}\frac{1}{h_{n}}\gamma_{n}^{
\varepsilon}C_{n}^{3/2}\left( \frac{y_{1}-y_{2}}{y_{1}+y_{2}}\right)
C_{n}^{3/2}(2x-1)\right] =\sum\limits_{m=0}^{k}B_{km}^{
\varepsilon}y_{1}^{m}y_{2}^{k-m}\,.  \label{int-CC-B}
\end{gather}

Taking $\phi (x)=(1-x)^{k}$ in identity (\ref{CC-sum-phi-action}) and
inserting the result into Eq. (\ref{int-CC-B}), we obtain 
\begin{gather}
(y_{1}+y_{2})^{k}\left\{ \left( 1-y\right)
^{k}-2\int_{0}^{1}dx\left\{ \left[ \delta _{\varepsilon ,-}\left(
1-x\right) ^{k}+\frac{\left( 1-x\right) ^{k}-\left( 1-y\right) ^{k}}{|x-y|}
\right] \right. \right. _{y=\frac{y_{1}}{y_{1}+y_{2}}}  \notag \\
\left. \left. \times \left[ \frac{x}{y}\theta (y-x)+\frac{1-x}{1-y}\theta
(x-y)\right] \right\} \right\} =\sum\limits_{m=0}^{k}B_{km}^{\varepsilon
}y_{1}^{m}y_{2}^{k-m}\,.
\end{gather}
Setting $y_{1}=0$ here, we find 
\begin{align}
B_{k0}^{\varepsilon }& =1-2\int_{0}^{1}dx\left[ \delta _{\varepsilon
,-}\left( 1-x\right) ^{k}+\frac{\left( 1-x\right) ^{k}-1}{x}\right] \left(
1-x\right)   \notag \\
& =1-\frac{2}{k+2}\delta _{\varepsilon ,-}+2\sum\limits_{j=2}^{k+1}\frac{1}{j
}\,.
\end{align}
Similarly taking $y_{2}=0$, we obtain 
\begin{equation}
B_{kk}^{\varepsilon }=-2\int_{0}^{1}dx\left[ \delta _{\varepsilon
,-}\left( 1-x\right) ^{k}+\left( 1-x\right) ^{k-1}\right] x=-\frac{2\delta
_{\varepsilon ,-}}{(k+1)(k+2)}-\frac{2}{k(k+1)}\,.
\end{equation}
Thus relations (\ref{B-kk}) and (\ref{B-k0}) are derived.

\end{document}